%% file: ANA-SUSY-2023-22-PAPER.tex
\documentclass[cernpreprint, LANGEDIT=true, UKenglish, texmf, orcidlogo]{atlasdoc}
\usepackage[subcaption,full]{atlaspackage}
\usepackage{atlasbiblatex}
\usepackage{pifont}

\usepackage[BSM,snippets]{atlasphysics}
\graphicspath{{logos/}{figures/}}

\usepackage{ANA-SUSY-2023-22-PAPER-defs}

\usepackage{tikz}
\usetikzlibrary{arrows,shapes}
\usetikzlibrary{trees}
\usetikzlibrary{matrix,arrows} 				%
\usetikzlibrary{positioning}				%
\usetikzlibrary{calc,through}				%
\usetikzlibrary{decorations.pathreplacing}  %
\usepackage{pgffor}							%

\usetikzlibrary{decorations.pathmorphing}	%
\usetikzlibrary{decorations.markings}
\usetikzlibrary{shapes.misc}

\tikzset{
arrow/.style={draw=black, postaction={decorate},
decoration={markings,mark=at position 1. with {\arrow[draw=black]{>}}}},
vector/.style={decorate, decoration={snake}, draw},
provector/.style={decorate, decoration={snake,amplitude=2.5pt}, draw},
antivector/.style={decorate, decoration={snake,amplitude=-2.5pt}, draw},
fermion/.style={draw=black, postaction={decorate},
decoration={markings,mark=at position .55 with {\arrow[draw=black]{>}}}},
fermionred/.style={draw=red, postaction={decorate},
decoration={markings,mark=at position .55 with {\arrow[draw=red]{>}}}},
fermion_/.style={draw=black, postaction={decorate} },
majorana/.style={draw=black},
tau/.style={draw=blue,thick, postaction={decorate},
decoration={markings,mark=at position .55 with {\arrow[draw=blue,thick]{>}}}},
fermionbar/.style={draw=black, postaction={decorate},
decoration={markings,mark=at position .55 with {\arrow[draw=black]{<}}}},
fermionnoarrow/.style={draw=black},
gluon/.style={decorate, draw=black,
decoration={coil,amplitude=4pt, segment length=5pt}},
smallgluon/.style={decorate, draw=black,
decoration={coil,amplitude=2pt, segment length=3pt}},
scalar/.style={dashed,draw=black, postaction={decorate}},
scalarbar/.style={dashed,draw=black, postaction={decorate},
decoration={markings,mark=at position .55 with {\arrow[draw=black]{<}}}},
scalarnoarrow/.style={dashed,draw=black},
electron/.style={draw=black, postaction={decorate},
decoration={markings,mark=at position .55 with {\arrow[draw=black]{>}}}},
bigvector/.style={decorate, decoration={snake,amplitude=4pt}, draw},
photon/.style={decorate, decoration={snake,amplitude=4pt, segment length=5pt}, draw=red},
particle/.style={draw=blue, postaction={decorate}, decoration={markings,mark=at position .5 with {\arrow[draw=blue]{>}}}},
antiparticle/.style={draw=blue, postaction={decorate}, decoration={markings,mark=at position .5 with {\arrow[draw=blue]{<}}}},
cross/.style={cross out, draw=black, minimum size=2*(#1-\pgflinewidth), inner sep=0pt, outer sep=0pt},
cross/.default={3pt}
}

\LEcontact{Bob Kowalewski bob.kowalewski@cern.ch}

\AtlasTitle{Search for new phenomena with top-quark pairs and large missing transverse momentum using 140 $\mathrm{fb}^{-1}$ of $pp$ collision data at $\sqrt{s}=13$ TeV with the ATLAS detector}

\AtlasAbstract{%
A  search is conducted for new phenomena in events with a top quark pair and large missing transverse momentum, where the top quark pair is reconstructed in final states with one isolated electron or muon and multiple jets.
The search is performed using the Large Hadron Collider proton--proton collision data sample at a centre-of-mass energy of  $\sqrt{s}=13$ TeV recorded by the ATLAS detector that corresponds to an integrated luminosity of 140 fb$^{-1}$.  An analysis based on neural network classifiers is optimised to search for directly produced pairs of supersymmetric partners of the top quark (stop), and to search for spin-0 mediators, produced in association with a pair of top quarks,  that decay into dark-matter particles.
In the stop search,  the analysis is designed to target models in which the mass difference between the stop and the neutralino from the stop decay is close to the top quark mass.  This new
search is  combined with previously published searches in final states with different lepton multiplicities.   No significant excess above the Standard Model background is observed, and
limits at 95\%
confidence level are set.
Models with neutralinos with masses up to 570 GeV are excluded,
while for small neutralino masses  models are excluded for stop masses up to 1230 GeV.
Scalar (pseudoscalar) dark matter mediator masses as large as 350 (370) GeV are excluded when the coupling strengths of the mediator to Standard Model and dark-matter particles are both set to one.  At lower mediator masses,  models with production cross-sections as small as 0.15 (0.16)  times the nominal predictions are excluded.  Results of this search are also used to set constraints on effective four-fermion contact interactions between top quarks and neutrinos.  %
}

\AtlasRefCode{SUSY-2023-22}

\PreprintIdNumber{CERN-EP-2024-003}

\AtlasJournal{JHEP}
\AtlasJournalRef{JHEP 03 (2024) 139}
\AtlasDOI{https://doi.org/10.1007/JHEP03(2024)139}

\hypersetup{pdftitle={ATLAS document},pdfauthor={The ATLAS Collaboration}}

\begin{document}

\maketitle

\section{Introduction}

The top quark, as the heaviest particle in the Standard Model (SM), is a key probe in the search for new particles predicted by several extensions of the SM.
Prominent examples of such extensions are Supersymmetry (SUSY) \cite{Golfand:1971iw,Volkov:1973ix,Wess:1974tw,  Wess:1974jb, Ferrara:1974pu,Salam:1974ig,Farrar:1978xj} and models with Dark Matter (DM) particles that interact with top quarks via a spin-0 mediator \cite{DAmbrosio2002,Isidori2012}.  In proton--proton ($pp$) collisions at the Large Hadron Collider (LHC),   both scenarios would produce final states with a pair of top quarks and missing transverse momentum ($\met$) from undetected new particles ($\ttbar+\met$), the signature investigated in this paper.
The $\ttbar$ pair is selected in final states with exactly one isolated charged lepton (electron or muon,\footnote{Electrons and muons from $\tau$-lepton decays are included.} henceforth referred to as `lepton’ or `$\ell$') from the decay of an on-shell $W$ boson, and jets.  Significant missing transverse momentum ($\vec{p}_{T}^\mathrm{miss}$, with magnitude \met), is also required.

Since the discovery of the Higgs boson at the LHC~\cite{HIGG-2012-27,CMS-HIG-12-028},
SUSY has been searched for with particular interest as it may offer a solution to the hierarchy problem~\cite{PhysRevD.13.974,PhysRevD.14.1667,PhysRevD.19.1277,PhysRevD.20.2619}.  SUSY extends the SM by introducing a supersymmetric partner  (sparticle) for each SM particle. The particle and sparticle quantum numbers are identical apart from a half-unit difference in spin.  The underlying supersymmetry is broken, so particle and sparticle masses differ.
In this theory,  the hierarchy problem may be solved by the presence of a light supersymmetric partner of the top quark,  referred to as the top squark or `stop',  which largely cancels the divergent loop corrections to the Higgs boson mass~\cite{DIMOPOULOS1981150,WITTEN1981513,DINE1981575,DIMOPOULOS1981353,Sakai,KAUL198236,BARBIERI198863,DECARLOS1993320}.
The superpartners of the left- and right-handed top quarks mix to form the mass eigenstates \stopone and \stoptwo,  where  $\tilde{t}_1$ is defined as the lighter of the two.
A generic $R$-parity-conserving\footnote{A multiplicative quantum number, referred to as $R$-parity, is introduced to conserve baryon and lepton numbers, where $R$-parity is $1 (-1)$ for all SM (SUSY) particles.} minimal SUSY extension of the SM (MSSM)~\cite{Farrar:1978xj,DIMOPOULOS1981150,Fayet:1976et,Fayet:1977yc,FAYET1979416} predicts pair production of SUSY particles and the existence of a stable lightest SUSY particle (LSP).  In the MSSM models considered in this search the LSP is assumed to be the lightest mass eigenstate of the superpartners of the Higgs boson and  neutral electroweak gauge bosons.  This LSP,  also referred to as the lightest neutralino or \ninoone, may be a DM candidate as it is stable and only interacts weakly with SM particles~\cite{Goldberg:1983nd,Ellis:1983ew}.  This paper presents a search for direct pair production of \stopone particles ($\stopone\stopone$)\footnote{When denoting pair production of a stop and an anti-stop, the charge-conjugation notation is omitted for notational simplicity.} in events with significant \met from the two weakly interacting LSPs that escape detection and the decay products of two on- or off-shell top quarks.
The search for a spin-0 mediator produced in association with top quarks that decays into a pair of DM particles ($\ttbar$+DM) is motivated by SM extensions that respect the principle of minimal flavour violation, which results in the interaction strength between the spin-0 mediator and the SM quarks being proportional to the quark masses via Yukawa-like couplings~\cite{ABERCROMBIE2020100371}.

Previous searches for direct \stopone pair production and for spin-0 mediators produced in association with top quarks and decaying into DM particles were reported by the ATLAS~\cite{SUSY-2018-12,SUSY-2018-07,SUSY-2018-08,SUSY-2019-12} and CMS~\cite{CMS-SUS-19-010,CMS-SUS-19-009,CMS-SUS-19-011,CMS-SUS-20-002} Collaborations using the full LHC Run~2 $pp$ recorded data sample.
This paper presents a new search in final states with exactly one lepton, which were already analysed by the ATLAS Collaboration~\cite{SUSY-2018-07} using the same data sample.
Compared with Ref.~\cite{SUSY-2018-07},  this analysis employs improved  object reconstruction and identification algorithms,  improved background simulations, and new neural-network-based classifiers for the reconstruction of the hadronically decaying top quark and for event discrimination.
In the $\stopone \stopone$ search, this analysis focuses on improving the sensitivity in the challenging scenario where the mass difference between the stop and the lightest neutralino is close to the top quark mass.  The results are statistically combined with those from an independent analysis of the $\ttbar+\met$ final state with no leptons~\cite{SUSY-2018-12} to provide improved sensitivity for scenarios with a heavy stop and a very light neutralino.  In the $\ttbar$+DM search, this analysis is designed to improve the sensitivity across the full range of mediator masses considered and is statistically combined with independent analyses in $\ttbar+\met$  final states with zero  and two leptons.
The data analysed in this search are also used to test for the presence of new interactions between top quarks and neutrinos that can be described by a four-fermion contact interaction (CI) in an Effective Field Theory (EFT) approach.

\section{Signal models and search strategy}

\begin{figure}[ht]
\centering
\begin{subfigure}[b]{0.24\textwidth}
\centering
\includegraphics[width=\textwidth]{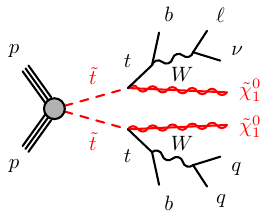}
\caption{$\stopone\stopone$ in two-body decays}
\end{subfigure}
\hfill
\begin{subfigure}[b]{0.24\textwidth}
\centering
\includegraphics[width=\textwidth]{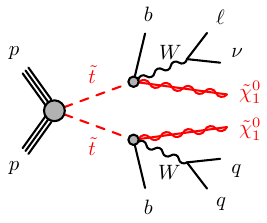}
\caption{$\stopone\stopone$ in  three-body decays}
\end{subfigure}
\hfill
\begin{subfigure}[b]{0.24\textwidth}
\centering
\includegraphics[width=\textwidth]{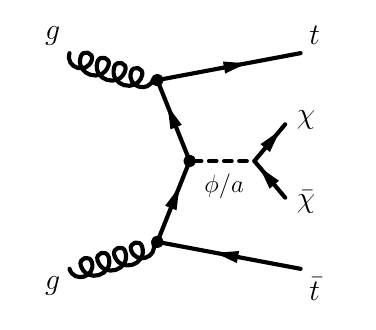}
\caption{$\ttbar+\mathrm{DM}$}
\end{subfigure}
\hfill
\begin{subfigure}[b]{0.24\textwidth}
\centering
\includegraphics[width=\textwidth]{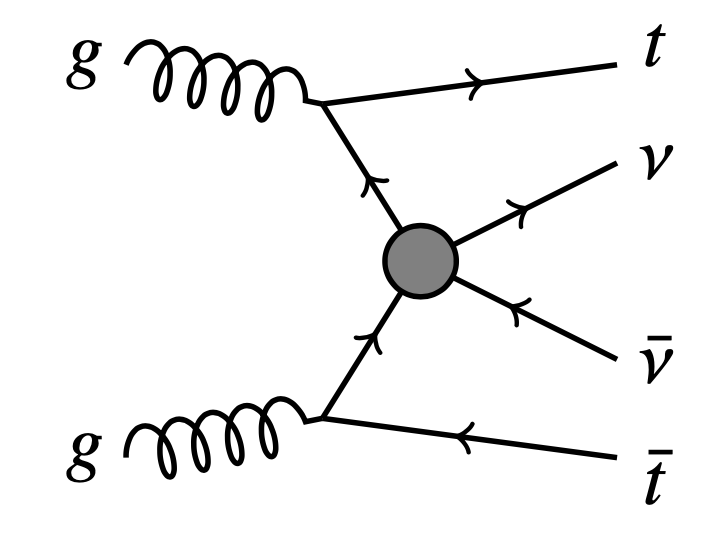}
\caption{$t\bar{t}\nu\bar{\nu}$~CI}
\end{subfigure}
\caption{Diagrams illustrating   direct \stopone pair production with (a) two-body ($\stopone \to t\ninoone$) or (b) three-body ($\stopone \to bW\ninoone$) decays,  (c) the production of DM fermions ($\chi$) in association with a pair of top quarks via a [pseudo] scalar mediator $\phi\,[a]$ and (d) the production of a $t\bar{t}\nu\bar{\nu}$ final state via an effective four-fermion contact interaction.  In the SUSY diagrams, the charge-conjugate symbols are omitted for simplicity.  %
}
\label{fig:diagrams}
\end{figure}

Leading-order  diagrams of the signal processes targeted in this search are shown in Figure~\ref{fig:diagrams}.

The $\stopone\stopone$ production is considered in a simplified SUSY model~\cite{Alwall:2008ve, Alwall:2008ag} where the only light sparticles are \stopone and \ninoone.
Each stop can decay promptly either via the two-body process $\stopone \to t\ninoone$, which produces an on-shell top quark, or via a three-body decay $\stopone \to bW\ninoone$ with an intermediate off-shell top quark.
These are assumed  to be the only possible stop decays.
The $\stopone$ is modelled to decay only in the two-body mode when $\Delta m(\stopone,\ninoone) = m(\stopone) - m(\ninoone) >m(t)$, and only in the
three-body mode when $m(W)<\Delta m(\stopone,\ninoone)<m(t)$.
Decays at smaller $\Delta m(\stopone,\ninoone)$ values, where both the top quark and the $W$ boson are off-shell were explored in Ref.~\cite{SUSY-2018-07}.  In this search, the $\stopone\stopone$ production is tested with a two-dimensional scan of the parameter space $(m(\stopone),  m(\ninoone))$.

The simplified benchmark model for the $\ttbar$+DM production~\cite{SUSY-2016-18,ABERCROMBIE2020100371} assumes the existence of a (pseudo)  scalar mediator $\phi\, (a)$, which can be produced in association with two top quarks and decay into a pair of SM-singlet DM particles ($\chi$).  The interactions of the mediator with SM particles are Yukawa-like.  This model has four parameters: the mass of the mediator $m(\phi/a)$,   the mass of the DM particle $m(\chi)$,  the strength of the DM--mediator interaction $g_\chi$,  and a universal scaling factor of the interactions of the mediator with SM fermions $g_q$.   To reduce the number of free parameters,  $g_\chi$ and $g_q$ are assumed to have the same value, $g=g_\chi=g_q$.
The results are presented as one dimensional scans of  either $m(\phi/a)$ or $m(\chi)$.
The $\ttbar+\met$ final state is mainly sensitive to the mediator mass range $m(\phi/a)<2m(t)$, where the $\phi/a \to \ttbar$ decay is kinematically forbidden.  %

Going beyond simplified models,  the search for new phenomena in $\ttbar+\met$ final states is generalised by searching for an effective $t\bar{t}\nu\bar{\nu}$ four-fermion CI.  Following the model developed in Refs.~\cite{AFIK2022115849, Afik:2020cvr, Afik:2019htr,Afik:2021jjh},  the CI is implemented via four-fermion operators acting on SM fields within the SMEFT theoretical framework~\cite{BUCHMULLER1986621,Arzt:1994gp,Einhorn:2013kja,Grzadkowski:2010es,Brivio:2017vri}.  When considering SM left-handed neutrinos,  out of the possible scalar, vector and tensor operators,  only vector operators can be implemented at dimension six.  The resulting effective Lagrangian is composed of two terms:
\begin{align*}
\mathcal{L}_{t\bar{t}\nu\bar{\nu}} = \frac{1}{\Lambda^2} \left[ V_{LL} (\bar{\nu}\gamma_\mu P_L \nu)(\bar{t}\gamma^\mu P_L t)  +  V_{LR} (\bar{\nu}\gamma_\mu P_L \nu)(\bar{t}\gamma^\mu P_R t  ) \right] \, \,  ,
\end{align*}
where $\Lambda$ is the energy scale of new physics,  $P_{L,R}=(1\mp \gamma_5)/2$ and $V_{LL}$ and $V_{LR}$ are the dimensionless Wilson coefficients of the vector four-fermion interactions with left-handed and right-handed top quarks,  respectively.
Operators of higher dimension, including tensor or scalar operators,  are  neglected\footnote{In this EFT approach with only SM fields,  the couplings of the operators to left-handed neutrinos and to left-handed charged leptons are the same,  so results of this search can combined with results from searches for $t\bar{t}\ell\bar{\ell}$ CIs.  If right-handed neutrinos were to be included,  scalar and tensor operators could have been implemented at dimension six.  These additional operators could be sensitive to  new physics not coupled to SM fields.}.  Results are presented by considering either the ${LL}$ or ${LR}$ operators separately.  The interference between the CI and the SM $\ttbar Z(\to\nu\bar{\nu})$ process is tested by considering the impact of both constructive and destructive interference hypotheses.

The search strategy is designed to retain sensitivity to a wide region of the signal parameter space by partitioning the $\ttbar+\met$ final state into orthogonal  event categories and by using neutral networks (NNs) to classify events.  Event categories are designed to target a specific final state,  rather than a signal  parameter region.    They differ in  the multiplicity of jets with $b$-hadrons  and in the presence or absence of jets produced by hadronic decays of  high-\pt top quarks.  Events from signals in different regions of the parameter space populate these  categories at different rates.
For instance,  events with a pair of top squarks with high mass are predominantly accepted in categories with high-\pt top quarks.  Moreover,   events accepted in the same category from different signal models feature similar kinematic properties.  Background processes also populate  event categories  at different rates, leading to a varying background composition across categories.
These features are exploited by employing several NNs to discriminate between signal and background events.

A key strategy implemented in this analysis is to avoid training different discriminators for different regions of the signal parameter space.  This simplifies the fit, the validation of the background modelling and future reinterpretations of these data using new signal models while retaining good sensitivity for the parameter regions studied.
In each event category,  one dedicated NN is trained with $\stopone\stopone$ or $\ttbar$+DM signal events from the full parameter space  accepted in that category.
The corresponding NN output distributions in each category are used as input to template fits across the full model parameter space.

The most important background processes for this search are
top-quark pair production ($\ttbar$),  the associated production of a top quark and a $W$ boson ($tW$),   $W$+jets production and the production of a $Z$ boson in association with a pair of top quarks and decaying into neutrinos ($\ttbar Z(\to\nu\bar{\nu})$).  Except for the irreducible $\ttbar Z$ background,  which has a small production cross-section and cannot be easily discriminated from signal,  all other major backgrounds are determined in situ in each event category using events classified at low NN output values.  The resulting background model is then validated with events classified using an intermediate range in NN output values where no significant potential signal contribution is expected.  Each event category is therefore split into three regions via orthogonal selections on the NN output value: a Control Region (CR) at low values,  a Validation Region (VR) at intermediate values,  and a Signal Region (SR) at high values.
A combined binned fit to the data in all CRs and SRs is then used to assess the presence of signal events. In this fit,  the distributions of the observed data and of the expected backgrounds remain the same,  while the signal templates change according to the point of the parameter space under test.
Separate NNs are used in the search for top squarks and in the search for DM particles.

\section{ATLAS detector and data collection}

\newcommand{\AtlasCoordFootnote}{%
ATLAS uses a right-handed coordinate system with its origin at the nominal interaction point (IP)
in the centre of the detector and the \(z\)-axis along the beam pipe.
The \(x\)-axis points from the IP to the centre of the LHC ring,
and the \(y\)-axis points upwards.
Cylindrical coordinates \((r,\phi)\) are used in the transverse plane,
\(\phi\) being the azimuthal angle around the \(z\)-axis.
The pseudorapidity is defined in terms of the polar angle \(\theta\) as \(\eta = -\ln \tan(\theta/2)\).
Angular distance is measured in units of \(\Delta R \equiv \sqrt{(\Delta\eta)^{2} + (\Delta\phi)^{2}}\).}

The ATLAS experiment~\cite{PERF-2007-01} at the LHC uses a multipurpose particle detector
with a forward--backward--symmetric cylindrical geometry and a near \(4\pi\) coverage in
solid angle.\footnote{\AtlasCoordFootnote}
The detector consists of an inner tracking detector (ID)  surrounded by a thin superconducting solenoid
that provides a \qty{2}{\tesla} axial magnetic field, electromagnetic (EM) and hadron calorimeters, and a muon spectrometer (MS).
The ID covers the pseudorapidity range \(|\eta| < 2.5\).
The high-granularity silicon pixel detector covers
the vertex region and typically provides four measurements (hits) per track, the first hit normally being in
the insertable B-layer (IBL) that was installed before Run 2~\cite{ATLAS-TDR-19,PIX-2018-001}.
It is surrounded by a silicon microstrip
detector and a straw tube transition-radiation tracking detector.
Lead/liquid-argon (LAr) sampling calorimeters provide EM energy measurements
with high granularity.
A steel/scintillator-tile hadron calorimeter covers the central pseudorapidity range (\(|\eta| < 1.7\)).
The endcap and forward regions are instrumented with LAr calorimeters
for both EM and hadronic energy measurements up to \(|\eta| = 4.9\).
The MS surrounds the calorimeters and is based on
three large superconducting air-core toroidal magnets with eight coils each.
The field integral of the toroids ranges between \num{2.0} and \qty{6.0}{\tesla\metre}
across most of the detector.
The MS includes a system of precision tracking chambers and fast detectors for triggering.
A two-level trigger system is used to select events~\cite{TRIG-2016-01}.
The first-level trigger is implemented in hardware and uses a subset of the detector information
to accept events at a rate up to \qty{100}{\kHz}.
This is followed by a software-based trigger that
reduces the accepted event rate to \qty{1}{\kHz} on average,
depending on the data-taking conditions.
An extensive software suite~\cite{ATL-SOFT-PUB-2021-001} is used in data simulation, in the reconstruction
and analysis of real and simulated data, in detector operations, and in the trigger and data acquisition
systems of the experiment.

The analysis uses the full Run~2 $pp$ data sample delivered by the LHC at $\sqrt{s}=\SI{13}{TeV}$ in the period between 2015 and 2018.
After the application of  data-quality requirements~\cite{DAPR-2018-01},  the total integrated luminosity is \SI{140.1 \pm 1.2}{fb^{-1}}.  The luminosity is measured with an accuracy of 0.83\% \cite{DAPR-2021-01} using the LUCID-2 detector~\cite{LUCID2}, complemented by measurements using the inner detector and calorimeters.  All analysed events were recorded by triggers that accepted events with large \met~\cite{TRIG-2019-01} or with an electron~\cite{TRIG-2018-05} or muon~\cite{TRIG-2018-01}.  The primary
vertex in these events,  defined as the reconstructed vertex with the highest $\sum_\mathrm{tracks}\pt^2$,  must have at least two associated
tracks with $\pt > \SI{500}{MeV}$.

\section{Simulated events}

Samples of Monte Carlo (MC) simulated events are used for the description of SM processes and for the hypothesized signals.
Details of the simulation samples used, including the matrix element (ME) event generator, its accuracy in the strong coupling constant (QCD accuracy),  the parton distribution function (PDF) set used, the parton shower (PS) and hadronisation model,  and the accuracy of the cross-section calculation,
are summarised in Table~\ref{tab:MC}.

\begin{table}[ht]
\caption{
Overview of the configurations used to simulate signal and background processes.  }
\label{tab:MC}
\resizebox{\textwidth}{!}{
\centering
\begin{tabular}{l|lllll}
\toprule
Process & ME event generator  & ME QCD accuracy & ME PDF & Parton shower & Cross-section  \\
& &  &  & and hadronisation & calculation \\
\midrule
SUSY signals & \MADGRAPH[2.8.1],  2.9.9~\cite{Alwall:2014hca} & 0,1,2j@LO & \NNPDF[2.3lo]  & \PYTHIA[8.240], 8.307 & NNLO+NNLL~\cite{Beneke:2009ye,Beenakker:2016lwe, Beenakker:1997ut, Beenakker:2010nq, Beenakker:2016gmf} \\
DM signals & \MADGRAPH[2.7.3] & 0,1j@LO & \NNPDF[2.3lo]  & \PYTHIA[8.244] & NLO~\cite{Mattelaer2015,backovic2015higherorder}   \\
CI signals & \MADGRAPH[2.9.9] & 0,1j@LO & \NNPDF[2.3lo]  & \PYTHIA[8.307] &   \\
\ttbar & \SHERPA[2.2.12]~\cite{Bothmann:2019yzt} & 0,1j@NLO & \NNPDF[3.0nnlo]~\cite{Ball:2014uwa} & \SHERPA~\cite{Schumann:2007mg,Hoeche:2011fd,Hoeche:2012yf,Catani:2001cc,Hoeche:2009rj} &  NNLO+NNLL~\cite{Beneke:2011mq,Cacciari:2011hy,Baernreuther:2012ws,Czakon:2012zr,Czakon:2012pz,Czakon:2013goa,Czakon:2011xx} \\
& & +2,3,4j@LO~\cite{Gleisberg:2008fv,Buccioni:2019sur,Cascioli:2011va,Denner:2016kdg} & & & \\
Single-top & \\
\,\,\, $tW$ & \POWHEGBOX[v2]~\cite{Re:2010bp,Nason:2004rx,Frixione:2007vw,Alioli:2010xd} & NLO & \NNPDF[3.0nlo] &  \PYTHIA[8.307]~\cite{Sjostrand:2014zea} &  NLO+NNLL~\cite{Kidonakis:2010ux,Kidonakis:2013zqa} \\
\,\,\,  s- and t-channel &  \POWHEGBOX[v2] &  NLO &  \NNPDF[3.0nlo] &   \PYTHIA[8.230] & NLO~\cite{Aliev:2010zk,Kant:2014oha} \\
$V$+jets $(V=Z,W)$ & \SHERPA[2.2.11] & 0,1,2j@NLO & \NNPDF[3.0nnlo] &  \SHERPA &  NNLO~\cite{Anastasiou:2003ds} \\
& & +3,4,5j@LO & & & \\
$\ttbar V$ & \MGNLO & NLO & \NNPDF[3.0nlo] &  \PYTHIA[8.210] & NLO QCD+EW~\cite{deFlorian:2016spz} \\
& 2.3.3~\cite{Alwall:2014hca} &  & & & \\
$VV^\prime$ & \SHERPA[2.2.1],  2.2.2 & 0,1j@NLO+2,3j@LO & \NNPDF[3.0nnlo] &  \SHERPA &   \\
\bottomrule
\end{tabular}
}
\end{table}

The samples produced with the \MGNLO~\cite{Alwall:2014hca} and \POWHEGBOX~\cite{Re:2010bp,Nason:2004rx,Frixione:2007vw,Alioli:2010xd} generators used the \EVTGEN~\cite{Lange:2001uf} software to model $b$-hadron decays.  In these samples, the \PYTHIA~\cite{Sjostrand:2014zea} library was used for the parton shower and hadronisation with the A14 tune~\cite{ATL-PHYS-PUB-2014-021} and the \NNPDF[2.3lo] set of PDFs~\cite{Ball:2012cx}.
In all samples,  the effect of multiple interactions in the same and neighbouring bunch
crossings (\pileup) was modelled by overlaying the simulated hard-scattering event with
inelastic \(pp\) collisions generated with \PYTHIA[8.186]~\cite{Sjostrand:2007gs}
using the \NNPDF[2.3lo] set of PDFs and the
A3 set of tuned parameters~\cite{ATL-PHYS-PUB-2016-017}.
All  samples were processed using the full simulation of the ATLAS
detector~\cite{SOFT-2010-01} based on the \GEANT~\cite{Agostinelli:2002hh} software.

The $\stopone\stopone$ production was simulated with the \MADGRAPH~\cite{Alwall:2014hca} package at leading order (LO) with up to two additional partons.  The simplified SUSY model used was the same as in Ref.~\cite{SUSY-2018-07}.
Stop decays were simulated with the \MADSPIN~\cite{Artoisenet:2012st} software, which emulates kinematic distributions without calculating the full ME.
The  $\stopone\stopone$ production cross-sections were calculated to approximate next-to-next-to-leading order in QCD, adding the resummation of soft gluon emission at next-to-next-to-leading-logarithm accuracy (approximate NNLO+NNLL)~\cite{Beneke:2009ye,Beenakker:2016lwe,Beenakker:1997ut, Beenakker:2010nq, Beenakker:2016gmf}.   The nominal cross-section and its uncertainty were derived using the PDF4LHC15\_mc PDF set, following the recommendations of Ref.~\cite{Butterworth:2015oua}.

The $\ttbar$+DM production was simulated with $g=g_\chi=g_q=1$ using \MADGRAPH at LO with up to one additional parton and with \MADSPIN for the decay of top quarks.  The kinematics of the mediator decay were found to not depend strongly on the values of the couplings; however, the particle kinematic
distributions are sensitive to the scalar or pseudoscalar nature of the mediator and to the mediator and DM particle masses.
The cross-sections were computed at next-to-leading order (NLO) in QCD~\cite{Mattelaer2015,backovic2015higherorder}.

The $\ttbar\nu\bar{\nu}$ CIs with either ${LL}$ and ${LR}$ operators were simulated with two samples,  one for the pure contribution from new physics (`NP') and one for the interference with the SM (`Int'),  so that the total contribution is estimated as
\begin{align*}
\sigma_{\ttbar\nu\bar{\nu}} =  \frac{V_{ij}}{\Lambda^2} \sigma_\mathrm{Int} +  \frac{V^2_{ij}}{\Lambda^4} \sigma_\mathrm{NP}  \,\,  ,
\end{align*}
where the sign of $V_{ij}$ determines whether the interference is constructive or destructive.
These samples were simulated with \MADGRAPH at LO with up to one additional parton using the model provided by  the authors of Ref.~\cite{AFIK2022115849}.  In simulation the model parameters were set to $\Lambda=\SI{1}{TeV}$ and $V_{ij}=1$.  These values scale the predicted cross-section,  but not the kinematics of the simulated events.  The ${LL}$ and ${LR}$ operators were simulated separately.  Assuming $\Lambda=\SI{1}{TeV}$,  $V_{ij}=\pm4\pi$ and considering all three generations of neutrinos,  the cross-sections of the NP and interference contributions for the ${LL}$ (${LR}$) operator were computed to be \SI{384}{fb} (\SI{365}{fb}) and $\pm\SI{17}{fb}$ ($\mp\SI{6.4}{fb}$),  respectively.

The  \ttbar  and \(W+\)jets processes were simulated with the
\SHERPA[2.2.12] and \SHERPA[2.2.11]~\cite{Bothmann:2019yzt}
generators,  respectively.  The simulation of \ttbar production used NLO MEs with up to one additional parton,  and  LO MEs
with up to four additional partons calculated using the Comix~\cite{Gleisberg:2008fv}
and \OPENLOOPS~\cite{Buccioni:2019sur,Cascioli:2011va,Denner:2016kdg} libraries.   Similarly,  the \(W+\)jets production used NLO MEs with up to two additional partons,  and  LO MEs
with up to five additional partons.
Approximate NLO EW corrections were included through event weights using the electroweak virtual approximation and the additive prescription.
These simulations were interfaced to the \SHERPA parton shower~\cite{Schumann:2007mg} using the \MEPSatNLO
prescription~\cite{Hoeche:2011fd,Hoeche:2012yf,Catani:2001cc,Hoeche:2009rj}
with the set of tuned parameters developed by the \SHERPA authors.
The \(tW\) production was modelled  by the \POWHEGBOX[v2]~\cite{Re:2010bp,Nason:2004rx,Frixione:2007vw,Alioli:2010xd}
generator at NLO in QCD using the five-flavour scheme. The diagram removal scheme~\cite{Frixione:2008yi} was used to
remove interference and overlap with \ttbar production. The  uncertainty related to this scheme was estimated by comparison with an alternative sample
generated using the diagram subtraction scheme~\cite{Frixione:2008yi,ATL-PHYS-PUB-2016-020}.
The production of a gauge boson with a top quark pair ($\ttbar W$ and $\ttbar Z$)  was modelled using the
\MGNLO[2.3.3]~\cite{Alwall:2014hca} generator at NLO.
Other sources of background events that are considered in this search are  diboson production ($VV^{\prime}$,  where $V=W,Z$),   Higgs boson production in association with a top quark pair,   $tZ$ and $tWZ$ production,  and the production of three and four top quarks.

Further details of the simulation  are provided in Table~\ref{tab:MC}.
A major difference from the simulations used in Ref.~\cite{SUSY-2018-07} is the use of \SHERPA instead of \POWHEGBOX[v2] for \ttbar production, which provides improved modelling of events in the high transverse momentum region relevant for this search.
Another difference is the use of dynamic rather than fixed scales in the MEs used for the simulation of the $tW$ production.

\section{Event reconstruction}

The event categorisation and discrimination is based on reconstructed objects:  leptons, jets (including those containing $b$-hadrons)  and \met.
Depending on object quality and kinematic requirements,  reconstructed electrons and muons are labelled
as either \emph{baseline} or \emph{signal}, where the latter is a subset of the former that passes tighter selection criteria.
Baseline leptons are used to classifying overlapping objects, to compute the missing
transverse momentum, and for background rejection.  Signal leptons are used to select events and construct discriminating variables.  %
Hadronically decaying top quarks are reconstructed using dedicated algorithms.

Electron candidates are reconstructed from an isolated electromagnetic calorimeter energy deposit
matched to a track in the ID~\cite{EGAM-2018-01}.  The pseudorapidity of the calorimeter energy cluster must
satisfy $ |\eta_\mathrm{cluster}| < 2.47$.  Baseline electrons are required to have $\pt>\SI{4.5}{GeV}$ and to satisfy loose
likelihood identification criteria.   Furthermore, their longitudinal impact parameter ($z_0$), defined as the distance along the beam direction between the primary vertex and the point of closest approach of the track to the beam axis,  must satisfy $|z_0 \sin \theta| < \SI{0.5}{mm}$,  where $\theta$ is
the polar angle of the track.  Signal electrons must satisfy, in addition to the baseline requirements, $\pt>\SI{20}{GeV}$ and $|d_0|/\sigma_{d_0} < 5$, where $d_0$ is the transverse
impact parameter with uncertainty $\sigma_{d_0}$.
Signal electrons are also required to be isolated and satisfy tighter identification criteria.  The isolation is defined as the sum of the transverse energy or momentum reconstructed in a cone of size $\Delta R =\sqrt{ (\Delta \eta)^2 + (\Delta\phi)^2}$ around the electron, excluding the energy of the electron itself.  The isolation criteria depend on the electron \pt.

Muon candidates are reconstructed either by combining tracks in the ID with tracks
in the MS or as stand-alone MS tracks~\cite{PERF-2015-10,MUON-2018-03}, and are required to have $|\eta| < 2.7$.
Baseline muons are required to have $\pt>\SI{4}{GeV}$,  $|z_0 \sin \theta| < \SI{0.5}{mm}$,  and to satisfy the `Medium' identification criterion.   In addition to all baseline requirements,   signal muons must  have $\pt>\SI{20}{GeV}$ and  $|d_0|/\sigma_{d_0} < 3$.  Signal muons must also be isolated,  based on criteria similar to those used for electrons.

Particle-flow (PFlow) jets in the region $|\eta|< 4.5$ are reconstructed using the anti-$k_t$ algorithm~\cite{Cacciari:2008gp} with
a radius parameter $R = 0.4$~\cite{Fastjet},  using neutral PFlow constituents and charged constituents associated
with the primary vertex as input~\cite{PERF-2015-09}.   They are referred to hereafter as `small-$R$ jets'.
These jets are calibrated to the particle level by applying a jet
energy scale derived from simulation with in situ corrections based on data~\cite{JETM-2018-05}.
Jets with $\pt > \SI{20}{GeV}$ are selected.
A cleaning procedure is used to identify and remove jets arising from calorimeter noise or
non-collision backgrounds. To suppress pile-up jets with $|\eta|<2.4$,  a discriminant called the `jet vertex tagger'
(JVT) is constructed using a likelihood method~\cite{PERF-2014-03}.
A similar discriminant, the `forward JVT' (fJVT), is used for jets with $|\eta|>2.5$~\cite{ATL-PHYS-PUB-2019-026}.
PFlow jets with $|\eta|< 2.5$ identified as containing $b$-hadrons are referred to as $b$-tagged jets.
This identification is based on the DL1r $b$-tagging algorithm~\cite{FTAG-2018-01,FTAG-2019-07} with a working point that corresponds to a $b$-tagging efficiency of 77\%\ as measured in a sample of
simulated \ttbar events.  The corresponding rejection factors are approximately 200 and 6 for light-quark
or gluon jets and $c$-jets, respectively.

To reconstruct high-\pt hadronic top quark decays,  `large-$R$ jets' are used~\cite{JETM-2018-06}.
These jets are reconstructed using the anti-$k_t$ algorithm with $R = 1.0$, taking as input topological clusters of calorimeter cells calibrated to the hadronic scale~\cite{ATL-LARG-PUB-2009-001-2}.
These jets are trimmed~\cite{Krohn:2009th} to remove energy deposits from pile-up and the underlying event.
The energy and mass of the large-$R$ jets are calibrated using data~\cite{JETM-2018-02,ATLAS-CONF-2020-022}.
Large-$R$ jets with $|\eta|<2$ are selected if they have a transverse momentum in the range $[600,2500]$~GeV and a mass in the range $[40,600]$~GeV.  Large-$R$ jets with $\pt<\SI{600}{GeV}$ are not considered, since low-\pt hadronic top quark decays are reconstructed from small-$R$ jets, as described in Section~\ref{sec:topNN}.
A multivariate classifier is used to identify the large-$R$ jets from hadronic top decays using a working point with a top-tagging efficiency of 80\%~\cite{JETM-2018-03, ATL-PHYS-PUB-2020-017}.  This classifier uses  jet substructure observables as input.
To improve the identification of hadronic top decays,  jets containing $b$-hadrons are reconstructed inside the large-$R$ jet using  a track-based jet reconstruction with a variable cone size~\cite{Krohn:2009zg} and are tagged with the DL1r  $b$-tagging algorithm.

Hadronically decaying $\tau$-leptons are reconstructed and used to veto background.
The reconstruction of jets from these $\tau$ decays ($\tau_\mathrm{had-vis}$)~\cite{ATL-PHYS-PUB-2022-044,ATLAS-CONF-2017-029} is seeded from  jets reconstructed from topological clusters using the anti-$k_t$ algorithm with $R = 0.4$.  Tracks that originate from the primary vertex and satisfy requirements on the impact parameter and on a multivariate discriminant  are associated with the  $\tau_\mathrm{had-vis}$ if they lie within a cone of angle $\Delta R<0.2$ around the jet axis.  Reconstructed $\tau_\mathrm{had-vis}$ candidates must have one or three associated tracks with a total charge equal to $\pm1$.  They must satisfy $\pt>\SI{20}{GeV}$ and be reconstructed in the pseudorapidity ranges $|\eta|<1.37$ or $1.52<|\eta|<2.5$.  Furthermore,  reconstructed $\tau_\mathrm{had-vis}$ need to satisfy loose identification criteria aimed at rejecting jets that originate from quarks, gluons or electrons.

The missing transverse momentum $\vec{p}_{T}^\mathrm{miss}$~\cite{ATLAS-CONF-2018-023} is reconstructed as the negative vector sum of the $\vec{p}_T$
of all selected electrons, muons and small-$R$ jets.  An extra track-based `soft term' is
built using additional tracks associated with the primary vertex, but not with any reconstructed object,  to improve the performance of the $\vec{p}_{T}^\mathrm{miss}$ reconstruction at high pile-up.  %
An object-based missing transverse momentum significance~\cite{ATLAS-CONF-2018-038} is used to identify events in which the reconstructed $\vec{p}_{T}^\mathrm{miss}$ comes from undetected or weakly interacting particles rather than from detected particles with mis-measured energies.

Correction factors are applied to  reconstructed objects to take account of  differences in energy scale and resolution and in trigger,  reconstruction,  identification and isolation efficiencies  between data and  simulation.
Ambiguities among  selected objects are resolved via an overlap removal procedure similar to the one used in Ref.~\cite{SUSY-2018-07}.
Given a set of baseline objects, the procedure checks for overlaps based on  shared tracks,
ghost-matching~\cite{MatteoCacciari2008}, or a minimum distance $\Delta R_y$ between pairs of objects.\footnote{
Rapidity $y \equiv \frac{1}{2} \ln \frac{E+p_z}{E-p_z}$ is used instead of pseudorapidity ($\eta$) when computing the distance $\Delta R_y$.}
Small-$R$ jets with $\Delta R<1.1$ from a large-$R$ jet are discarded and  no overlap removal is performed between leptons and large-$R$ jets.

\subsection{Resolved hadronic top decays}
\label{sec:topNN}

Jets in signal events can arise from top quark decays, additional radiation, pile-up and the underlying event.
A NN-based algorithm,  hereafter referred to as the `top-NN',  is used to identify sets of small-$R$ jets associated with the hadronic decay of a top quark.
This algorithm targets resolved decays, in which the jets from the top-quark decay are spatially separated and can be reconstructed as individual small-$R$ jets.  The target \pt range is below the $600$~GeV threshold used for the large-$R$--jet reconstruction of hadronically decaying top candidates.
The momentum reconstructed using the top-NN algorithm is used in the event classification.

The top-NN evaluates all two- and three-jet combinations in an event.  While  a fully reconstructed resolved decay would result in three jets,
jet pairs are also considered to allow for cases in which one jet is outside acceptance or where the two jets from a high-\pt $W$ boson are reconstructed as a single small-$R$ jet.  The two- and three-jet combinations are formed from a set of jets that includes up to two $b$-tagged jets and up to four not-$b$-tagged jets (referred to as `light jets' hereafter) in the event.  Each candidate jet combination must contain exactly one $b$-tagged jet.
To facilitate the use of jet four-momentum components in the top-NN and remove spurious degrees of freedom a set of transformations is used.
These transformations include a Lorentz boost into the rest frame of the jet combination and a rotation such that the $b$-tagged jet is aligned along the $z$-axis and the leading-\pt  light jet lies in the $xz$-plane with a positive $p_x$ component.  After these transformations, six non-trivial jet momentum components together with the \pt of the original jet combination are used as NN inputs.  The top-NN is a binary feed-forward NN trained with \textsc{Keras}~\cite{chollet2015keras} with the Tensorflow backend.  It is trained using SUSY signal events drawn from all simulated points of the parameter space in order to include a large range of top quark \pt.   Jet combinations that are kinematically matched to the partons from the top-quark decay are labelled signal and all other combinatorics of jets are labelled background.
In the training,  weights are assigned so that both signal and background candidates are uniformly distributed in \pt. In each event the top-NN assigns an output value to all combinations of selected jets.

The combination with the highest top-NN output value in each event is chosen as the hadronic top quark candidate.  According to simulation,  the top-NN reaches the highest efficiency  in finding the correct jet combination for top quark \pt between 200 and \SI{600}{GeV}.  In this \pt range,  the top-NN identifies the right daughter jets in about 70\% of the events in which a correctly tagged $b$-tagged jet and at least one light jet are kinematically matched to the hadronically decaying top quark.
Above this range the performance worsens, as the decay products are less likely to be reconstructed as individual small-$R$ jets.  These high \pt decays are better reconstructed using large-$R$ jets.  Therefore the top-NN is used to reconstruct the hadronic top candidate only in events that have no large-$R$ jet with $\pt>\SI{600}{GeV}$.  Events selected by the top-NN are further required to have an output value above a minimum threshold to ensure the presence of a good hadronic top quark candidate.
\section{Event classification}
\label{sec:classification}

Events are collected using both \met and  single lepton triggers, and different thresholds on the reconstructed \met and signal lepton \pt are applied to select events in the kinematic range where these triggers reach  high efficiency.  Reconstructed events with $\met>\SI{230}{GeV}$ are required to have satisfied the \met trigger.  Events with \met in the range $[70,230]$~GeV are instead required to have satisfied a single lepton trigger.  For these events, the signal electron (muon) associated with the lepton selected at trigger level must have $\pt>27\, (27.5)$~GeV and $|\eta|<2.47 \, (2.4)$.  All events are required to have exactly one or three signal leptons and no additional baseline lepton or $\tau_\mathrm{had-vis}$ candidate.  Events with three signal leptons are only used for the validation of the $\ttbar Z$ background.

Events that satisfy the selection are classified into mutually exclusive categories: high-\met events with $\met>\SI{230}{GeV}$, and `boosted' events with a large-$R$ jet.  If an event satisfies the criteria for both categories, it is assigned to the boosted category.  These  categories target events with top-quark decays at different transverse momenta.  Events are further categorised according to the number of $b$-tagged jets and the presence of a top-tagged large-$R$ jet.  The requirements for these categories are summarised in Table~\ref{tab:categories}.  Events not accepted in any of these categories are not further considered.

\begin{table}[ht]
\caption{Summary of the selections for each event category  together with the definitions of the hadronic and leptonic top quark candidates.  Trigger,  lepton and \met selections are described in the text.  Large-$R$ jets are referred to as `\textit{lR}' for brevity. }%
\label{tab:categories}
\centering
\resizebox{\columnwidth}{!}{
\begin{tabular}{l|c|c|c|c|c|c|c|c}
\toprule
\multirow{2}{*}{ Analysis Category}  &  \multicolumn{2}{c|}{High-\met} & \multicolumn{6}{c}{Boosted} \\
&    1b & 2b & 1b-lep-0t & 1b-had-0t & 2b-0t & 1b-lep-1t & 1b-had-1t & 2b-1t\\
\midrule
$N$(\textit{lR} jet,  $\pt>\SI{600}{GeV}$) & \multicolumn{2}{c|}{0}  & \multicolumn{6}{c}{  $\geq 1$} \\
$N$(top-tagged \textit{lR} jet) & \multicolumn{2}{c|}{-}   & \multicolumn{3}{c|}{0} & \multicolumn{3}{c}{$\geq 1$}\\
$N_{b\mathrm{-jet}}$ with $\Delta R(b,\mathrm{\textit{lR}\ jet})<1.1$  & \multicolumn{2}{c|}{-} &  0 & $\geq1$ & $\geq1$ & 0 & $\geq1$ & $\geq1$\\
$N_{b\mathrm{-jet}}$ with $\Delta R(b,\mathrm{\textit{lR}\ jet})>1.1$  & \multicolumn{2}{c|}{-} &    $\geq1$ & 0 & $\geq1$ &  $\geq1$ & 0 & $\geq1$ \\
\midrule
top-NN-tagged multiplet &    \multicolumn{2}{c|}{\checkmark}  &\multicolumn{6}{c}{-}\\
$N_{b\mathrm{-jet}}$ &    1 &   $\geq2$ & \multicolumn{6}{c}{-} \\
$N_\mathrm{light-jet}$ &   $\geq2$ & $\geq1$ & \multicolumn{6}{c}{-} \\
\midrule
$\mathrm{top}_\mathrm{had}$ candidate & \multicolumn{2}{c|}{top-NN multiplet} & \multicolumn{6}{c}{\textit{lR} jet} \\
$\mathrm{top}_\mathrm{lep}$ candidate &   $\ell+\mathrm{j}$ & $\ell+b$ & $\ell+b$ & $\ell(+\mathrm{j})$ & $\ell+b$ & $\ell+b$ & $\ell(+\mathrm{j})$ & $\ell+b$\\
\midrule
Event NN selection & \multicolumn{8}{c}{See Table~\ref{tab:CRthr}} \\
\bottomrule
\end{tabular}
}
\end{table}

High-\met events are placed in `1b' or `2b' categories according to the number of $b$-tagged jets.  A minimum number of light jets is required so that a top-NN-tagged jet combination can be defined by at least two jets and an additional jet can be used for the reconstruction of the leptonically decaying top quark.  %
In the 2b (1b) categories,  the hadronic top quark candidate is formed by the combination of jets that give the highest top-NN output value,  while the  leptonic  top quark candidate comprises the lepton and the $b$-tagged (light) leading-\pt jet that is not associated with the hadronic top quark candidate.  In the 1b category,  the leading-\pt jet is used to build the leptonic top quark candidate to retain a good kinematic description of the events in which the jet from the $b$-quark in the leptonic top-quark decay is not correctly tagged.

Boosted events must have at least one large-$R$ jet.  In events with more than one, only the leading-\pt large-$R$ jet is considered for event selection and classification.  These boosted events are first categorised according to whether the large-$R$ jet is top-tagged (`1t') or not (`0t'), and then according to the numbers of $b$-tagged jets inside and outside the large-$R$ jet.  The `2b' events have at least one $b$-tagged jet inside the large-$R$ jet ($\Delta R(b,\mathrm{large\textnormal{-}}R\ \mathrm{jet})<1.1$) and at least one $b$-tagged jet outside.
The `1b-had' events have at least one $b$-tagged jet inside the large-$R$ jet and none outside.  Finally,  the  `1b-lep' events have at least one $b$-tagged jet outside the large-$R$ jet and none inside.  This classification defines six orthogonal boosted categories.  In these events,  the hadronic top quark candidate is defined by the large-$R$ jet momentum,  while the  leptonic top quark momentum is the vectorial sum of the momenta of the lepton and the  leading-\pt $b$-tagged jet outside the large-$R$ jet.  In cases where this $b$-tagged jet is not present,  the leading-\pt light jet,  if present, is used.

Overall,  eight orthogonal categories (two high-\met and six boosted) in different kinematic regimes and with different reconstructed objects are used to maximise the signal acceptance.
In each high-\met category,  two NN-based event classifiers,  one for the $\stopone\stopone$ search (`stop-NN') and one for the $\ttbar$+DM search (`DM-NN'),  are  employed to discriminate signal from background events.
In each boosted category,  only a dedicated stop-NN is employed.
This is because $\ttbar$+DM signals in the mediator mass range targeted in this search are not expected to yield sizeable fractions of events in the boosted regions.
Each NN is trained with all simulated events accepted in a given category, including background processes and signals from across the parameter space.  For the stop-NNs,  only signal events from two-body decays are used in the training.  For the DM-NNs,  signal events with both scalar and pseudoscalar mediators are used together.  Each event is weighted by the predicted cross-section of the parent process.  The  distribution  of the output value of either the stop-NN or the DM-NN in each category is used in a fit to the observed data to assess the presence of signal events across the parameter space.
Events generated from different signal model parameters populate different event categories at different rates, but events accepted into a particular category have similar kinematic properties.
For instance,  $\stopone\stopone$ events with large $\Delta m(\stopone,\ninoone)$ are accepted at similar rates in the boosted and high-\met categories, while events with small $\Delta m(\stopone,\ninoone)$ are accepted almost exclusively in the high-\met categories.
When compared with more complex approaches, in which  NNs are trained for specific signal model parameters,
this approach retains high classification power across the parameter space with the simplicity of having only one distribution per event category to fit.

\begin{figure}[ht]
\centering
\includegraphics[width=0.48\textwidth]{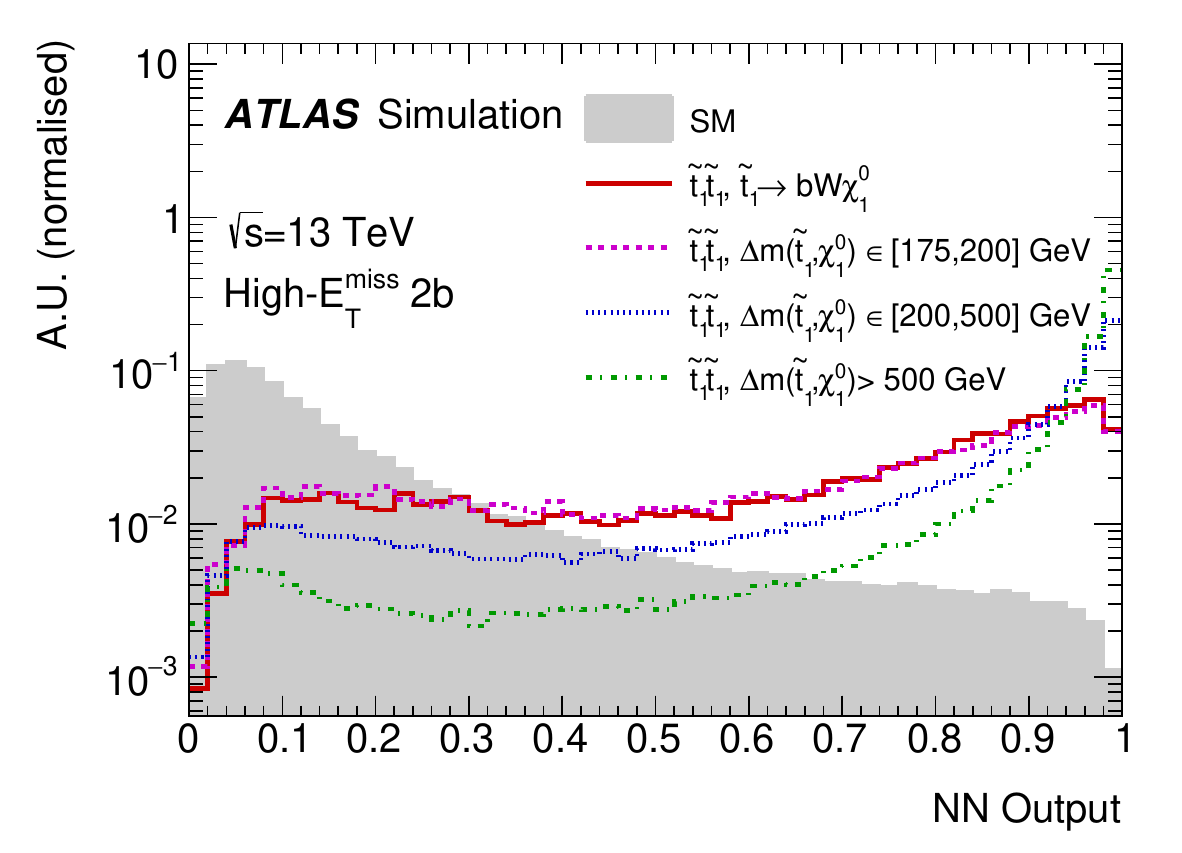}
\includegraphics[width=0.48\textwidth]{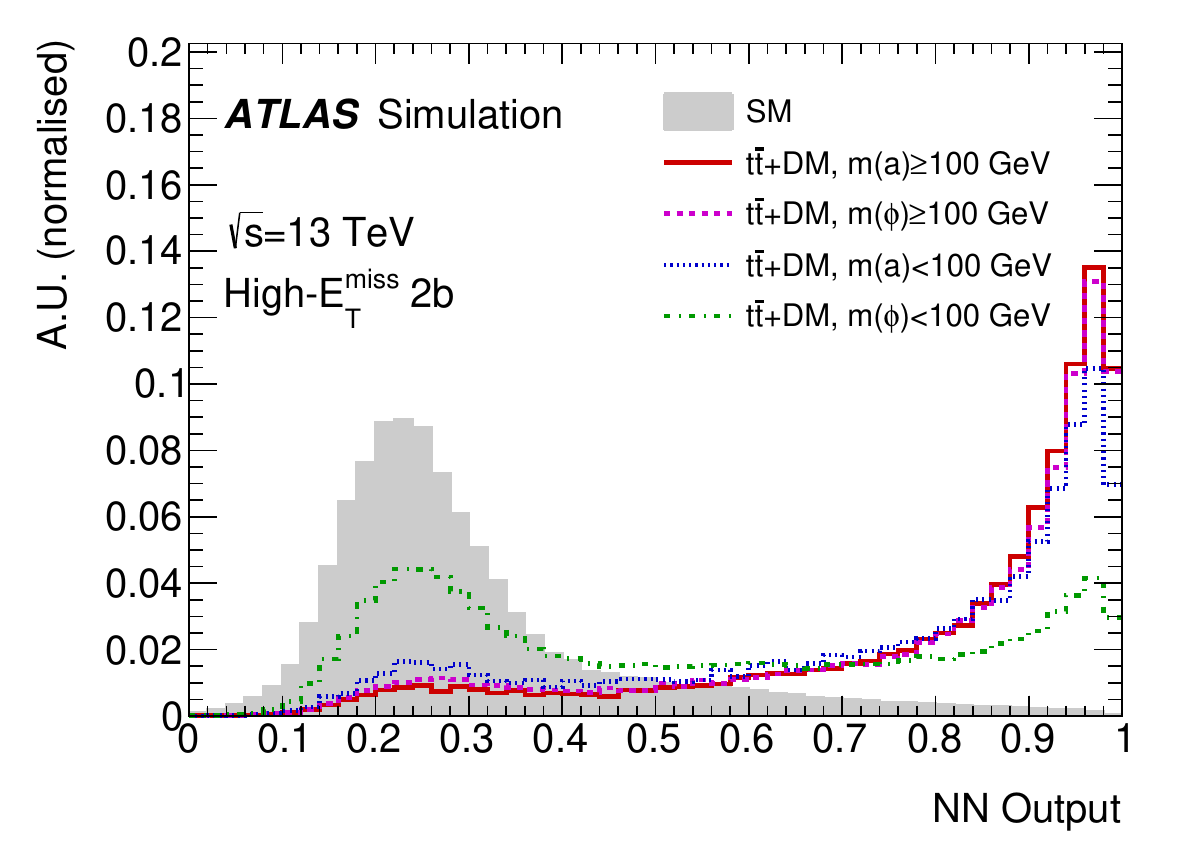}
\caption{Expected distributions of events as a function of the output value of the stop-NN (left) and of the DM-NN (right) in the high-\met 2b category.  The distribution of the SM background processes is compared with classes of signal models.  In the left plot,  $\stopone\stopone$ models are grouped according to $\Delta m(\stopone,\ninoone)$ and the \stopone decay mode.  In the right plot,  $\ttbar$+DM models with a scalar $\phi$ or pseudoscalar $a$ mediator are shown for mediator masses larger or smaller than \SI{100}{GeV} and $m(\chi)=\SI{1}{GeV}$.    Distributions are normalised to the same integral.}
\label{fig:NNscore}
\end{figure}

The stop- and DM-NNs are binary feed-forward NNs.  In each event category the sets of inputs are the same for the stop- and DM-NNs,  but these sets differ across categories.   A set of `low-level' inputs is used by all NNs in all categories.  It includes the momentum components of the hadronic and leptonic top quark candidates,  the momentum components of $\vec{p}_{T}^\mathrm{miss}$,  and the momentum components of the lepton and of the $b$-tagged jets associated with the two  top quark candidates.\footnote{In the high-\met 1b category,  the momentum components of the light jet associated with the leptonic  top candidate are used.  In the boosted 1b-lep categories,  the $b$-tagged jet associated with the hadronic  top quark candidate is omitted.} Transformations in the plane transverse to the beam direction are used to remove rotational symmetries and reduce the input dimensionality.  Additional `high-level' variables include the \met significance,  the top-NN output value (only for the high-\met categories),  and kinematic variables such as the transverse mass  $m_T(\vec{p}_T^\ell,\vec{p}_T^\mathrm{miss})=\sqrt{2 | \vec{p}_T^\ell | |\vec{p}_T^\mathrm{miss}|(1-\cos\Delta\phi(\vec{p}_T^\ell, \vec{p}_T^\mathrm{miss}))}$ and the stransverse mass~\cite{Cheng:2008hk}:
\begin{equation*}
m_{T2}(\vec{p}_T^1,  \vec{p}_T^2, \vec{p}_T^\mathrm{miss})= \min_{\vec{q}_T^1+\vec{q}_T^2=\vec{p}_T^\mathrm{miss}} \left\lbrace
\max \left[  m_T(\vec{p}_T^1,\vec{q}_T^1)  ,    m_T(\vec{p}_T^2,\vec{q}_T^2)    \right]
\right\rbrace   \,\, ,
\end{equation*}
where $\vec{p}_T^i$ is the transverse momentum of a visible particle and $\vec{q}_T^i$ is the transverse momentum of an undetected particle,  assumed to be massless,  which contributes to \met.
The variants of $m_{T2}$ used in this search are $m_{T2}(b,b,\met)$ and $m_{T2,\mathrm{min}}(b+\ell,b,\met) = \min \left[ m_{T2}(b_1+\ell,b_2,\met), m_{T2}(b_1,b_2+\ell,\met)  \right]$,  where $b$ indicates the $b$-tagged (or light) jet associated with one of the two top quark candidates.  The distributions of these variables are expected to have endpoints around $m(W)$ or $m(t)$ in \ttbar events,  while reaching higher values in signal events with more undetected particles.  The $m_{T2}$ variables are not used as NN inputs in the boosted categories.  In total, depending on the event category,  the number of NN inputs varies from 19 to 26.
The classification power in each event category depends on the signal source.
SUSY signals at  large stop masses have the largest classification power, with output values  close to unity,  while  signals with small $\Delta m(\stopone,\ninoone)$ or from the three-body regime have lower classification power.   Among DM signals,  events with a pseudoscalar mediator have higher classification power than events with a scalar mediator for the same mediator mass.  Moreover,  when considering mediators with the same parity,  signals with small $m(\phi/a)$ have a lower classification power than signals with large $m(\phi/a)$.
To illustrate these features,  Figure~\ref{fig:NNscore} shows the  expected distributions of events as a function of the NN output value in the high-\met 2b category for SM background processes and for classes of signal models.
The analysis retains sensitivity to all classes of signal events by performing a binned fit to the data distribution of the NN output value,  as described in Section~\ref{sec:bkg}.

The  stop-NN and DM-NN output values are used to split each event category into an SR,  a CR and a VR.  The SR contains events at high output values with the highest purity of signal events.  The CR contains events at low  output values with negligible contributions from potential signals.  These CR events are used to improve the modelling of the background normalisations and templates in situ in each event category.  The VR contains events in the intermediate range  which are used to check the validity of the background model, derived at low output values, when applied to background events at higher values.  %
For each SR,   the lower threshold on the NN output value is set based on its binned distribution and on  the maximum of the  signal-to-background ratio ($s/b^\mathrm{max}_\mathrm{bin}$) expected in each bin for all signals not excluded by Ref.~\cite{SUSY-2018-07}.\footnote{For the DM signals,  given the previously excluded cross-sections,  the expected $s/b^\mathrm{max}_\mathrm{bin}$ in each bin is evaluated assuming $g=0.5$.}
As the $s/b^\mathrm{max}_\mathrm{bin}$ increases monotonically,  the minimum threshold for the SR is chosen as the maximum output value below which all bins have $s/b^\mathrm{max}_\mathrm{bin}<0.1$.
The lower threshold for the VR is chosen in the same way with $s/b^\mathrm{max}_\mathrm{bin}<0.05$.  In case the resulting expected number of events in the VR is less than 40,  this threshold is lowered in order to reach a number of events suitable to validate the background model.
Events with NN output values below the VR lower threshold are accepted into the CR.  In some categories,   a lower threshold is set for the CR, and events below that threshold are rejected and not used for further analysis.  These rejected events have background compositions that are not representative of the background events in the SR and may contain multijet events, which are not simulated.
With this approach,  all CRs,  VRs and SRs are orthogonal and span the full range of the NN output value above a minimum threshold.  Table~\ref{tab:CRthr} reports a summary of these selections, together with the signal acceptance efficiencies in the SRs.

\begin{table}[ht]
\caption{Summary of the selections on the stop-NN and DM-NN output values that define  CRs, VRs and SRs.  Signal  efficiencies, computed as the fraction of signal events in a given  category with a NN output value in the range accepted in the SR, are also reported.
The quoted range encompasses efficiencies estimated for all signals across the simulated parameter space.  In boosted categories,  only  efficiencies for $\stopone\stopone$ signals with $\Delta m(\stopone,\ninoone)>\SI{500}{GeV}$ are quoted.
}%
\label{tab:CRthr}
\centering
\resizebox{\columnwidth}{!}{
\begin{tabular}{l|l|l|ll|l|l|ll}
\toprule
& \multicolumn{4}{c|}{stop-NN} &  \multicolumn{4}{c}{DM-NN} \\
Category & CR & VR & \multicolumn{2}{l|}{SR} & CR & VR & \multicolumn{2}{l}{SR} \\
& Range & Range & Range & Eff.   & Range & Range & Range & Eff.    \\
\midrule
High-\met 1b & $[0.2,0.64)$ & $[0.64, 0.79)$ & $[0.79, 1.0]$ & 0.4-0.9 &$[0.3,0.69)$ & $[0.69,0.87)$ & $[0.87, 1.0]$ & 0.3-0.4\\
High-\met 2b & $[0.1,0.56)$ & $[0.56, 0.70)$ & $[0.70, 1.0]$ & 0.5-0.9 &$[0.3,0.60)$ & $[0.60,0.76)$ & $[0.76, 1.0]$ & 0.6-0.8\\
Boosted 1b-lep-1t & $[0.0,0.65)$ & $ [0.65,0.80)$ & $[0.80, 1.0]$ & 0.5-0.9 & \multicolumn{4}{c}{ } \\
Boosted 1b-had-1t & $[0.0,0.65)$ &  $[0.65,0.85)$ & $[0.85, 1.0]$ & 0.6-0.9 & \multicolumn{4}{c}{ }\\
Boosted 2b-1t & $[0.0,0.75)$ & $ [0.75,0.95)$ & $[0.95, 1.0]$ & 0.6-0.8 &\multicolumn{4}{c}{ }\\
Boosted 1b-lep-0t & $[0.0,0.70)$ & $ [0.70,0.85) $ & $[0.85, 1.0]$ & 0.6-0.8 &\multicolumn{4}{c}{ }\\
Boosted 1b-had-0t & $[0.0,0.75)$ & $ [0.75,0.95) $ & $[0.95, 1.0]$ & 0.4-0.8 & \multicolumn{4}{c}{ }\\
Boosted 2b-0t & $[0.0,0.65)$ & $ [0.65,0.80) $ & $[0.80, 1.0]$ & 0.6-0.9&\multicolumn{4}{c}{ } \\
\bottomrule
\end{tabular}
}
\end{table}

\section{Background estimation and statistical model}
\label{sec:bkg}

The dominant sources of background events are processes that produce \ttbar, single-top (including $tW$) and $W$+jets.
Despite its small cross-section, $\ttbar Z(\to\nu\bar{\nu})$ production constitutes another important background as it yields events with  signal-like kinematic properties.
All these backgrounds,  along with other minor contributions,  are estimated from simulation and corrected based on a fit to CRs in data.
According to simulation,  the fraction of background events  in which the selected signal lepton is not a real prompt lepton from a $W$ boson or $\tau$ lepton decay is  small.  This contribution is estimated from simulation with a conservative systematic uncertainty.
Multijet events are expected to yield negligible contributions in all categories after the selection on the minimum NN output value and are therefore neglected in the background estimate.

Events from  \ttbar production are modelled via two  classes: the semileptonic \ttbar events with up to one prompt lepton (`\ttbar-1L')
and the fully leptonic \ttbar events with at least two prompt leptons (`\ttbar-2L').  The \ttbar-1L events have the same final state as the signal events,  but a lower \met due to the presence of only one neutrino from  $W$-boson decays.
The  \ttbar-2L events,  in which  only one  lepton is selected,   have a similar level of \met to signal events,  but differ in the properties of the hadronic top quark candidate, which is built from the additional jet activity in the event.
Given the different kinematics and the higher sensitivity of the \ttbar-2L events to the modelling of additional radiation,  the \ttbar-1L and the \ttbar-2L backgrounds are estimated separately.  Their prediction is based on the same simulation,  but their in situ estimates are independent.

The predicted event yields from \ttbar, single-top and $W$+jets are corrected in situ by fitting the observed events in  CRs.  This correction is determined jointly with the fit to the observed events in SRs to test the presence of signal events.  This fit uses a binned maximum-likelihood approach implemented in the statistical analysis packages \textsc{RooFit}~\cite{RooFit}, \textsc{RooStats}~\cite{RooStats} and \textsc{HistFitter}~\cite{Baak:2014wma}.  The statistical model  is composed of observed and expected binned distributions in CRs and SRs.
In  SRs,  events are binned by the output values of either the stop-NNs or the DM-NNs,  depending on the type of signal being tested.  In  CRs,  events are binned by \mt multiplied by the lepton electric charge ($\mt \times q(\ell)$) as this variable allows  the individual contributions of the \ttbar-1L,  \ttbar-2L, and $W$+jets backgrounds to be determined simultaneously.
The \mt variable is used because events with one leptonically decaying $W$ boson (eg,  \ttbar-1L and $W$+jets events) are distributed differently from events with more (eg,  \ttbar-2L) as the former feature an endpoint around $m_W$,  while the latter extend to higher values.
The lepton charge is  used to discriminate  between  \ttbar-1L and $W$+jets events by exploiting the difference in the production cross-section for $W^+$ and $W^-$ bosons.  %
The contribution from single-top events is determined mainly in dedicated CRs.  The boosted-2b-1t and boosted-2b-0t CRs are split  with a selection on $m_{T2,\mathrm{min}}(b+\ell,b,\met)$ at \SI{300}{GeV} into `low-$m_{T2}$' and `high-$m_{T2}$' regions,  with the high-$m_{T2}$ region being enhanced in single-top events.
Given the small number of events in the high-$m_{T2}$ regions,  only their total event yields are considered in the fit.

Overall,  the statistical model  used for the $\stopone\stopone$ search includes ten CRs and
eight SRs, as summarised in Table~\ref{tab:NF}.  In the $\ttbar$+DM search, only the high-\met SRs are considered since no sizeable signal yield is expected in the boosted SRs.  However,  the boosted categories are still used to improve the determination of the single-top background.  Therefore the statistical model for the $\ttbar$+DM includes the two high-\met SRs with their associated CRs and the eight boosted CRs used in the $\stopone\stopone$ search.
In these fits,  the normalisations of the \ttbar-1L,  \ttbar-2L, single-top and $W$+jets contributions are scaled by independent unconstrained fit parameters (normalisation factors or `NFs'),  as reported in Table~\ref{tab:NF}.  %
Via these NFs and the experimental and theoretical systematic uncertainties described in Section~\ref{sec:sys},  the background predictions based on simulation are corrected in situ with the data in  CRs and then used to estimate the background contributions in VRs and SRs.
The samples of events in VRs are useful to validate the background predictions since they have background compositions  similar to those in SRs and only small potential signal contributions.  The observed and predicted distributions in CRs, VRs and SRs are shown in Section~\ref{sec:result}.

\begin{table}[ht]
\caption{Summary of the statistical model  in terms of event categories,  category subdivisions, observables  used in the template fit and normalisation factors (`NFs') applied to background contributions in each region.  `NN' stands for the NN output value.  The background process to which a NF applies is indicated by the name of the NF.  All ten CRs and eight SRs are used in the fit for the $\stopone\stopone$ search.  The SRs in the boosted categories are not used in the statistical model for the $\ttbar$+DM search. }%
\label{tab:NF}
\centering
\resizebox{\columnwidth}{!}{
\begin{tabular}{l|ll|ccccccc}
\toprule
Category &  \multicolumn{2}{c|}{Fitted observable}  & \multicolumn{7}{c}{Normalisation Factors} \\
&  CR  & SR  & $\mathrm{NF}^\mathrm{High-met}_\mathrm{top-1L}$ & $\mathrm{NF}^\mathrm{Boosted}_\mathrm{top-1L}$ & $\mathrm{NF}^\mathrm{High-met}_\mathrm{top-2L}$ & $\mathrm{NF}^\mathrm{Boosted}_\mathrm{top-2L}$ & $\mathrm{NF}^\mathrm{High-met}_W$ & $\mathrm{NF}^\mathrm{Boosted}_W$ & $\mathrm{NF}_\mathrm{singletop}$ \\
\midrule
High-\met 1b    &   $\mt \times q(\ell)$ & NN       & \checkmark &                      & \checkmark &                       &  \checkmark &                       & \checkmark \\
High-\met 2b    &   $\mt \times q(\ell)$  & NN         & \checkmark &                      & \checkmark &                        &  \checkmark &                        & \checkmark \\
Boosted 1b-lep-1t    &  $\mt \times q(\ell)$  & NN &                       & \checkmark &                      & \checkmark &                        & \checkmark & \checkmark \\
Boosted 1b-had-1t   & $\mt \times q(\ell)$  & NN &                      & \checkmark &                       & \checkmark &                        & \checkmark & \checkmark\\
\multirow{ 2}{*}{Boosted 2b-1t}           & $\mt \times q(\ell)$  (low-$m_{T2}$)   & \multirow{ 2}{*}{NN} &                      & \multirow{ 2}{*}{\checkmark} &                      & \multirow{ 2}{*}{\checkmark} &                          & \multirow{ 2}{*}{\checkmark} & \multirow{ 2}{*}{\checkmark} \\
& Yield  (high-$m_{T2}$)  &   &             &   &                      &  &                          &   &  \\
Boosted 1b-lep-0t    &   $\mt \times q(\ell)$  & NN &                      & \checkmark &                      & \checkmark &                          & \checkmark & \checkmark \\
Boosted 1b-had-0t   &   $\mt \times q(\ell)$  & NN &                      & \checkmark &                      & \checkmark &                          & \checkmark & \checkmark\\
\multirow{ 2}{*}{Boosted 2b-0t}           & $\mt \times q(\ell)$  (low-$m_{T2}$)   & \multirow{ 2}{*}{NN} &                      & \multirow{ 2}{*}{\checkmark} &                      & \multirow{ 2}{*}{\checkmark} &                          & \multirow{ 2}{*}{\checkmark} & \multirow{ 2}{*}{\checkmark} \\
& Yield  (high-$m_{T2}$)    &   &             &   &                      &  &                          &   &  \\
\bottomrule
\end{tabular}
}
\end{table}

\begin{figure}[ht]
\centering
\includegraphics[width=0.48\textwidth]{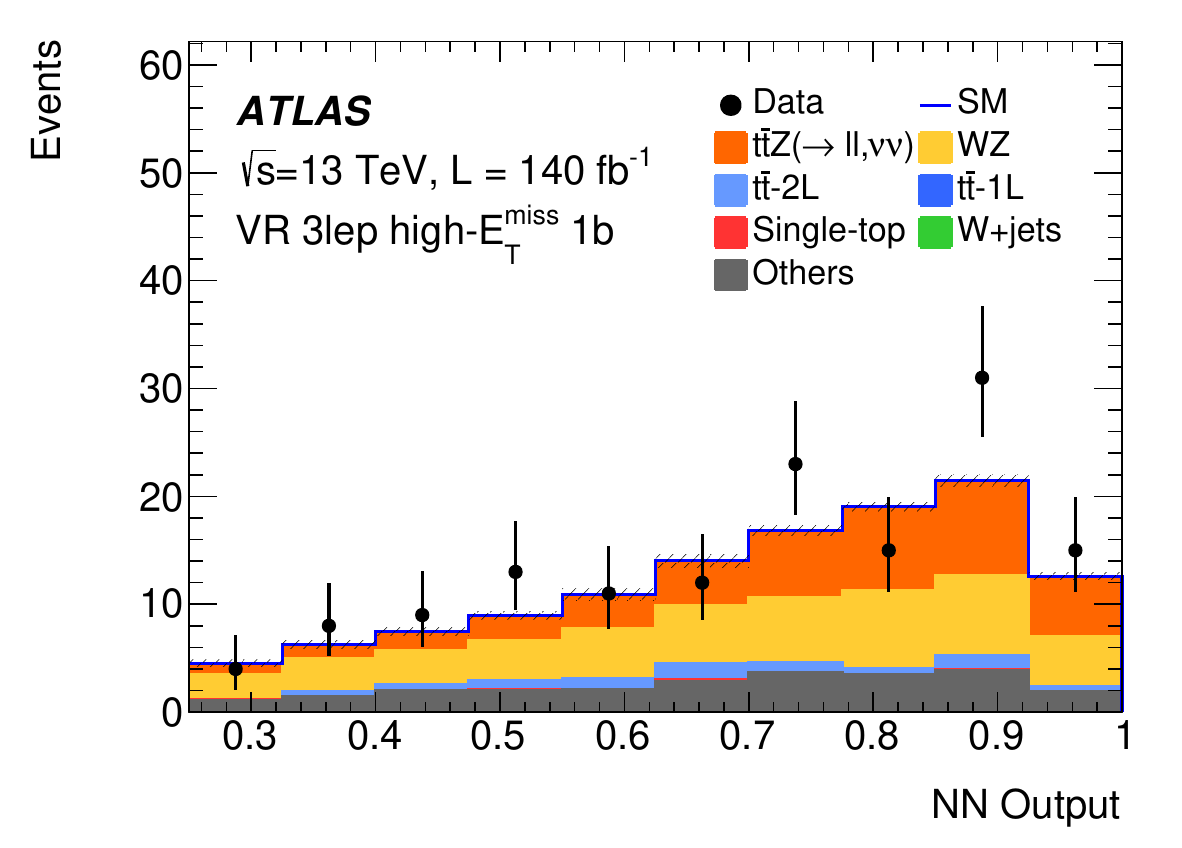}
\includegraphics[width=0.48\textwidth]{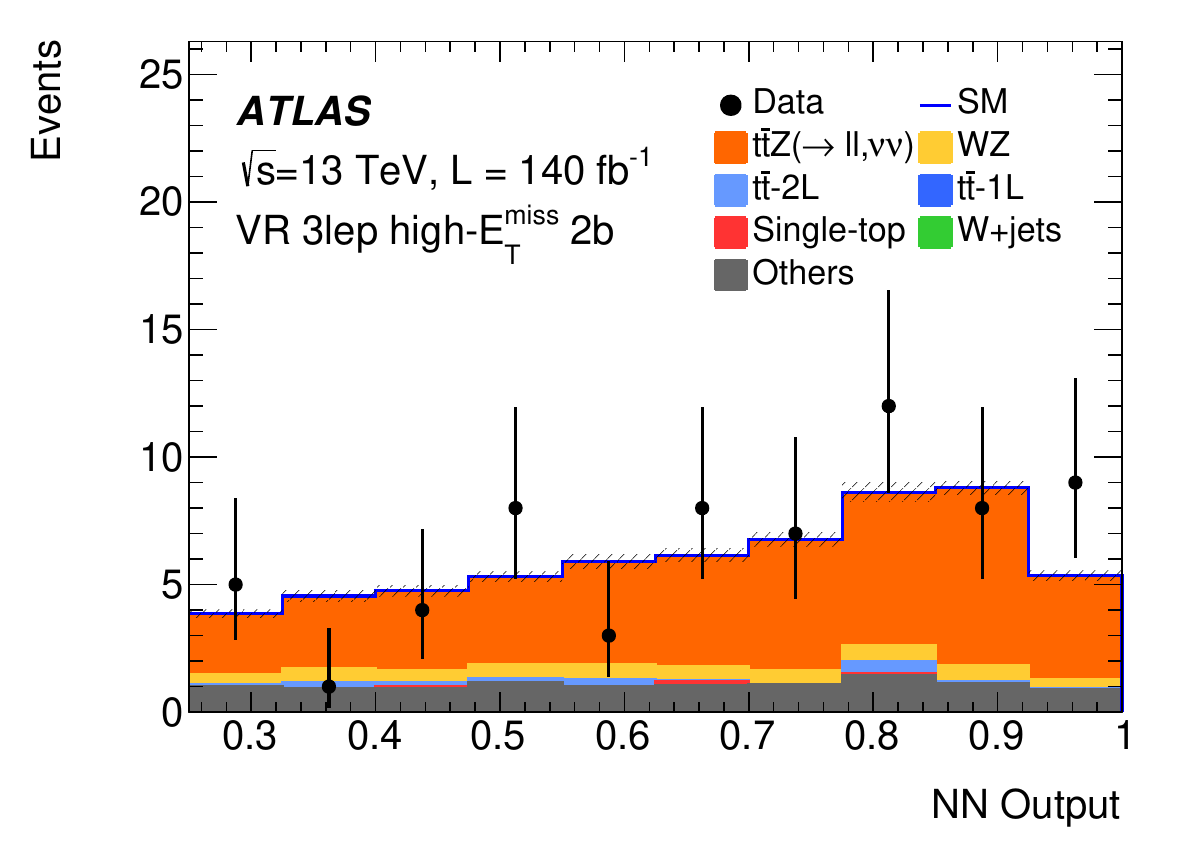}
\caption{Observed and expected distributions of the stop-NN output values in the  high-\met three-lepton events with one (left) and two or more (right) $b$-tagged jets.  Only statistical uncertainties are shown and no data-driven correction is applied to  background predictions.   }
\label{fig:VR3l_NN}
\end{figure}

The validation of the predictions for the irreducible $\ttbar Z(\to\nu\bar{\nu})$ background is done using events enhanced in  $\ttbar Z(\to\ell\ell)$.  These events are subject to the selection described in Section~\ref{sec:classification},  except  that exactly three signal leptons and no additional baseline lepton are required.  From these three leptons,  two are required to be of the same flavour and have opposite electric charges.  The invariant mass of these two leptons is required  to be compatible with the leptonic decay of a $Z$ boson,  $m_{\ell\ell}\in[81,101]$~GeV.   To select $\ttbar Z(\to\ell\ell)$ events with kinematic properties similar to those of the $\ttbar Z(\to\nu\bar{\nu})$ events accepted in the SRs,  the three-lepton events are categorised and classified as one-lepton events using a modified missing transverse momentum which incorporates the vectorial sum of the two leptons from the $Z$-boson decay.\footnote{The \met significance used to classify  three-lepton events is not computed with the modified missing transverse momentum, but rather with the original $\vec{p}_{T}^\mathrm{miss}$.} As shown in Figure~\ref{fig:VR3l_NN},
the observed distributions of events as a function of the stop-NN output value in the high-\met three-lepton regions are in good agreement with predictions.  These theoretical predictions do not include data-driven corrections.
Similar agreement is observed for the DM-NN output values.
The three-lepton selection in the boosted regions is not used as it does not yield enough events to provide a useful validation.

The distributions of NN output values in the three-lepton categories are also useful to validate the NN modelling of signal-like events in the simulation.
\section{Systematic uncertainties}
\label{sec:sys}

The main experimental systematic uncertainties on the signal and background predictions include uncertainties on the determination of the energy scale and resolution of jets and of the \met soft-term,  and on the modelling of the efficiencies of $b$-tagging, top-tagging and the (f)JVT selection.  Other systematic uncertainties include uncertainties on the modelling of lepton energy scale and reconstruction,  trigger,  identification and isolation efficiencies.  Uncertainties in the integrated luminosity and on the reweighting of the simulation as a function of the number of interactions per bunch crossing in data are also considered.
The predicted yield of background events with a non-prompt lepton is assigned a  50\% uncertainty, which has minimal impact on the results of the searches.

Theoretical uncertainties on the differential cross-sections of the main sources of background are also considered.  For \ttbar and $W$+jets events,  these include uncertainties related to  factorisation and renormalisation scales,  PDF uncertainties,   uncertainties on the matching of partons between  ME and  PS,  uncertainties on the resummation scale,  and uncertainties related to electroweak NLO corrections.  For \ttbar, uncertainties on the fragmentation model are also considered.
For $W$+jets,  a relative uncertainty of 30\% between the $W$+jets predictions in event categories with one and two or more $b$-tagged jets is assigned to cover uncertainties on the production of a $W$ boson in association with heavy-flavour quarks~\cite{ATL-PHYS-PUB-2017-006}.
To match the NF scheme used in the statistical model,  these theoretical uncertainties are considered uncorrelated between \ttbar-1L and \ttbar-2L events,  as well as between the predictions in the  high-\met and boosted categories.
Theoretical uncertainties on single-top production include uncertainties on the amount of initial-state radiation, on PDFs,  and on factorisation and renormalisation scales.  An uncertainty on the removal of the on-shell contributions in the $tW$ process is estimated as the difference in predicted yields when using the diagram removal or the diagram subtraction prescriptions.
For  $\ttbar Z$,  uncertainties related to PDFs and to factorisation and normalisation scales are considered.  Other minor background contributions,   mostly $VV$ and $ttV$ (apart from $\ttbar Z(\to \nu\bar{\nu},\ell\bar{\ell})$),  are predicted with a conservative uncertainty of 30\% on the total cross-section~\cite{ATL-PHYS-PUB-2016-002,  ATL-PHYS-PUB-2016-005}.
Theoretical uncertainties on signal predictions from factorisation and renormalisation scales and PDF uncertainties are also accounted for.

In the statistical model,  systematic uncertainties are associated with nuisance parameters (NPs) whose values are constrained with Gaussian probability density functions.  Statistical uncertainties on simulated events are also accounted for.  The impact of these uncertainties on the sensitivity of the searches is discussed in Section~\ref{sec:interpretations}.

\section{Results}
\label{sec:result}

Multiple types of fits,  either for the $\stopone\stopone$ or the $\ttbar$+DM  search,  are performed to analyse the observed data.  First a `background-only' fit that uses only data in CRs is done to predict the background in the VRs.  The data in the VRs is then used to validate the background modelling.   A background-only fit using data from CRs and SRs is done to check the compatibility of the data in SRs with background-only contributions.  Finally, `full' fits, in which data in CRs and SRs are fitted with contributions from background processes and from a signal model, is done.
These fits are performed separately for the $\stopone\stopone$ and $\ttbar$+DM searches, with all associated regions fit simultaneously.

\begin{figure}[htp]
\centering
\includegraphics[width=0.42\textwidth]{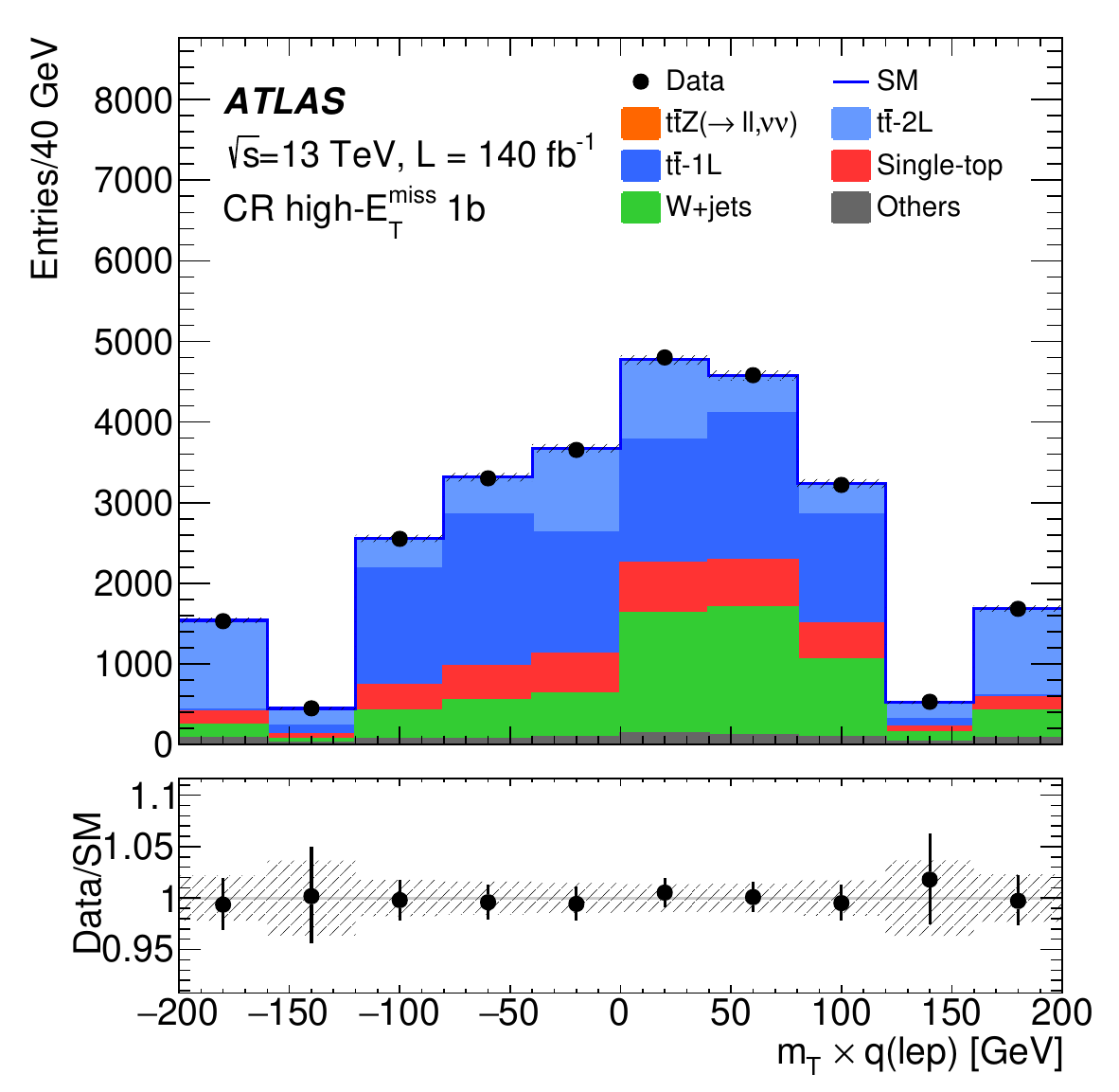}
\includegraphics[width=0.42\textwidth]{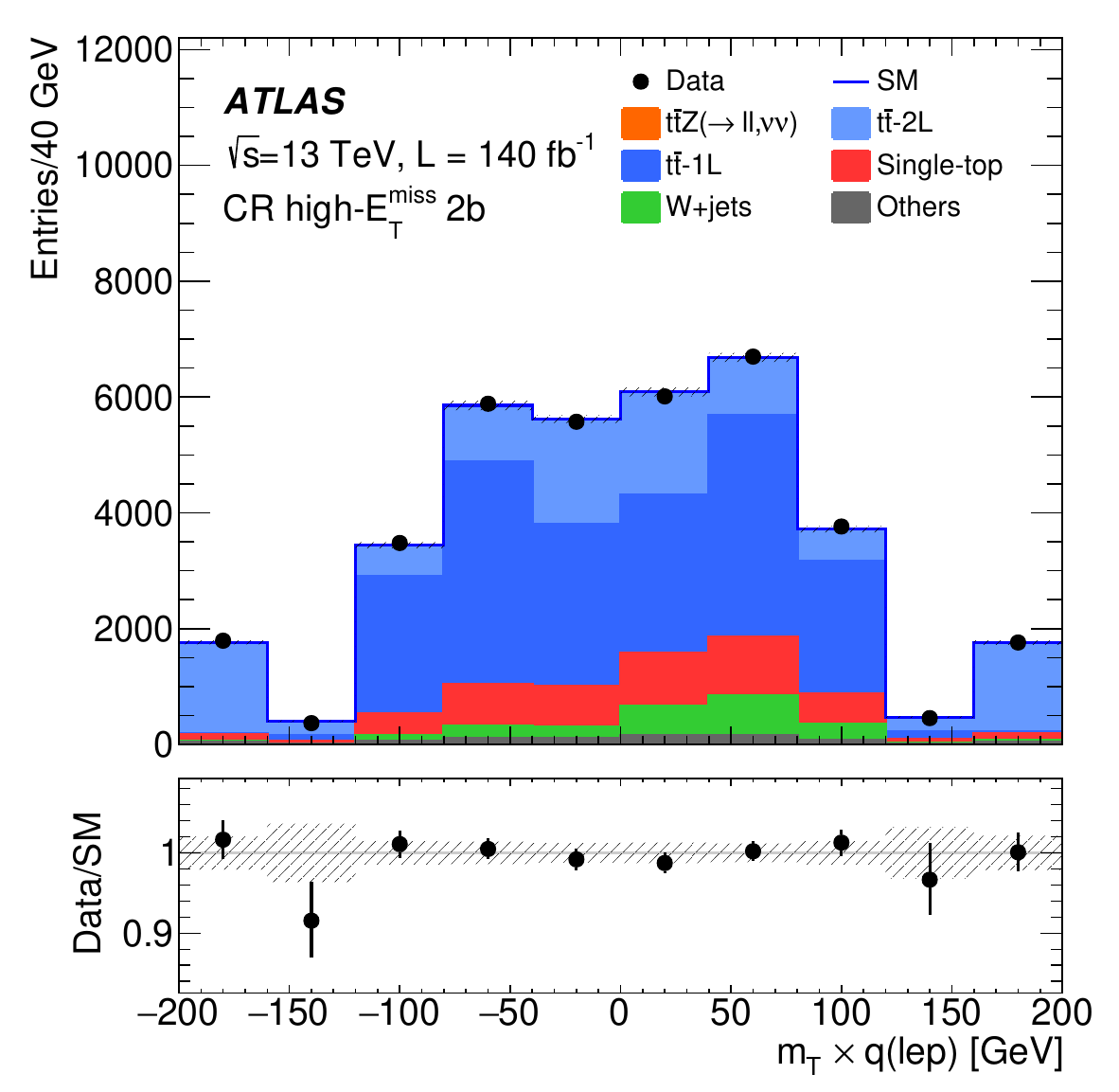}
\includegraphics[width=0.42\textwidth]{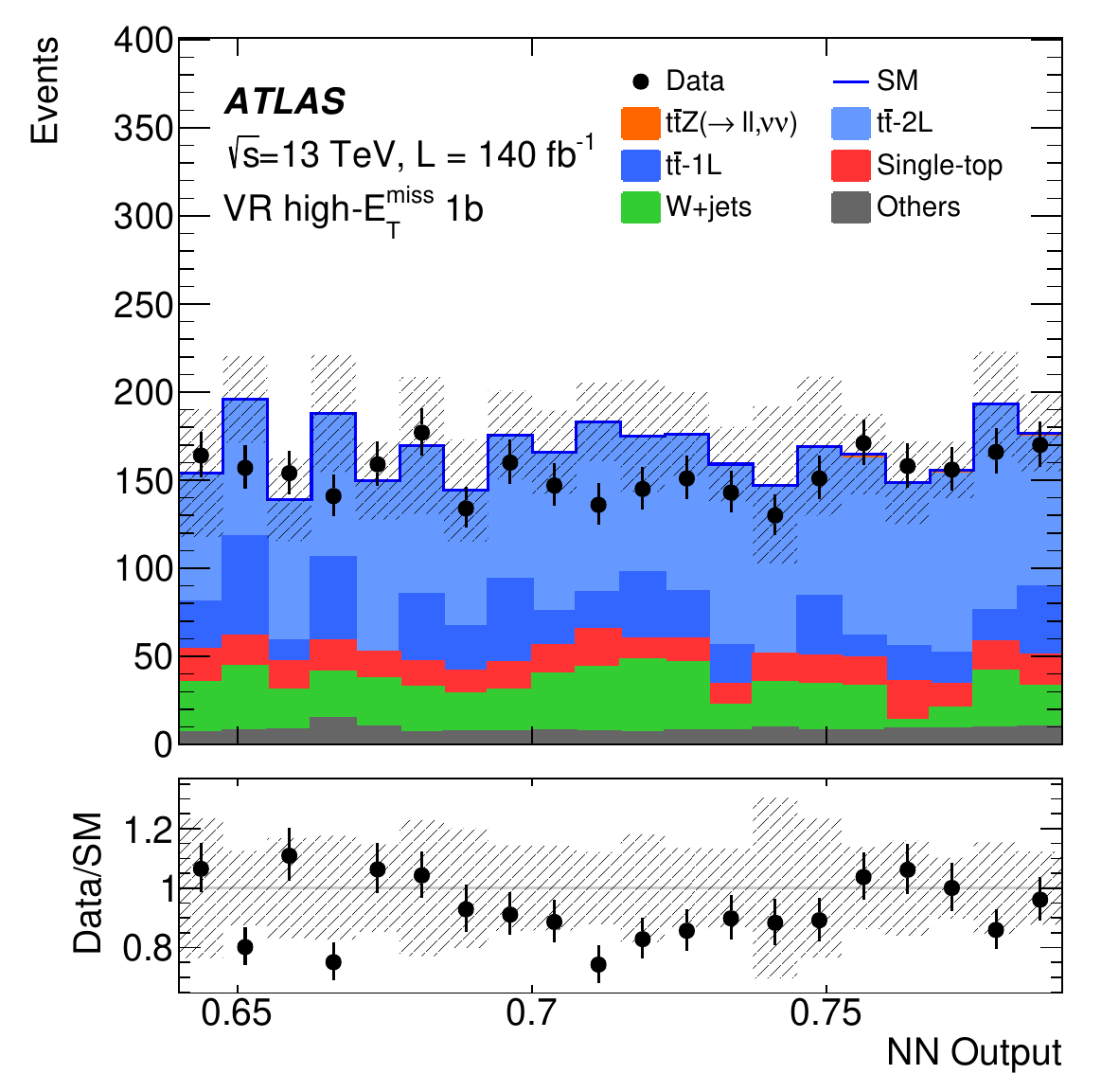}
\includegraphics[width=0.42\textwidth]{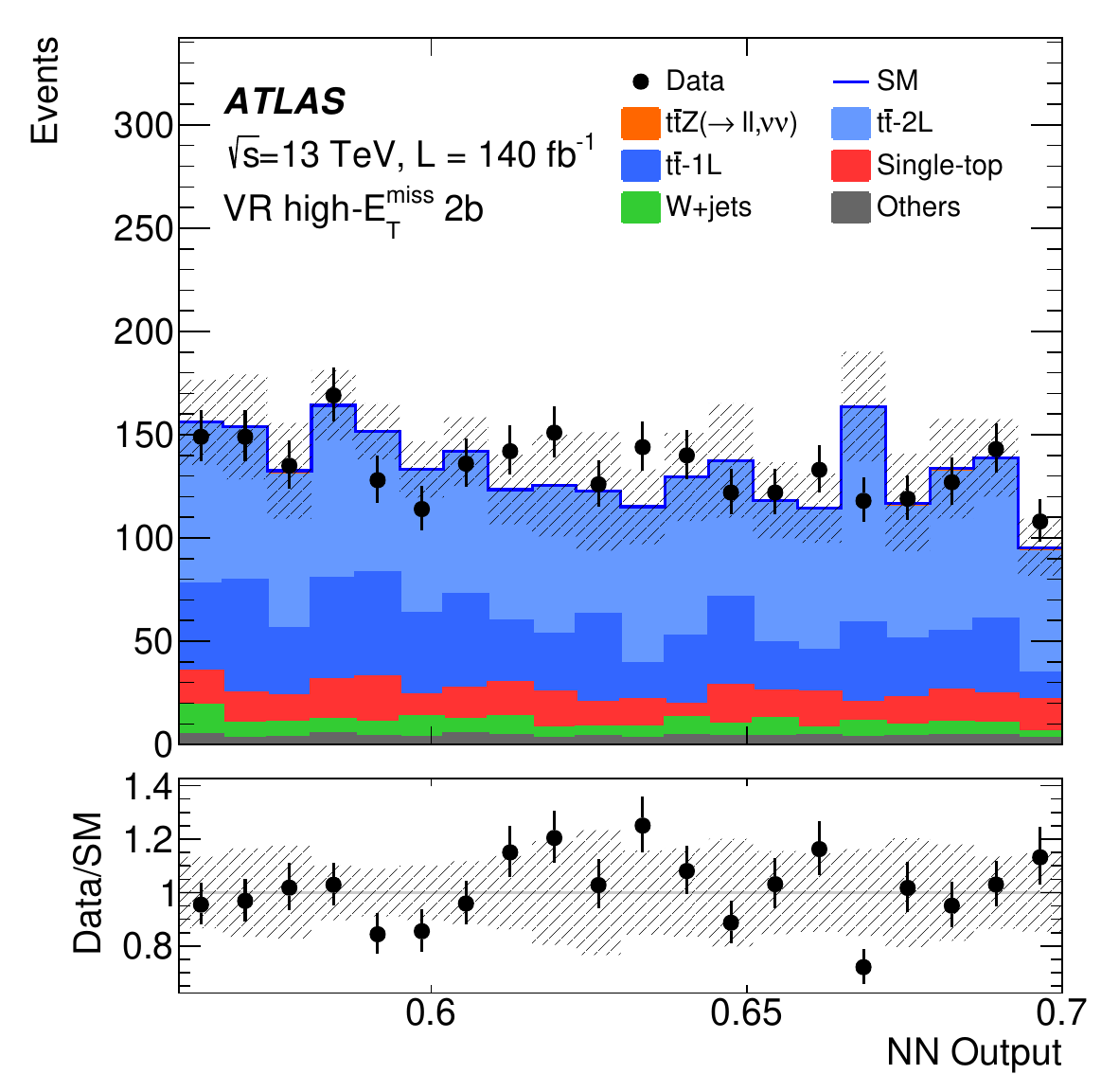}
\includegraphics[width=0.42\textwidth]{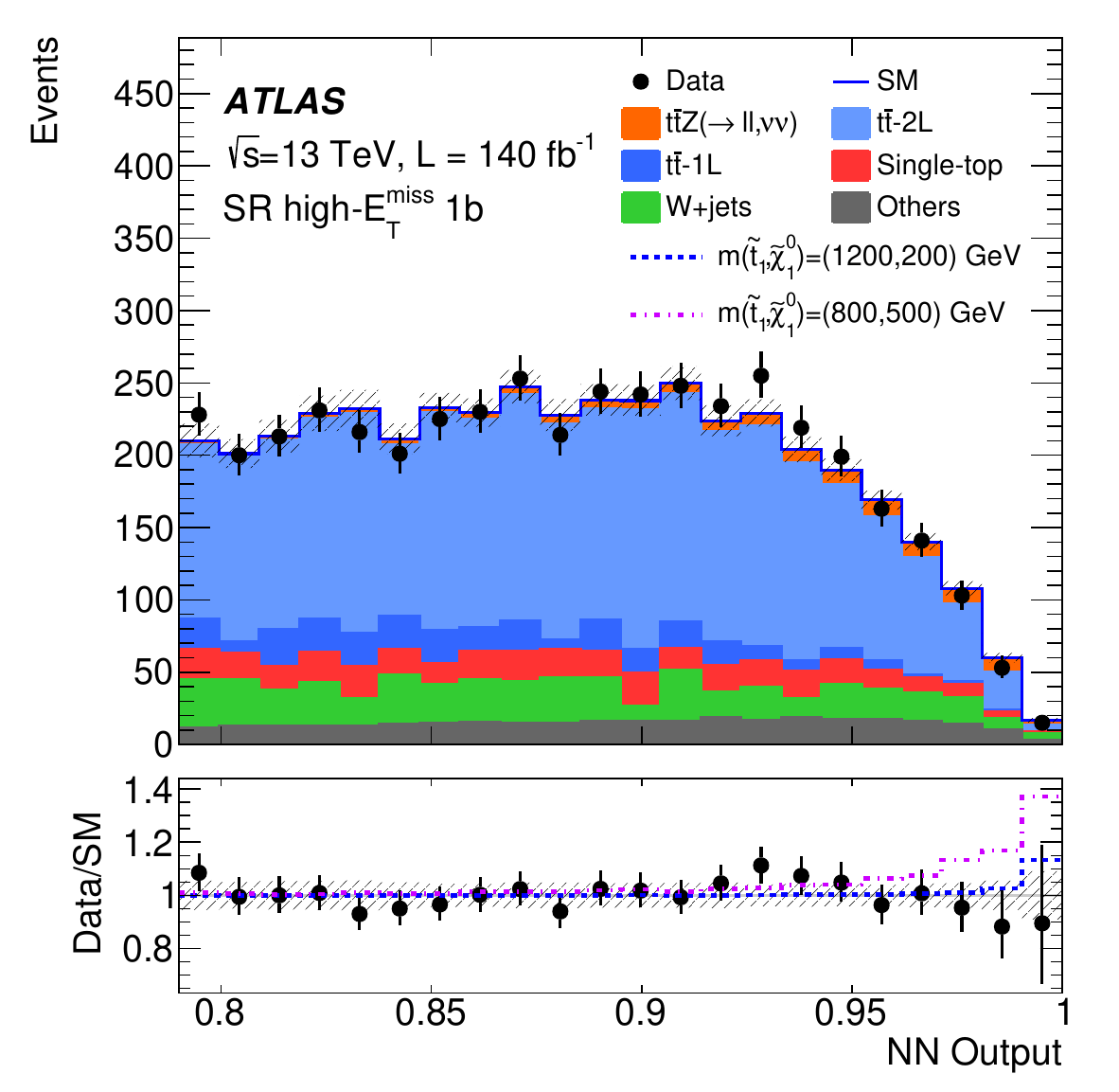}
\includegraphics[width=0.42\textwidth]{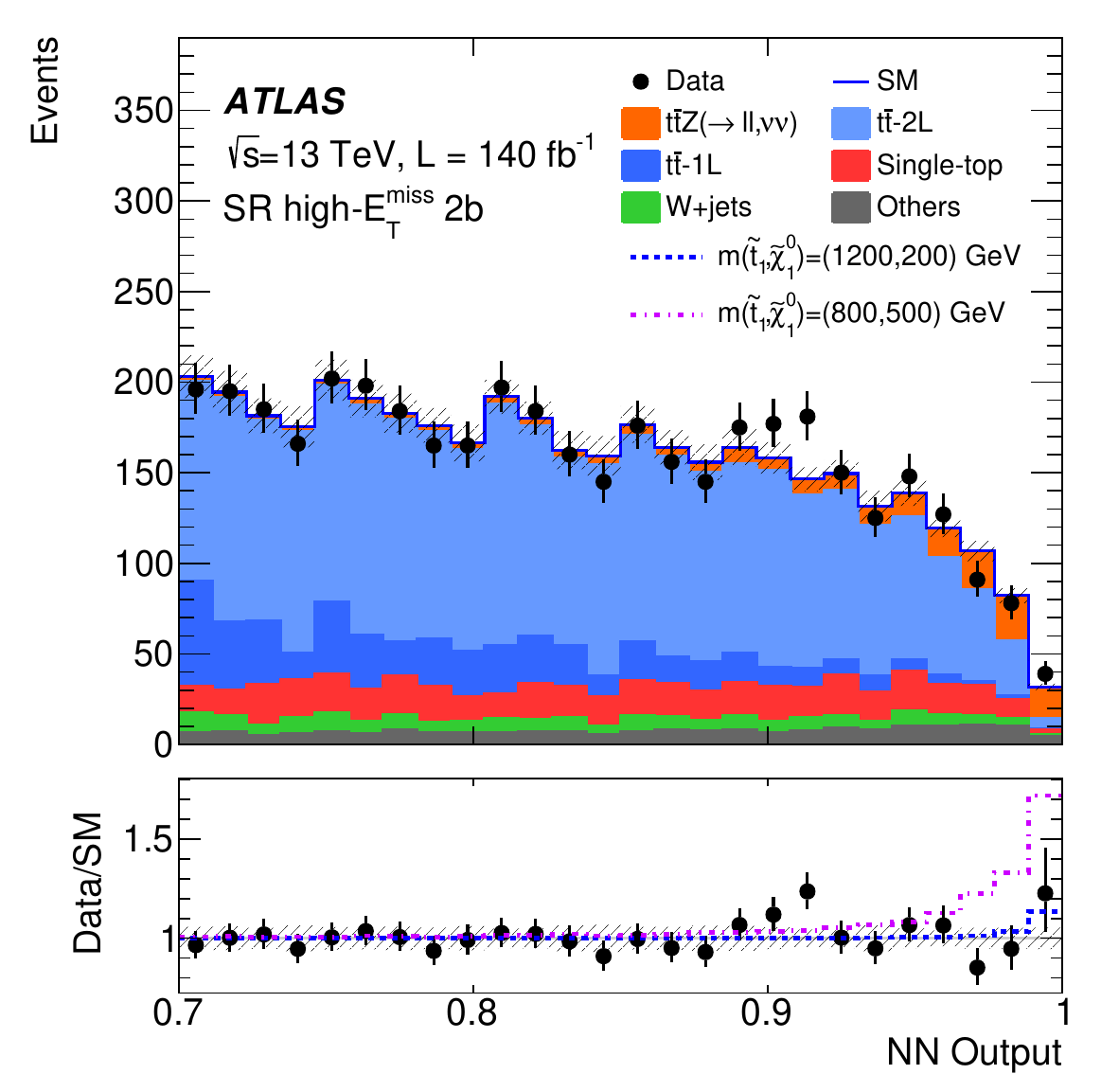}
\caption{Observed and expected events as a function of $\mt \times q(\ell)$  in the CR (top) and of the stop-NN output value in the VR (middle row) and the SR (bottom) in the high-\met 1b (left) and 2b (right) stop regions.  The expected distributions are as determined in the background-only fit with data in CRs and SRs for the $\stopone\stopone$ search.  The hatched area around the total SM prediction includes statistical and systematic uncertainties.  Uncertainties on predictions in VRs are dominated by  statistical uncertainties on  simulated   samples.
The lower panels show the ratio of the observed data to the expected SM events and the ratio of the sum of the SM and expected signal  events to the SM events.
In CRs,  the last (first) bin contains overflows (underflows).    }
\label{fig:postfit_stop_highMET}
\end{figure}

\begin{figure}[htp]
\centering
\includegraphics[width=0.42\textwidth]{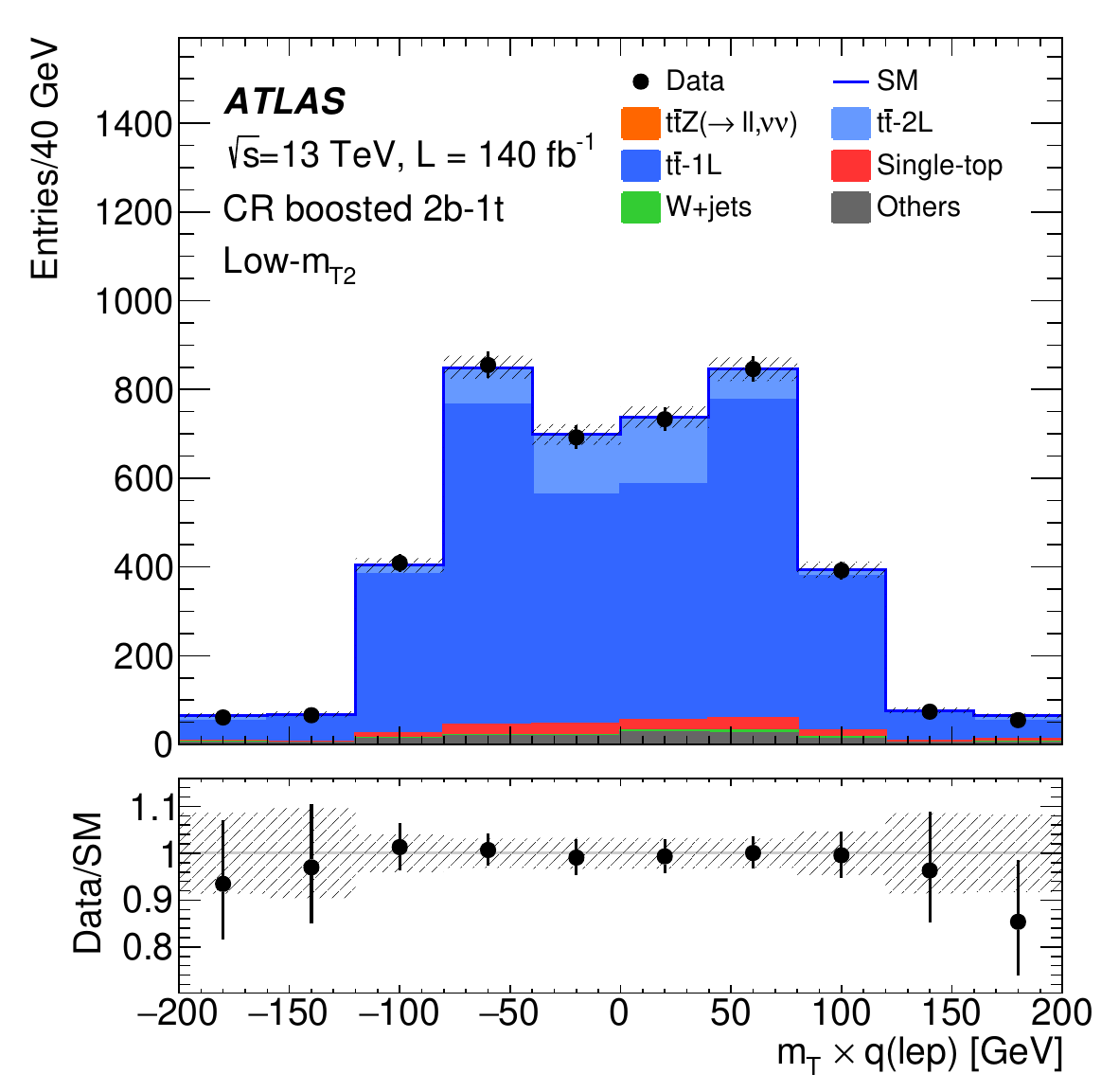}
\includegraphics[width=0.42\textwidth]{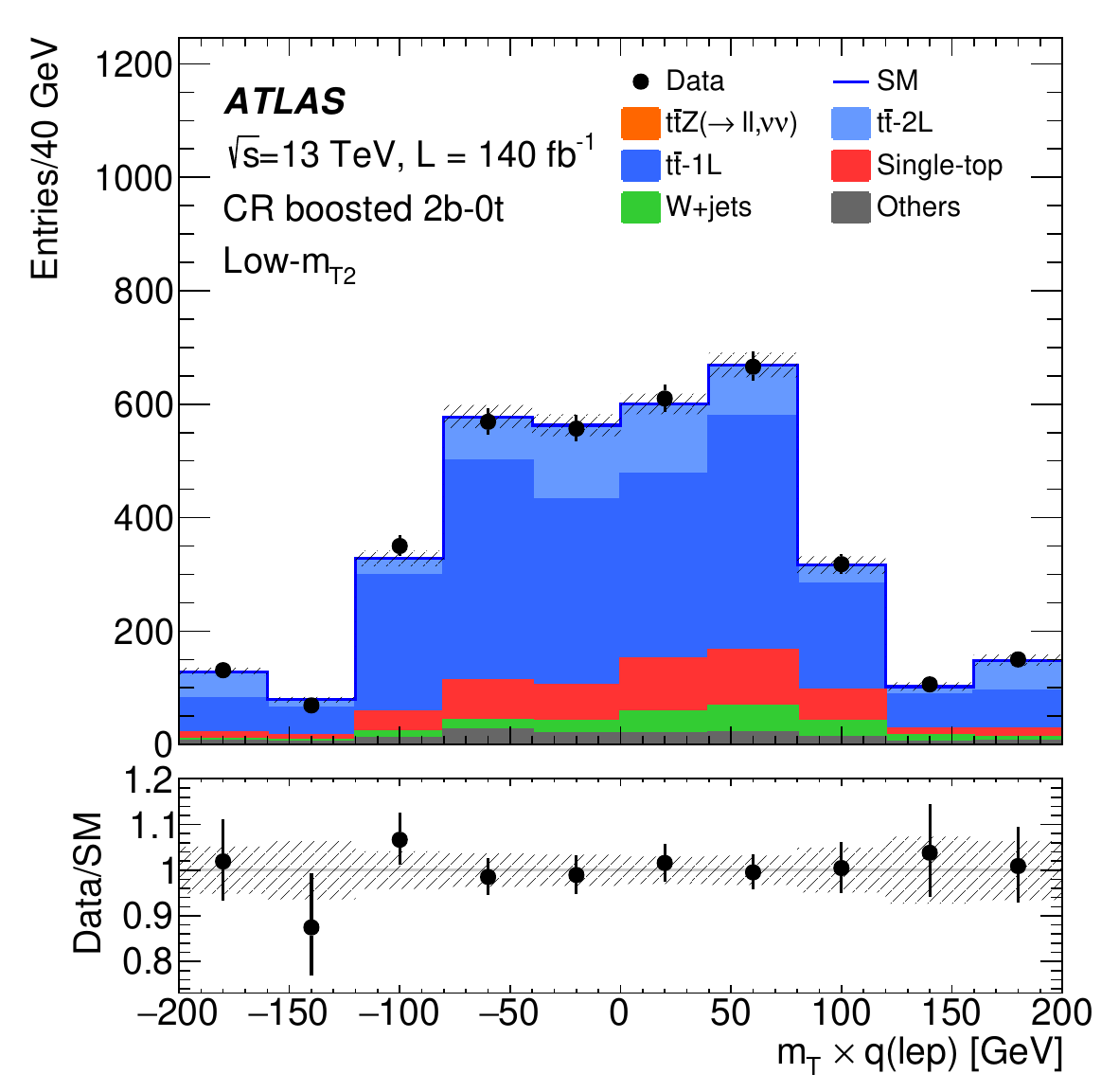}
\includegraphics[width=0.42\textwidth]{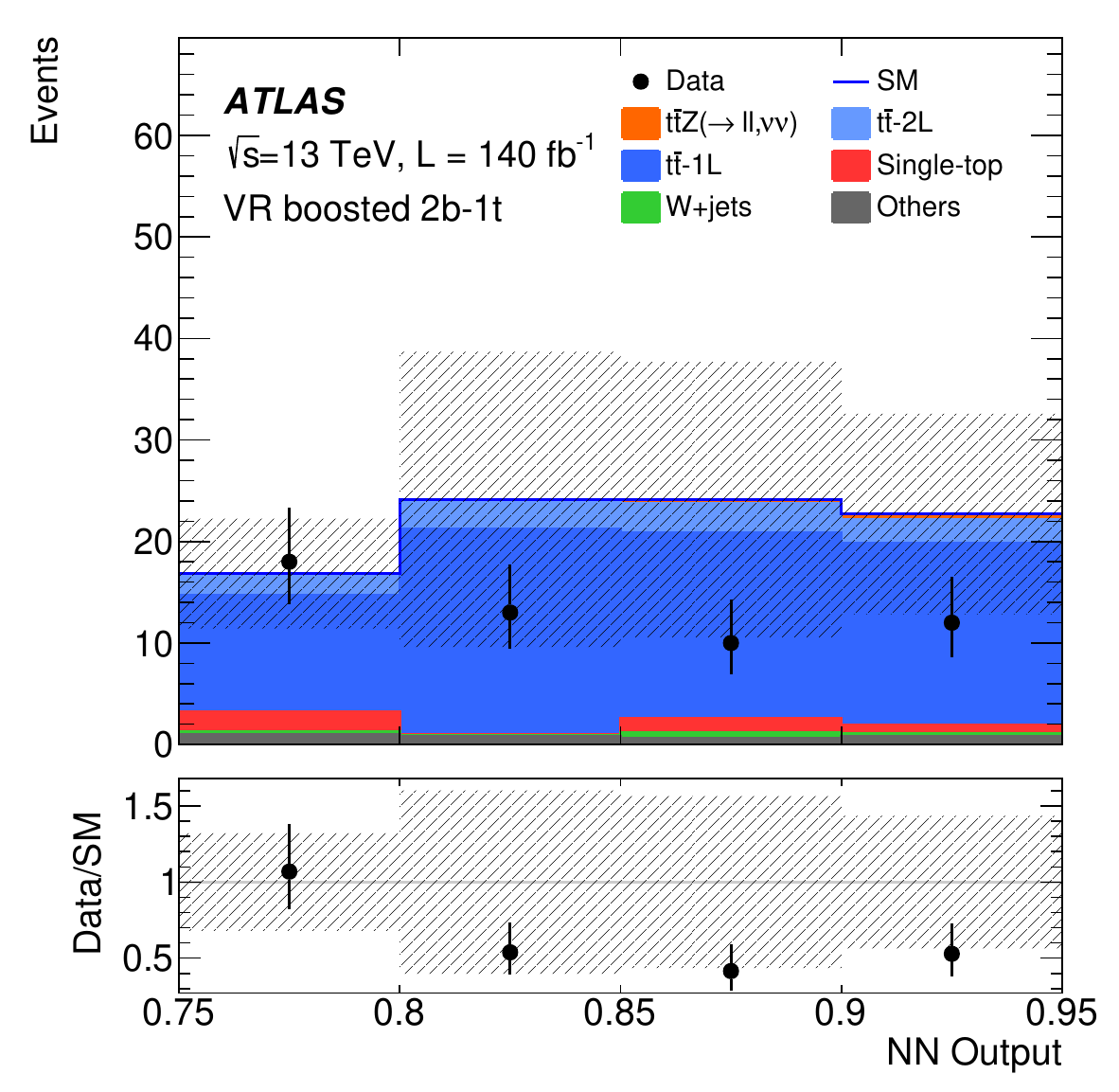}
\includegraphics[width=0.42\textwidth]{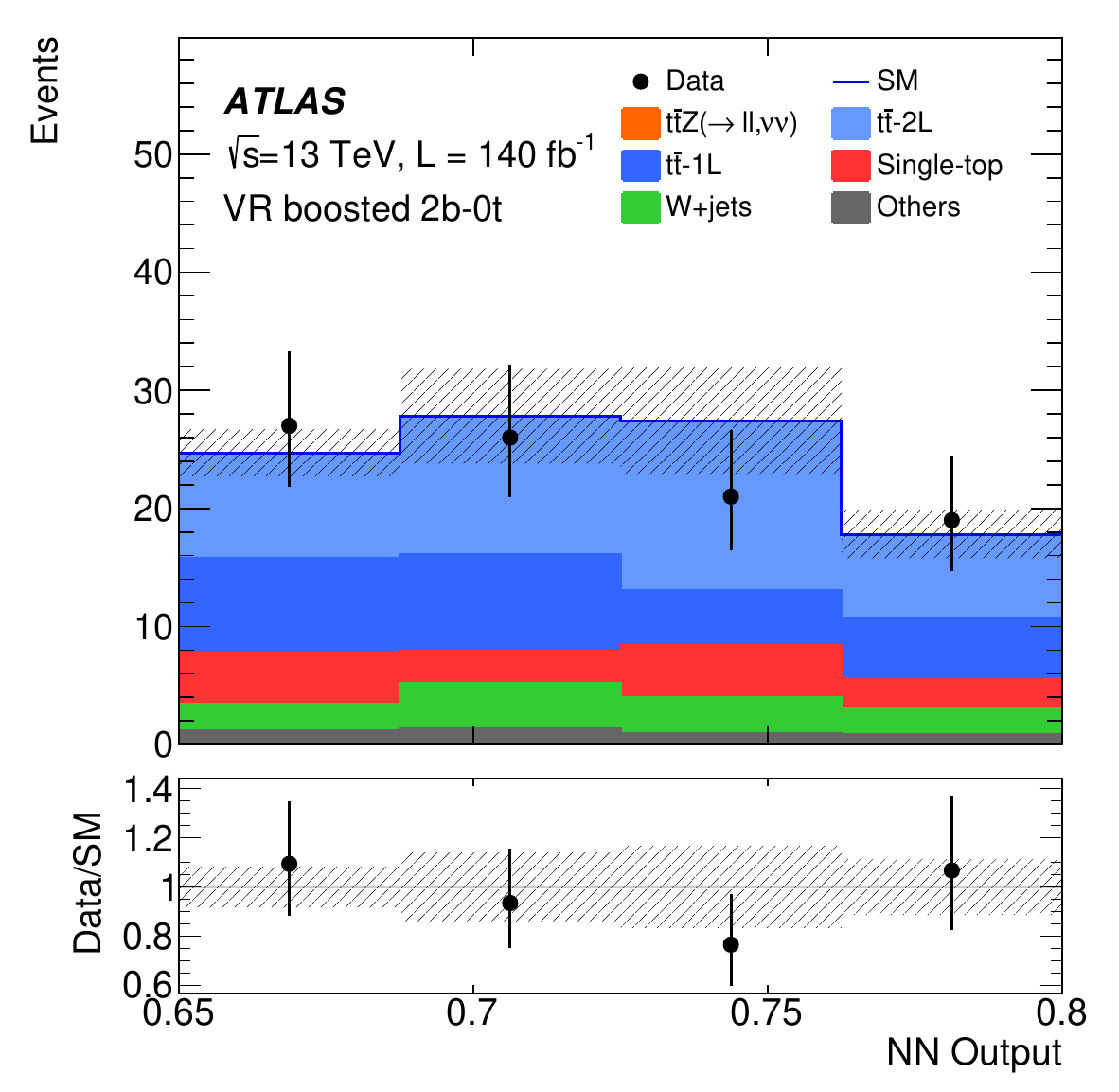}
\includegraphics[width=0.42\textwidth]{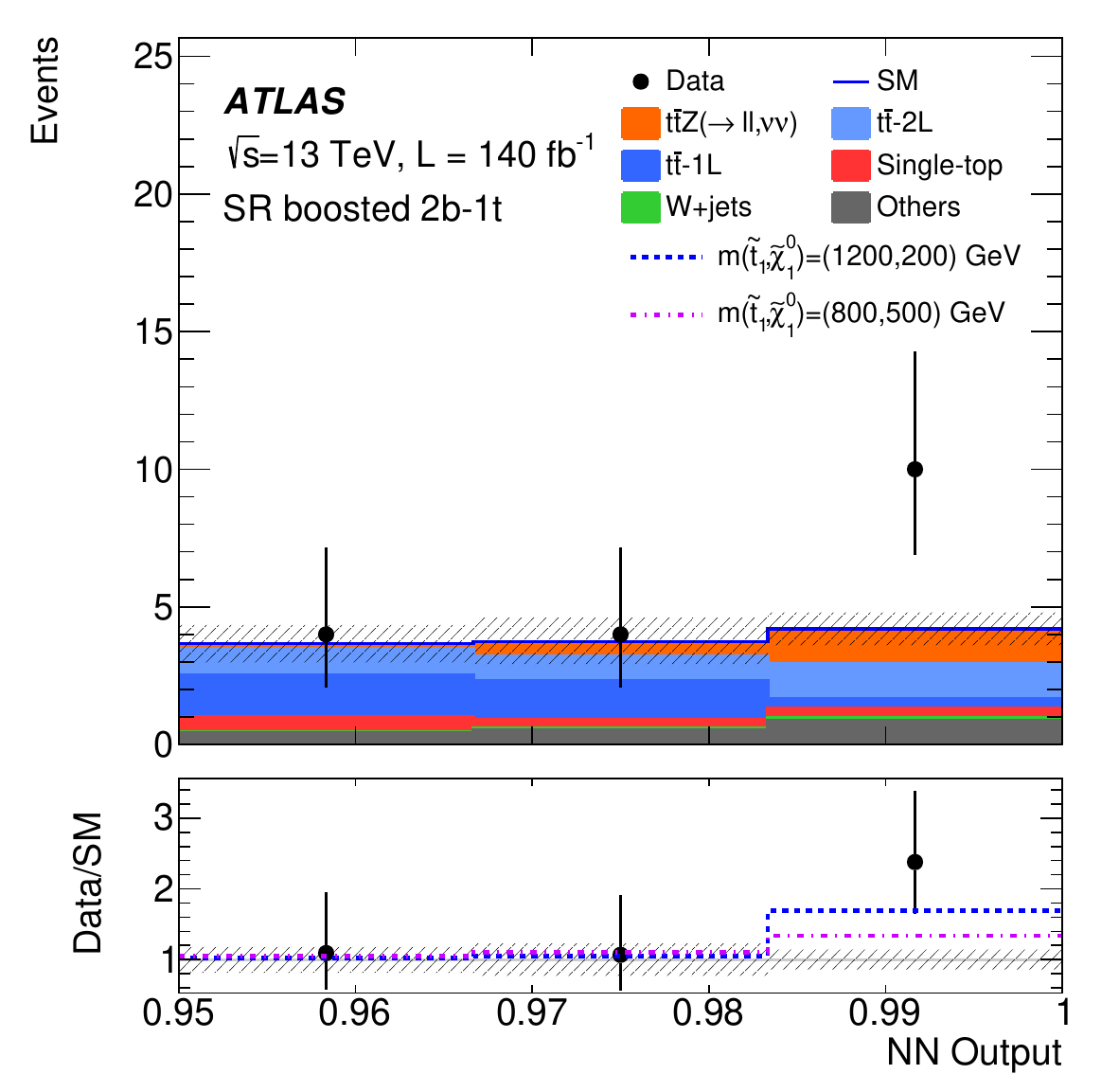}
\includegraphics[width=0.42\textwidth]{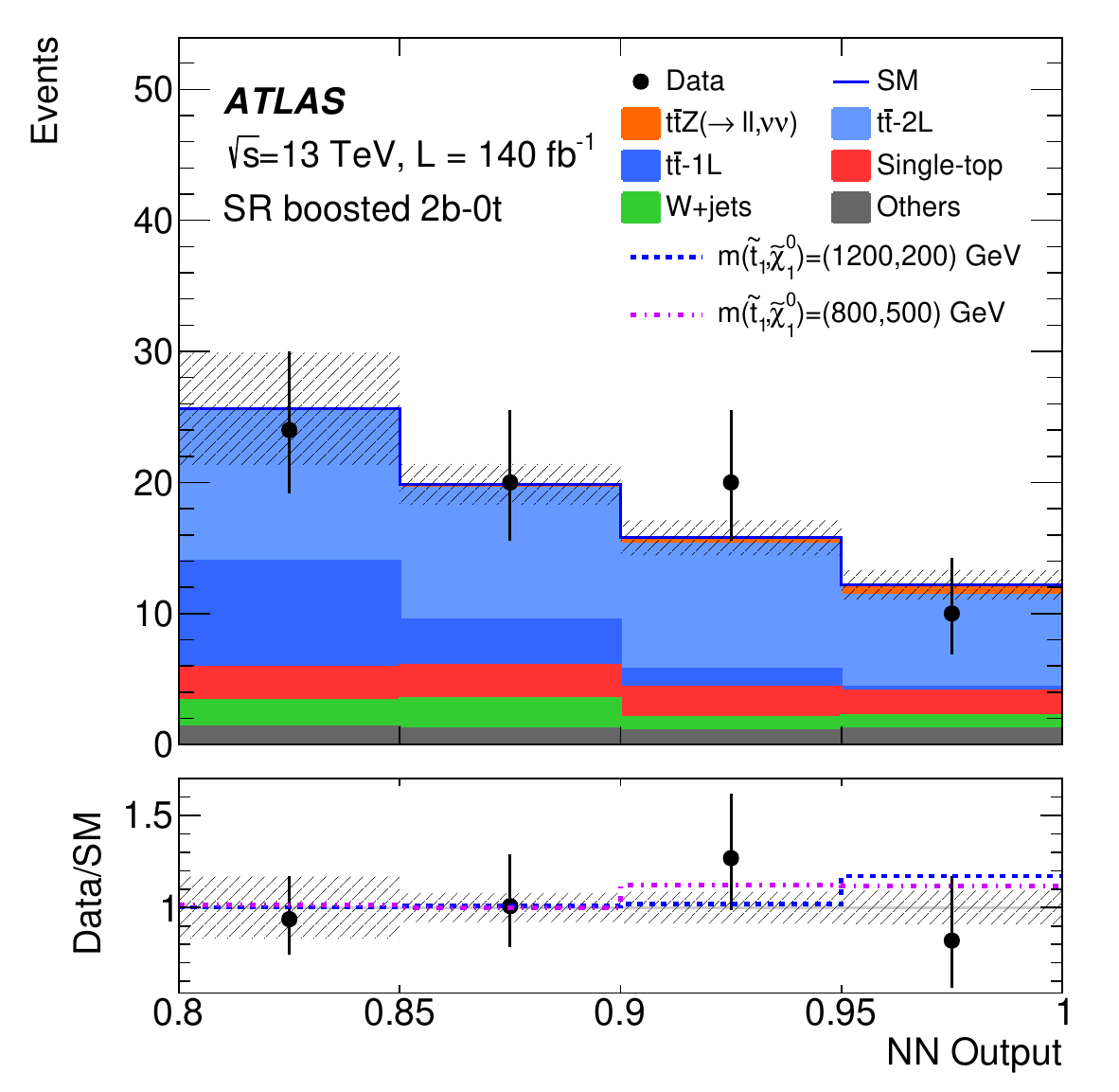}
\caption{Observed and expected  events distributed
as a function of $\mt \times q(\ell)$  in the low-$m_{T2}$ CR (top) and of the stop-NN output value in the VR (middle row) and SR (bottom) in the boosted 2b-1t (left) and 2b-0t (right) stop regions.  The expected distributions are as determined in the background-only fit with data in CRs and SRs for the $\stopone\stopone$ search.  The hatched area around the total SM prediction
includes statistical and systematic uncertainties.   Uncertainties on predictions in VRs are dominated by  statistical uncertainties on  simulated   samples. The lower panels show the ratio of the observed data to the expected SM events and the ratio of the sum of the SM and expected signal  events to the SM events.
In CRs,  the last (first) bin contains overflows (underflows).   }
\label{fig:postfit_stop_boosted1}
\end{figure}

\begin{figure}[htp]
\centering
\includegraphics[width=0.42\textwidth]{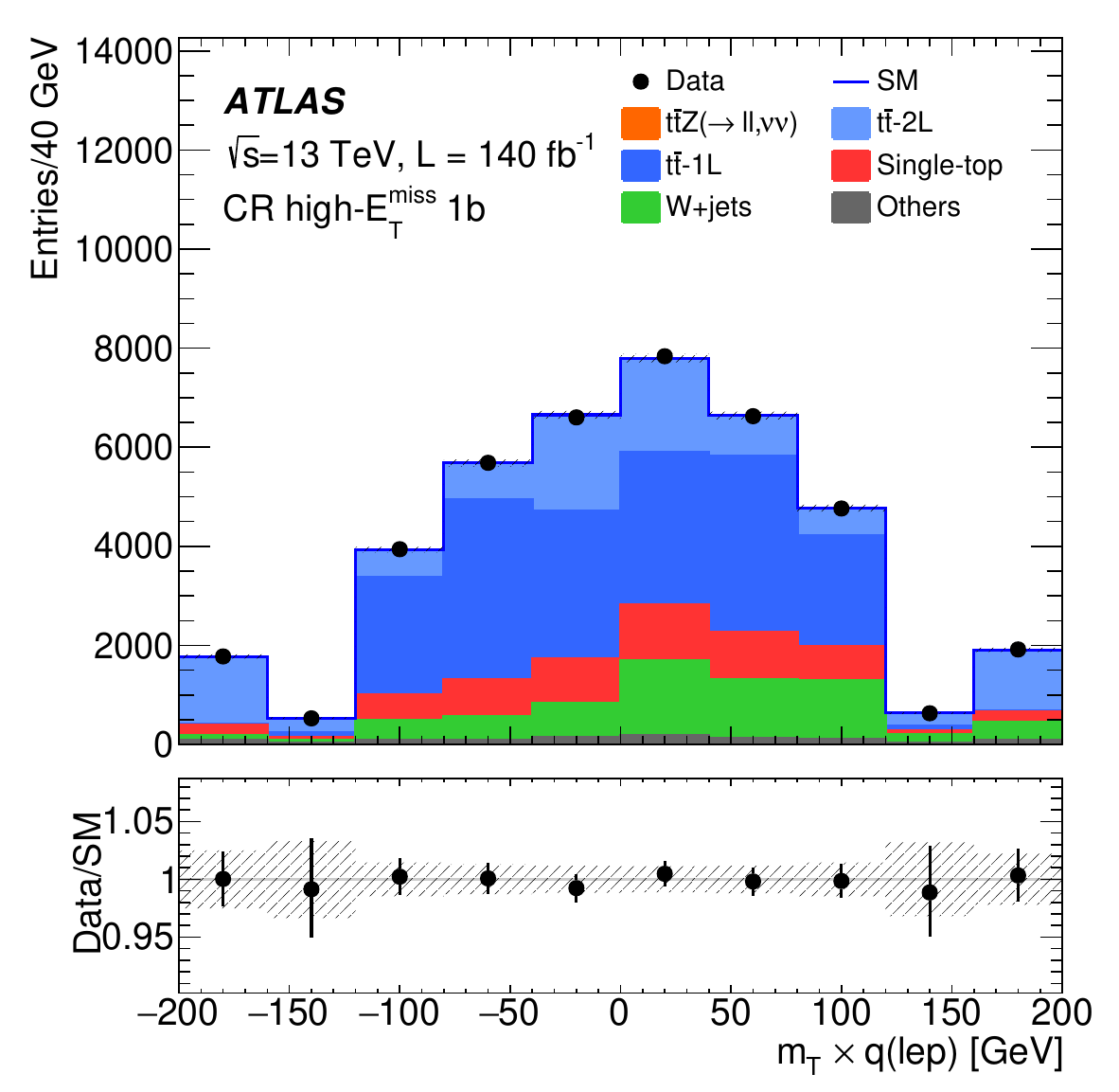}
\includegraphics[width=0.42\textwidth]{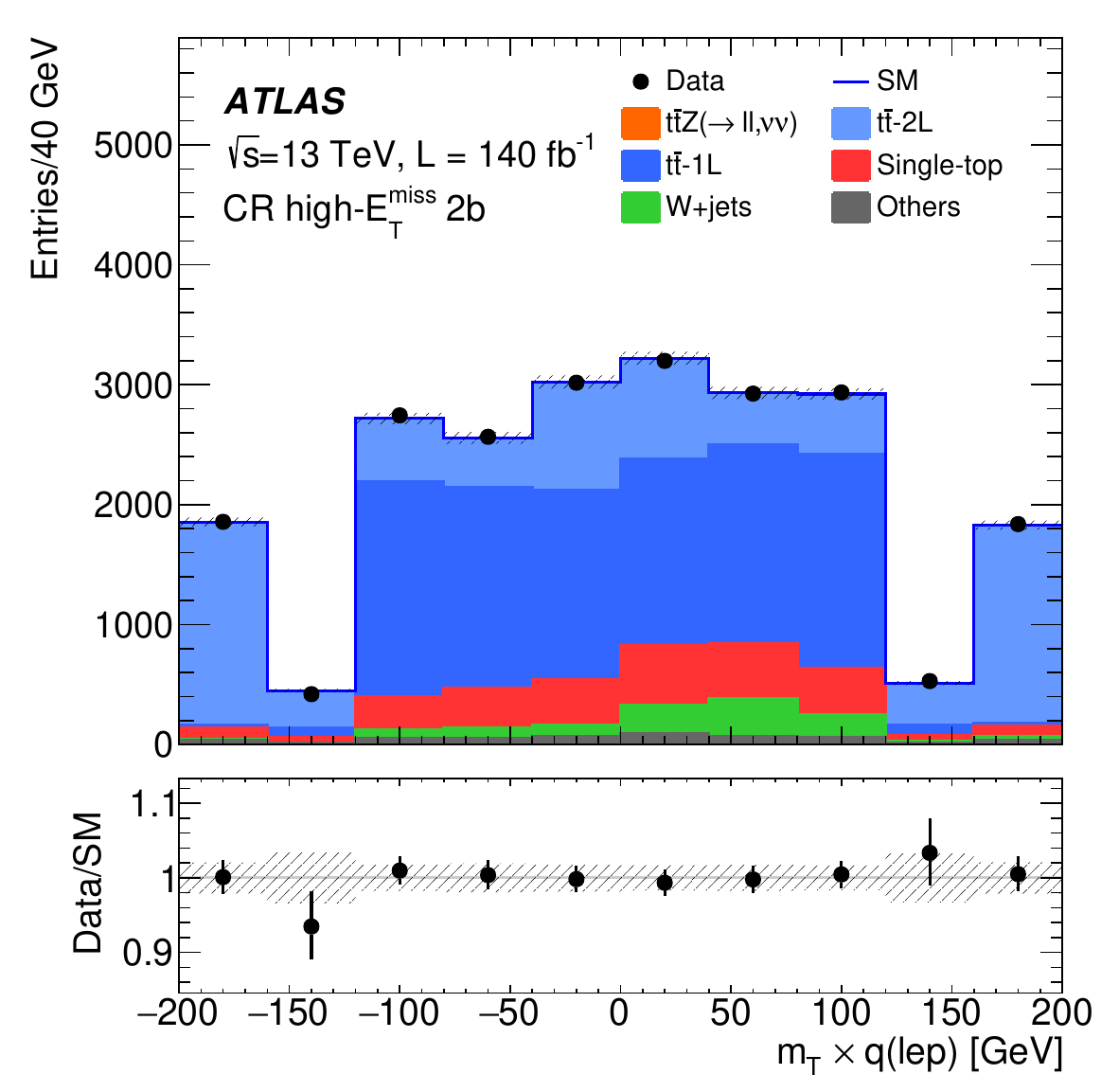}
\includegraphics[width=0.42\textwidth]{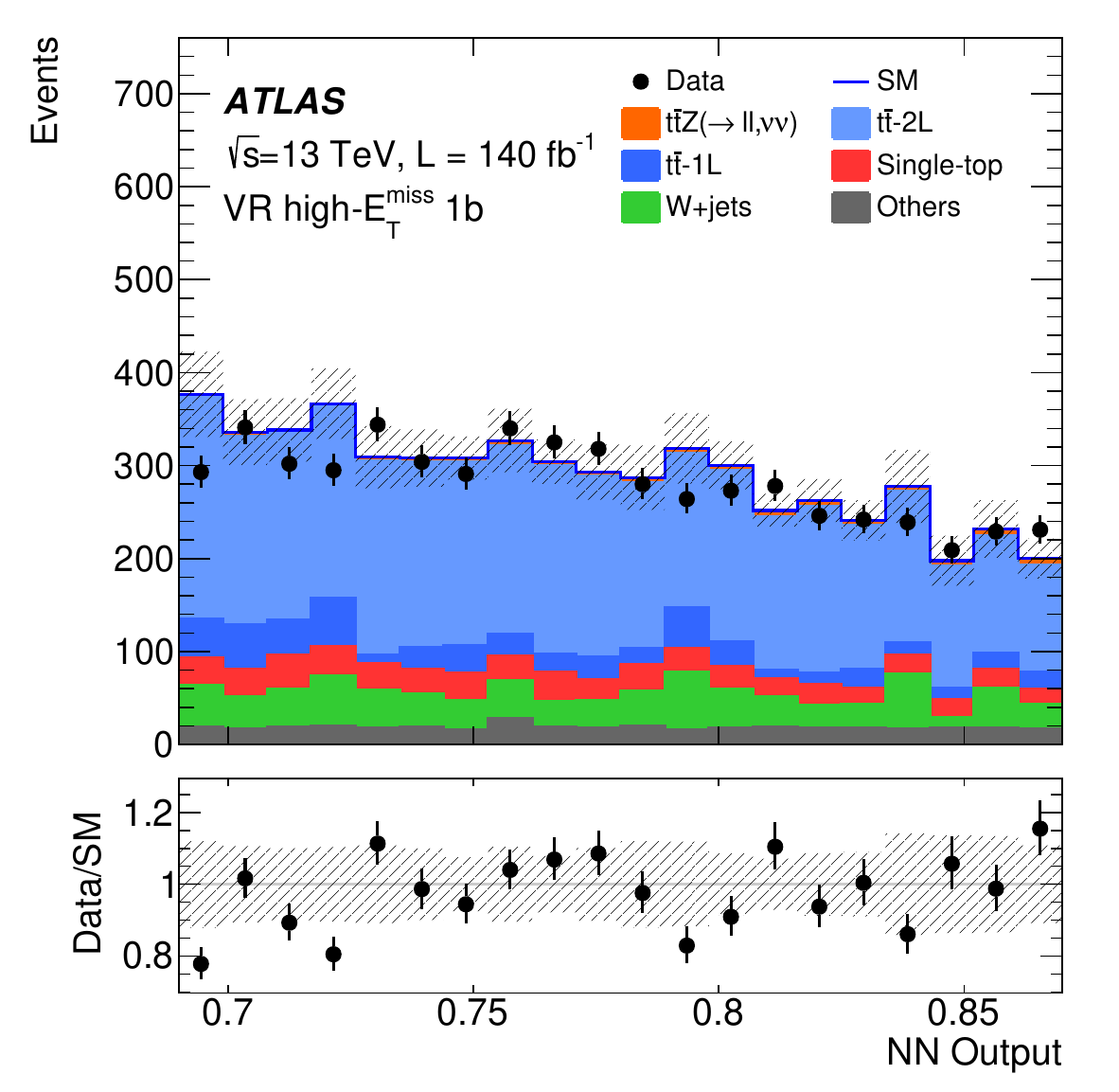}
\includegraphics[width=0.42\textwidth]{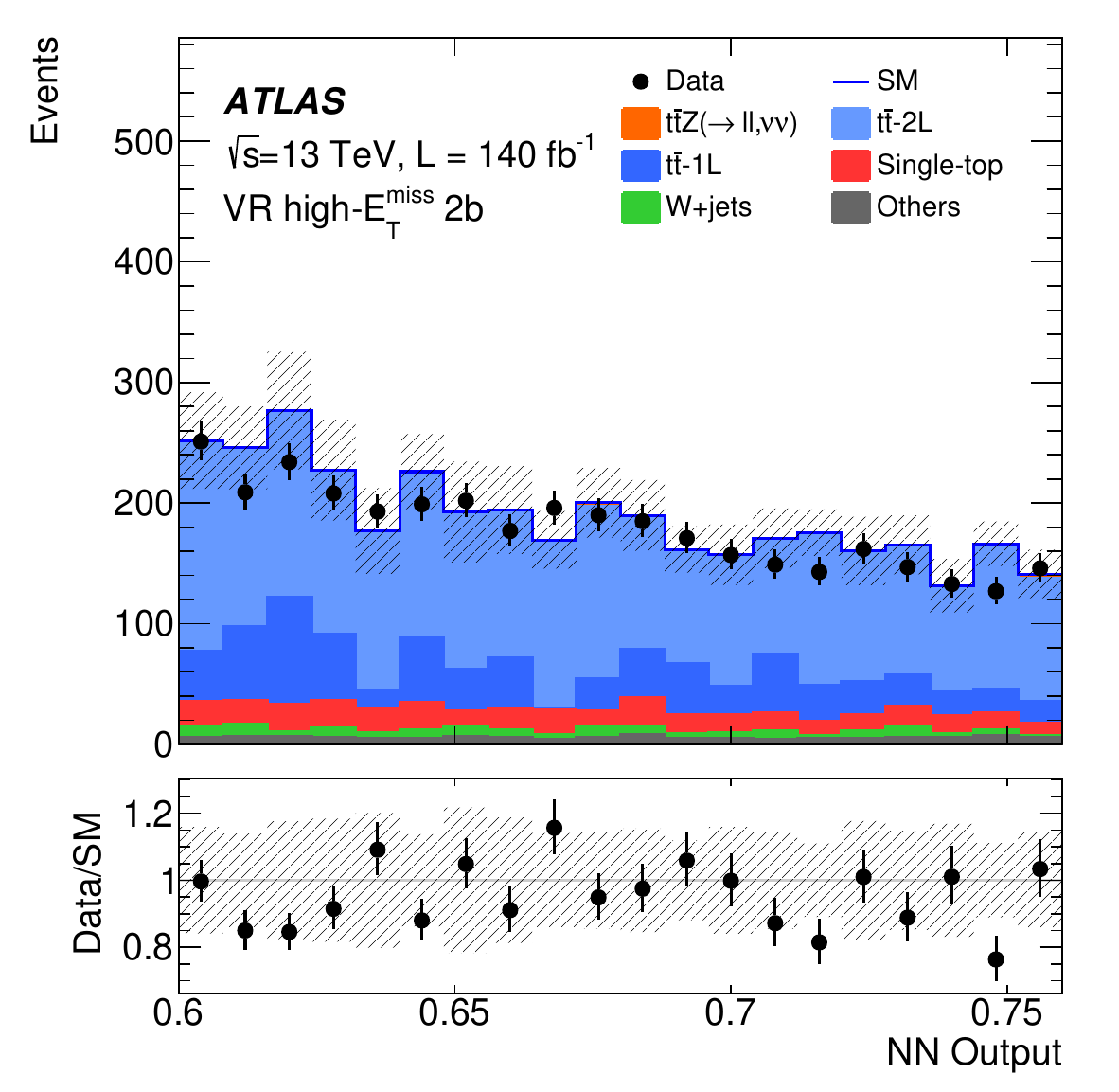}
\includegraphics[width=0.42\textwidth]{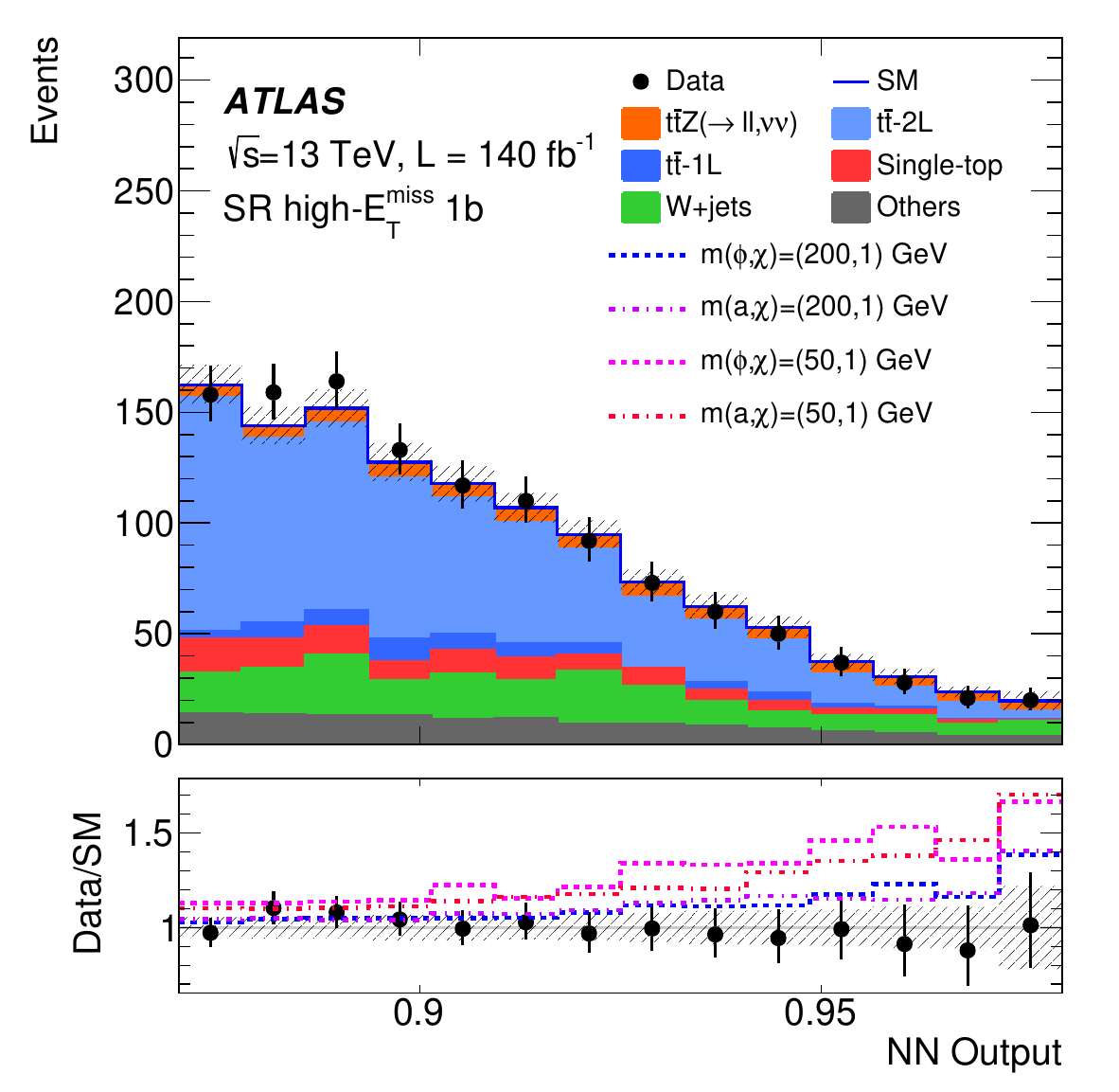}
\includegraphics[width=0.42\textwidth]{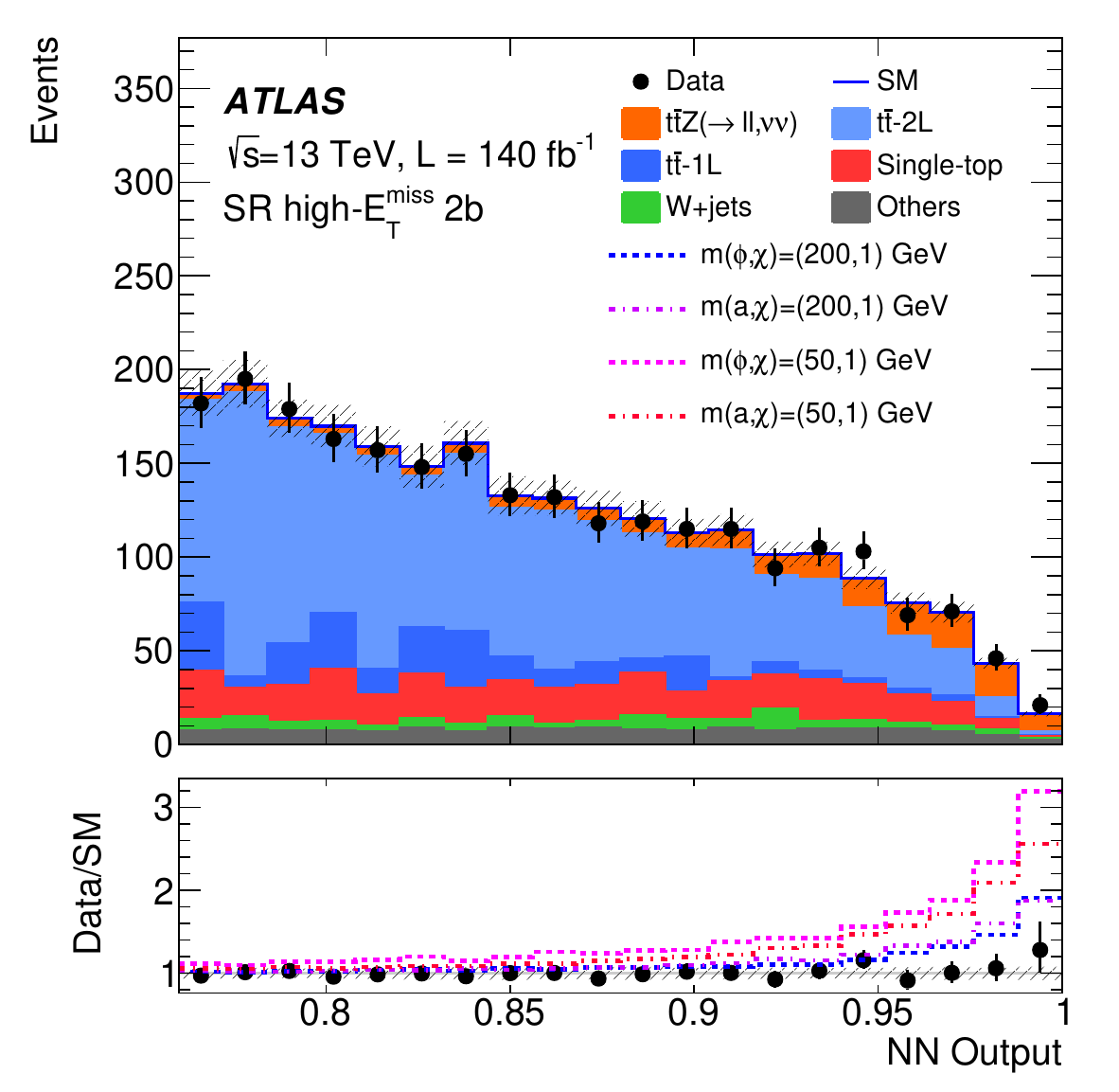}
\caption{Observed and expected  events as a function of   $\mt \times q(\ell)$ in the CR (top) and of the DM-NN output value in the VR (middle row) and the SR (bottom) in the high-\met 1b (left) and 2b (right) DM regions.  The expected distributions are as determined in the background-only fit with data in CRs and SRs for the $\ttbar$+DM search.  The hatched area around the total SM prediction
includes statistical and systematic uncertainties.   Uncertainties on predictions in VRs are dominated by  statistical uncertainties on  simulated  samples.  The lower panels show the ratio of the observed data to the expected SM events, and  the ratio of the sum of the SM and expected signal  events to the SM events.
In CRs,  the last (first) bin contains overflows (underflows).    }
\label{fig:postfit_DM_highMET}
\end{figure}

In the first background-only fit,  background predictions were fitted to data in all CRs  and then applied via the fit model to VRs,  where predictions were found to agree with data.  After this validation of the background modelling,  the background-only fit including data from SRs was performed.  Observed and expected distributions of $\mt \times q(\ell)$ and total yields in CRs  and of the NN output values in VRs and SRs after this fit are shown in Figs.~\ref{fig:postfit_stop_highMET}--\ref{fig:postfit_DM_highMET}.\footnote{Distributions in the boosted CRs as determined in the  fit for the $\ttbar$+DM search are not shown as they are similar to those obtained in the fit for the stop search. }
The expected distributions in SRs from benchmark signal models are also shown.
In the $\stopone\stopone$ search,  the NFs of the \ttbar,  $W$+jets and single-top contributions are generally compatible with unity within uncertainties.   The NFs for the \ttbar-1L (-2L) background are $1.02\pm 0.06$ ($1.09\pm 0.05$) and $1.07\pm 0.06$ ($1.29\pm 0.13$) in the high-\met and boosted regions, respectively.  The NFs for the $W$+jets backgrounds are  $1.30\pm 0.15$ and $1.33\pm 0.11$ in the same regions.  The single-top NF is $1.4\pm 0.4$,  with the large uncertainty due in part to the correlation with the uncertainty related to the removal of the on-shell $tW$ contributions.  Similar values are obtained in the fit for the $\ttbar$+DM  search.  Compared with Ref.~\cite{SUSY-2018-07},  the overall agreement between data and predictions is better due to improved background modelling,   namely the simulation of the \ttbar production with \SHERPA  and  the simulation of the $tW$ production with dynamic scales.  The results of the background-only fits with and without data in SRs are statistically compatible with each other.

\begin{figure}[htp]
\centering
\includegraphics[width=0.42\textwidth]{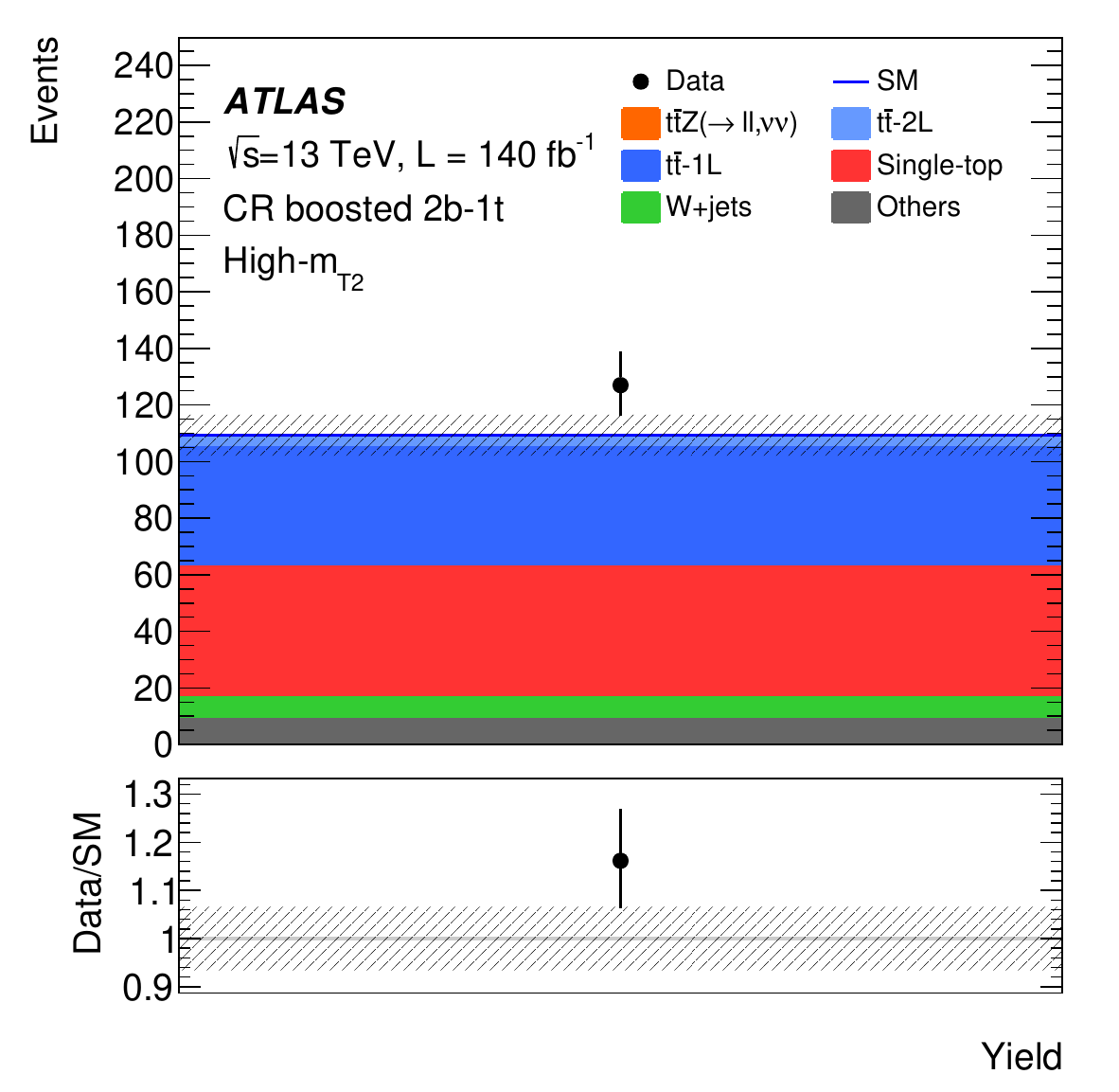}
\includegraphics[width=0.42\textwidth]{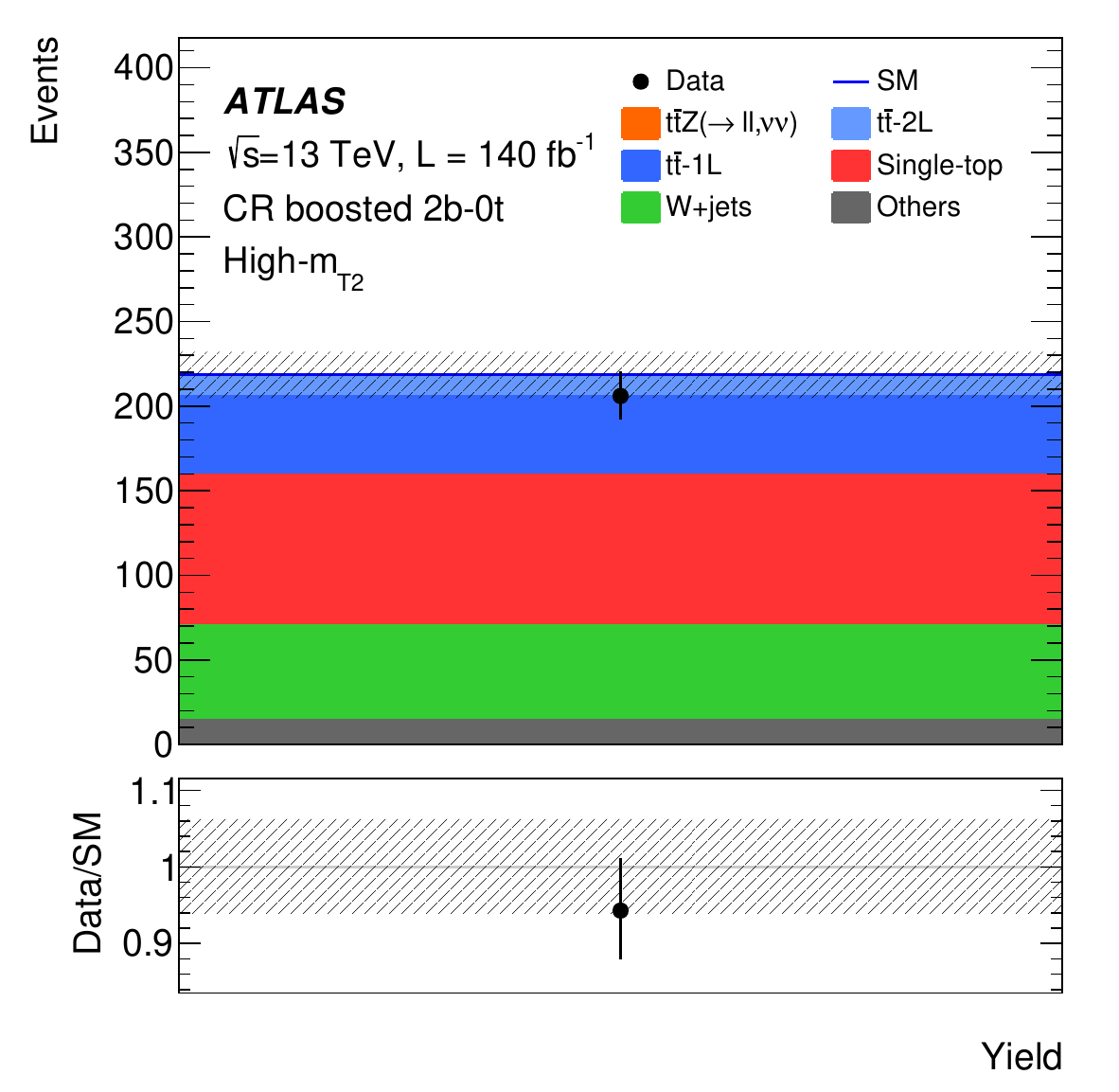}
\caption{Observed and expected  event yields  in the high-$m_{T2}$ CR in the boosted 2b-1t (left) and 2b-0t (right) stop regions.  The expected distributions are as determined in the background-only fit with data in CRs and SRs for the $\stopone\stopone$ search.  The hatched area around the total SM prediction
includes statistical and systematic uncertainties.  The lower panels show the ratio of the observed data to the expected SM events.   }
\end{figure}

\begin{figure}[htp]
\centering
\includegraphics[width=0.42\textwidth]{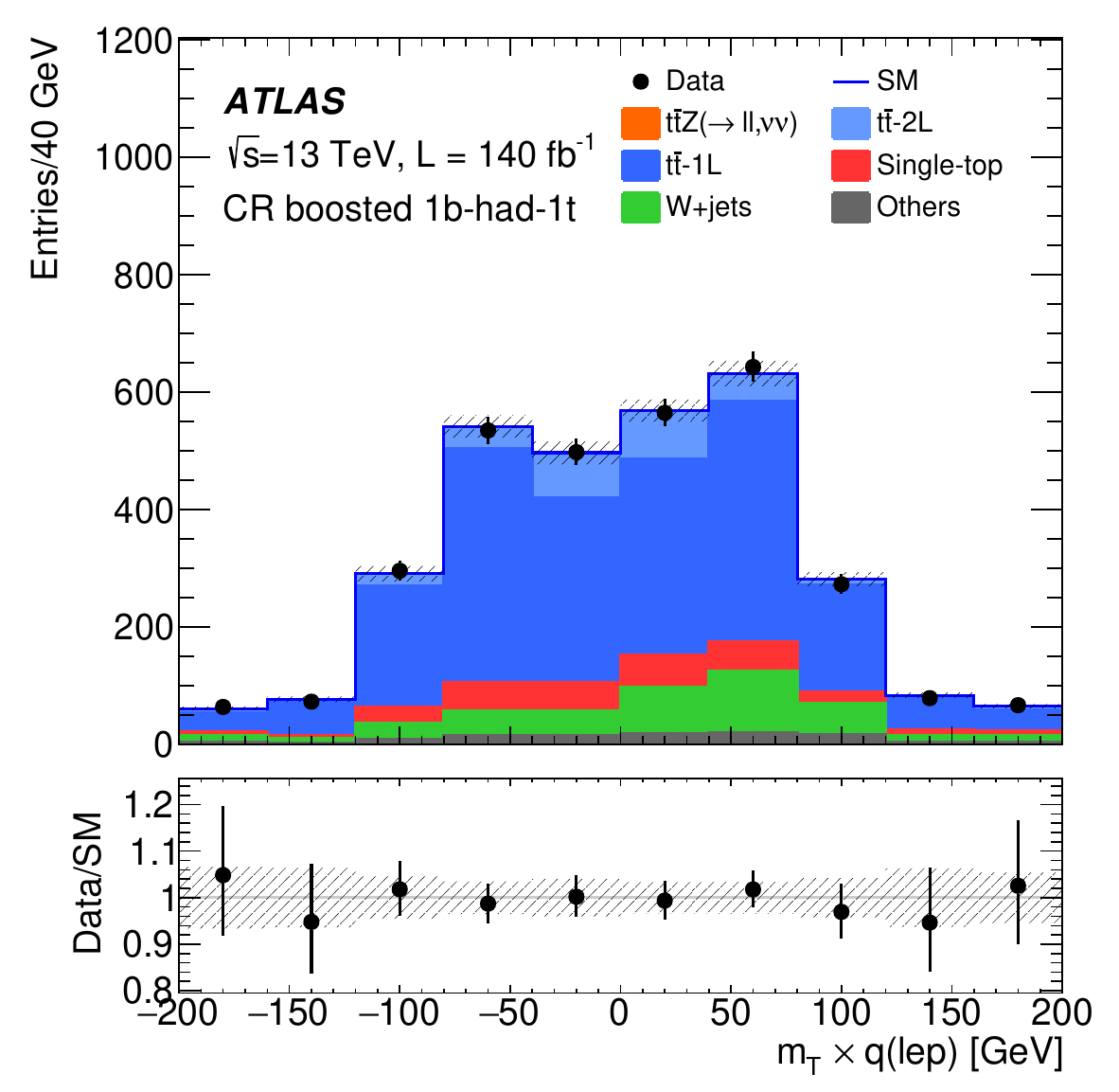}
\includegraphics[width=0.42\textwidth]{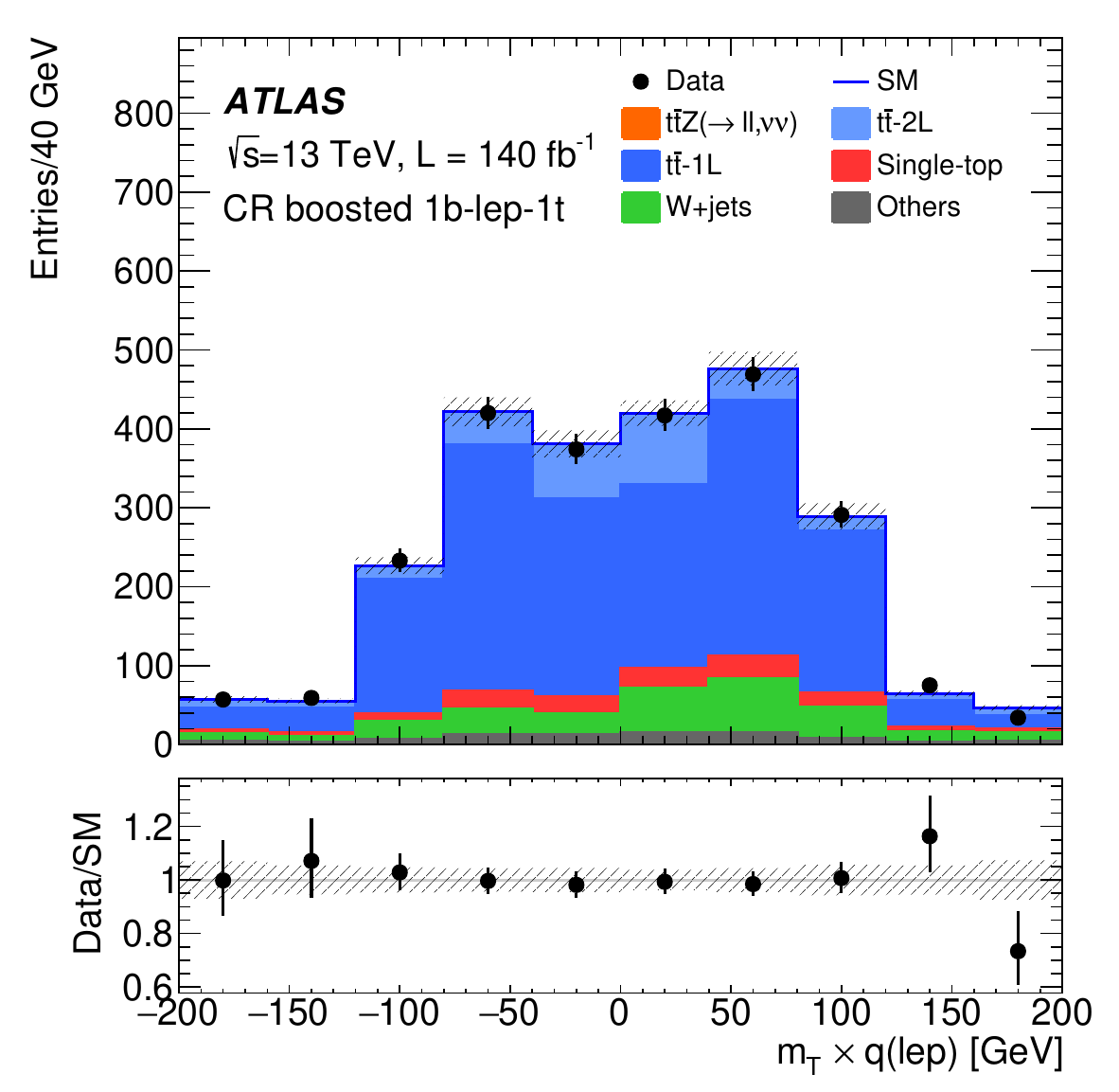}
\includegraphics[width=0.42\textwidth]{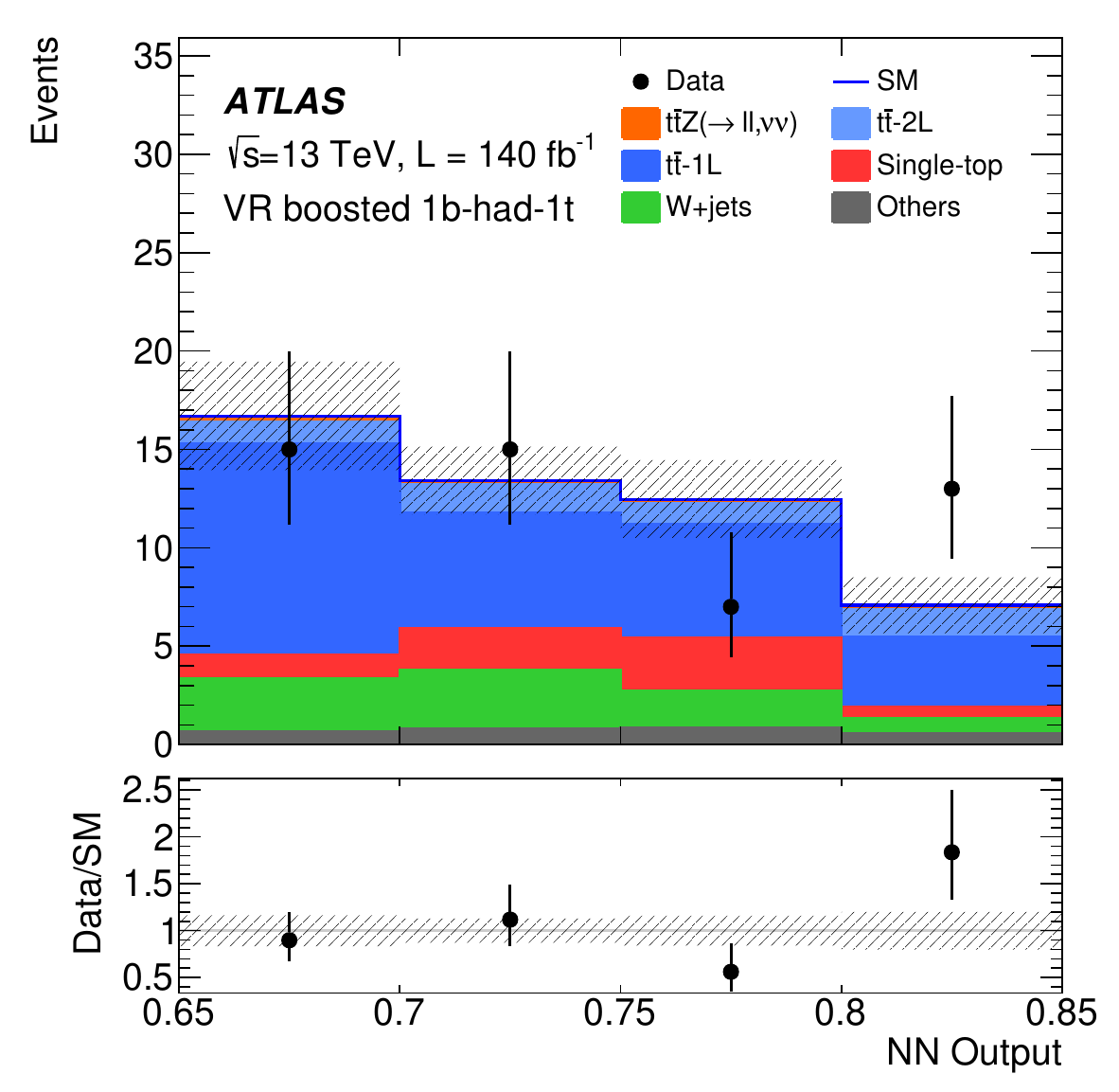}
\includegraphics[width=0.42\textwidth]{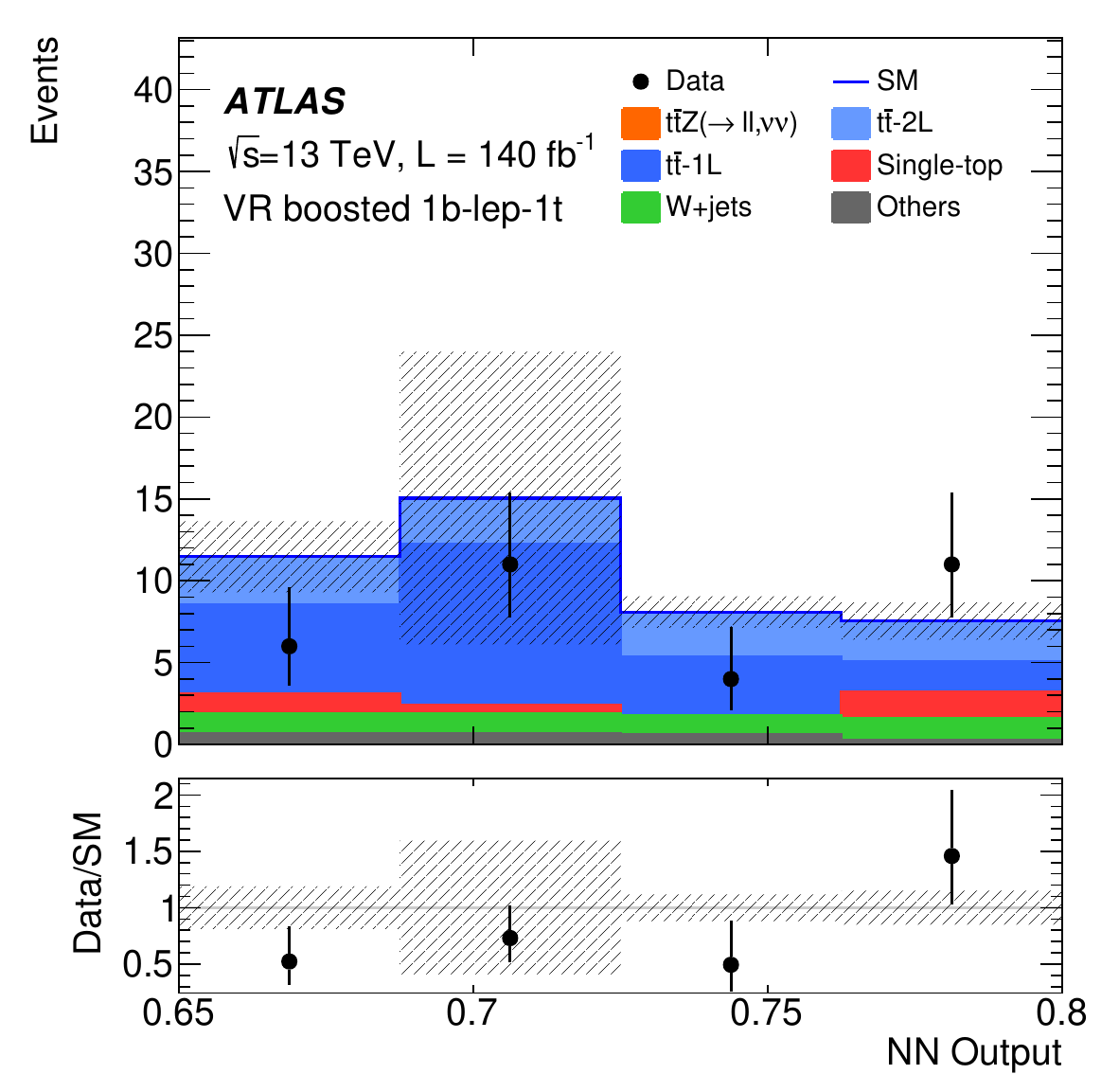}
\includegraphics[width=0.42\textwidth]{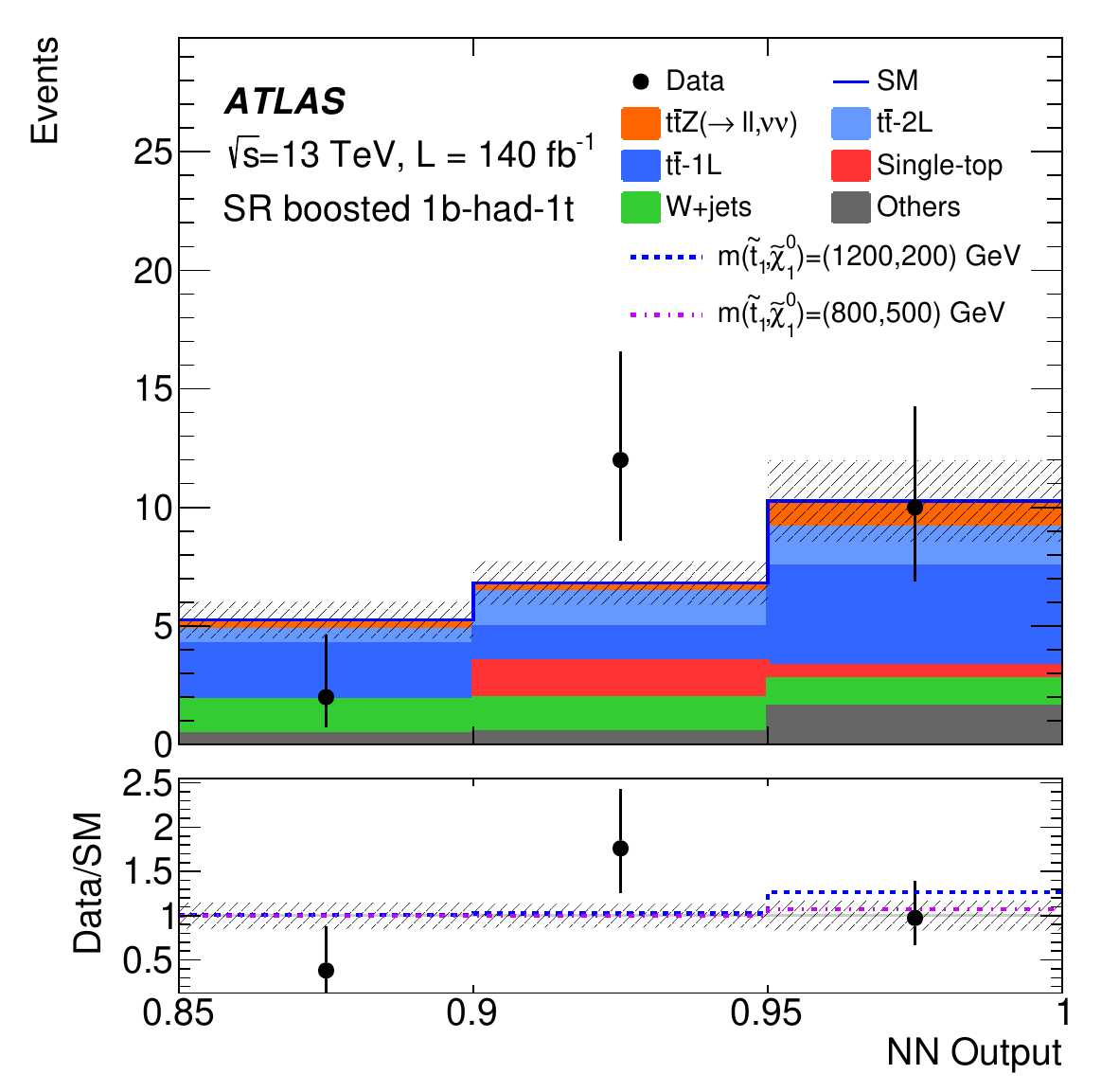}
\includegraphics[width=0.42\textwidth]{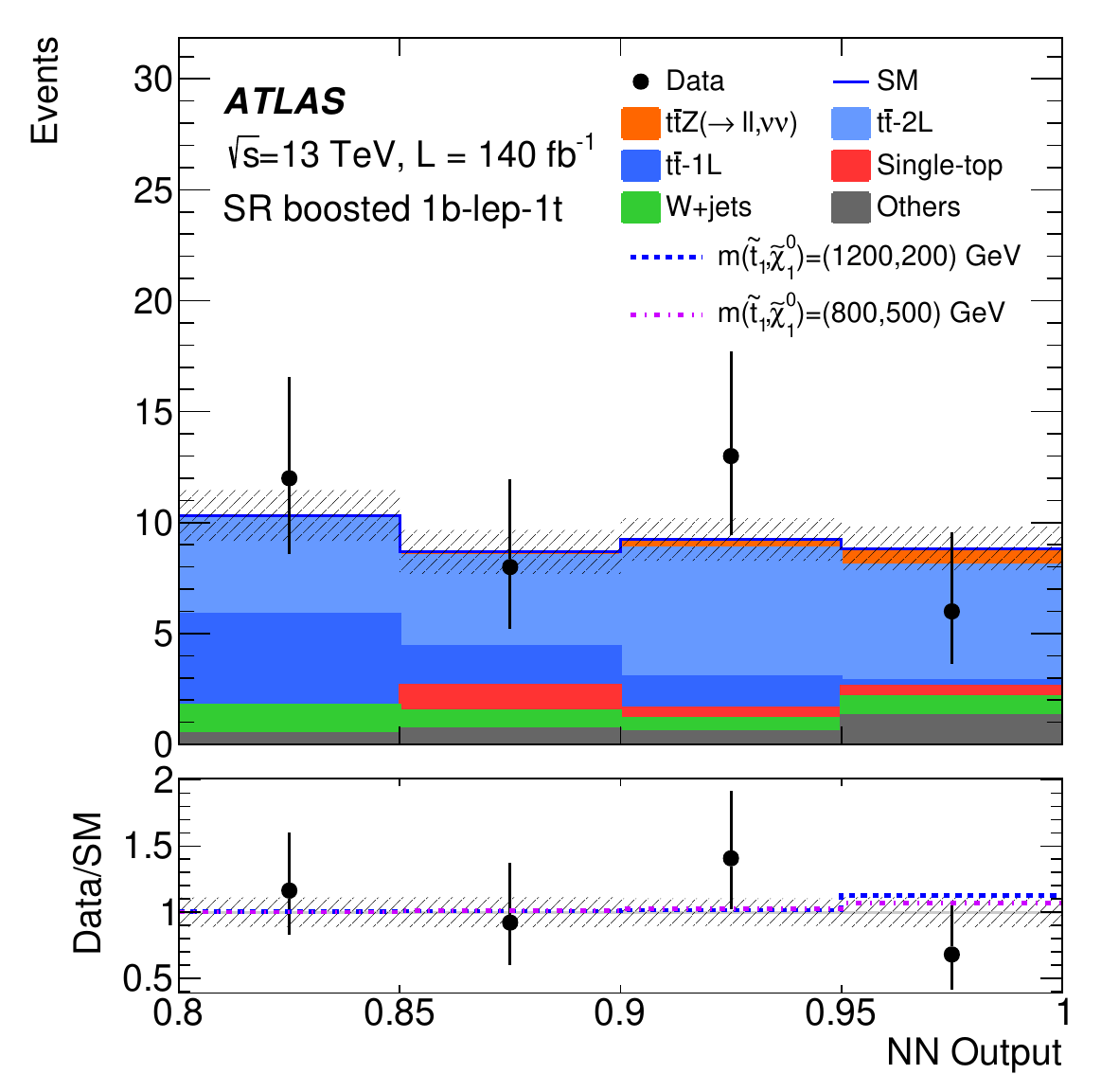}
\caption{Observed and expected  events as a function of $\mt \times q(\ell)$  in the CR (top) and of the stop-NN output value in the VR (middle row) and the SR (bottom) in the boosted 1b-had-1t (left) and  1b-lep-1t (right) stop regions.  The expected distributions are as determined in the background-only fit with data in CRs and SRs  for the $\stopone\stopone$ search.  The hatched area around the total SM prediction
includes statistical and systematic uncertainties.  Uncertainties on predictions in VRs are dominated by  statistical uncertainties on  simulated  samples.  The lower panels show the ratio of the observed data to the expected SM events  and the ratio of the sum of the SM and expected signal  events to the SM events.
In CRs,  the last (first) bin contains overflows (underflows).    }
\end{figure}

\begin{figure}[htp]
\centering
\includegraphics[width=0.42\textwidth]{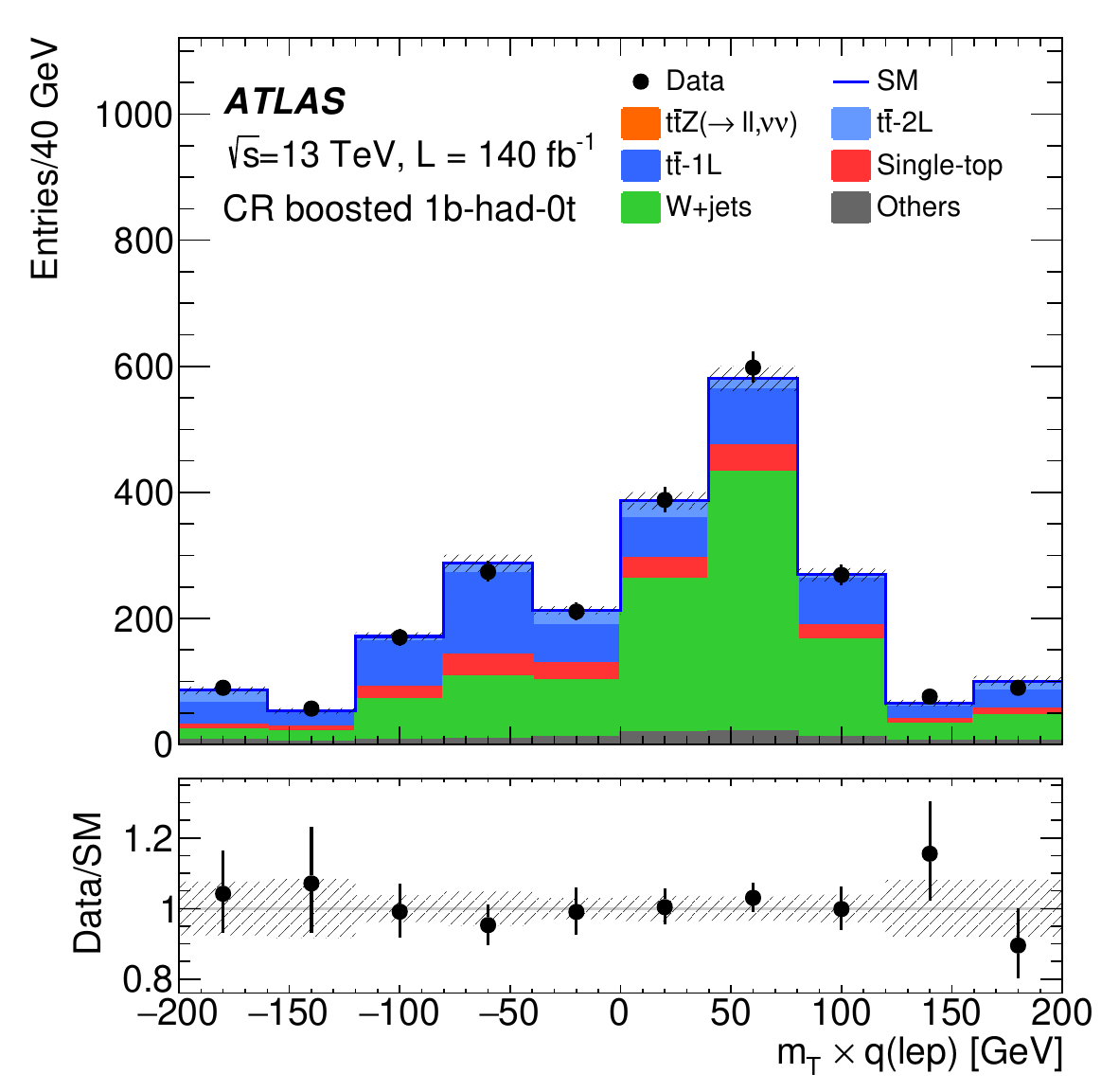}
\includegraphics[width=0.42\textwidth]{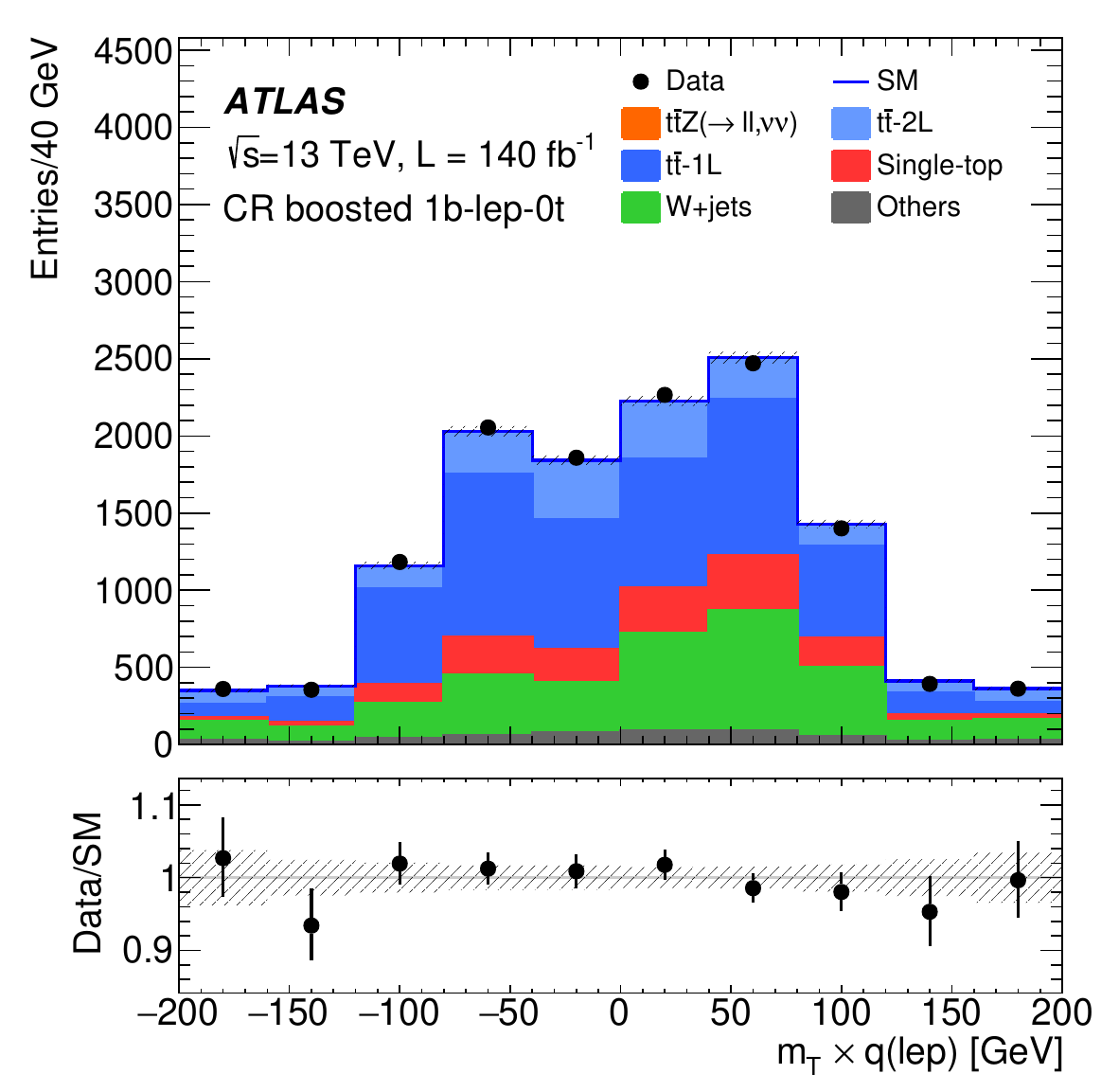}
\includegraphics[width=0.42\textwidth]{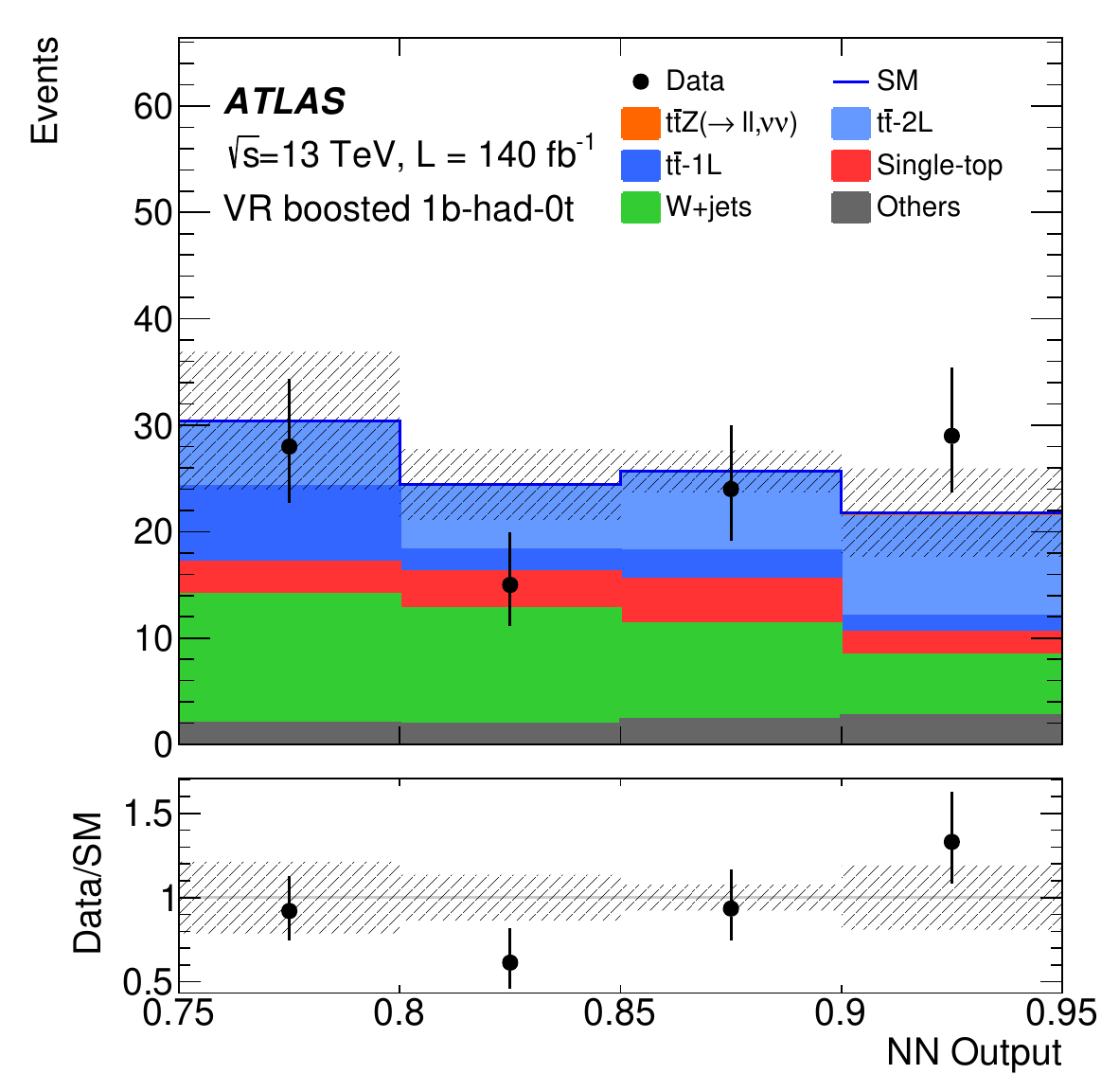}
\includegraphics[width=0.42\textwidth]{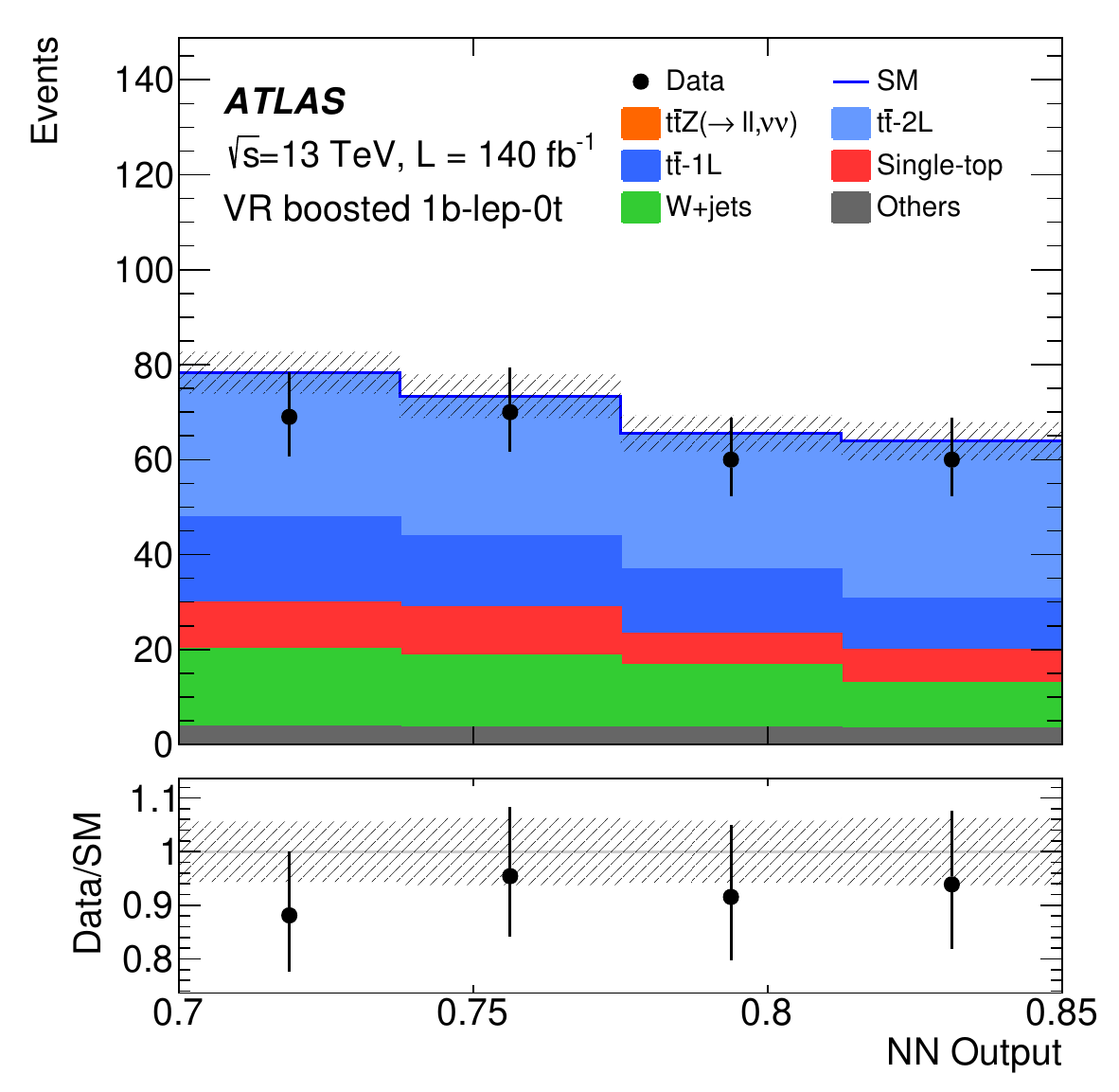}
\includegraphics[width=0.42\textwidth]{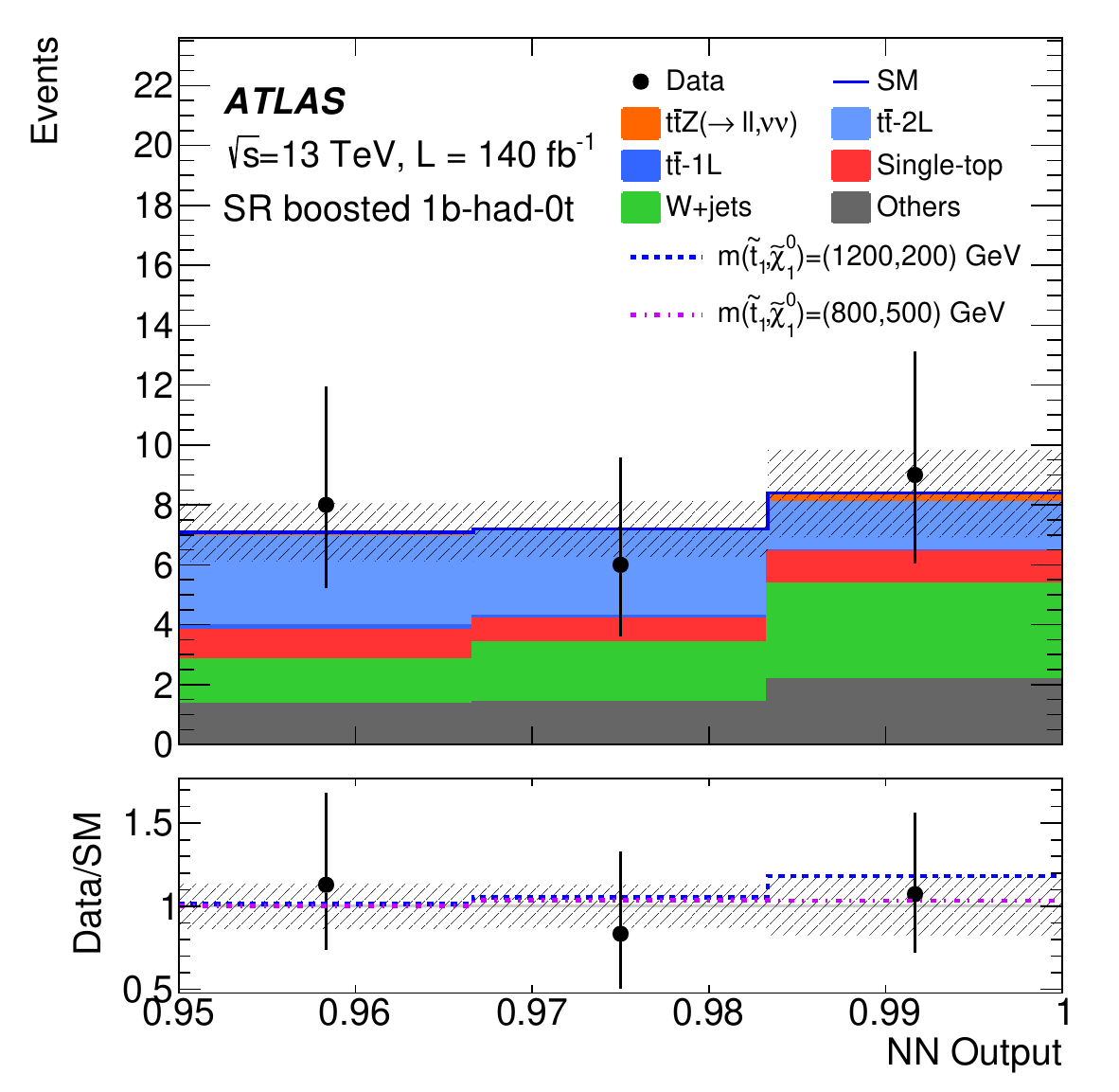}
\includegraphics[width=0.42\textwidth]{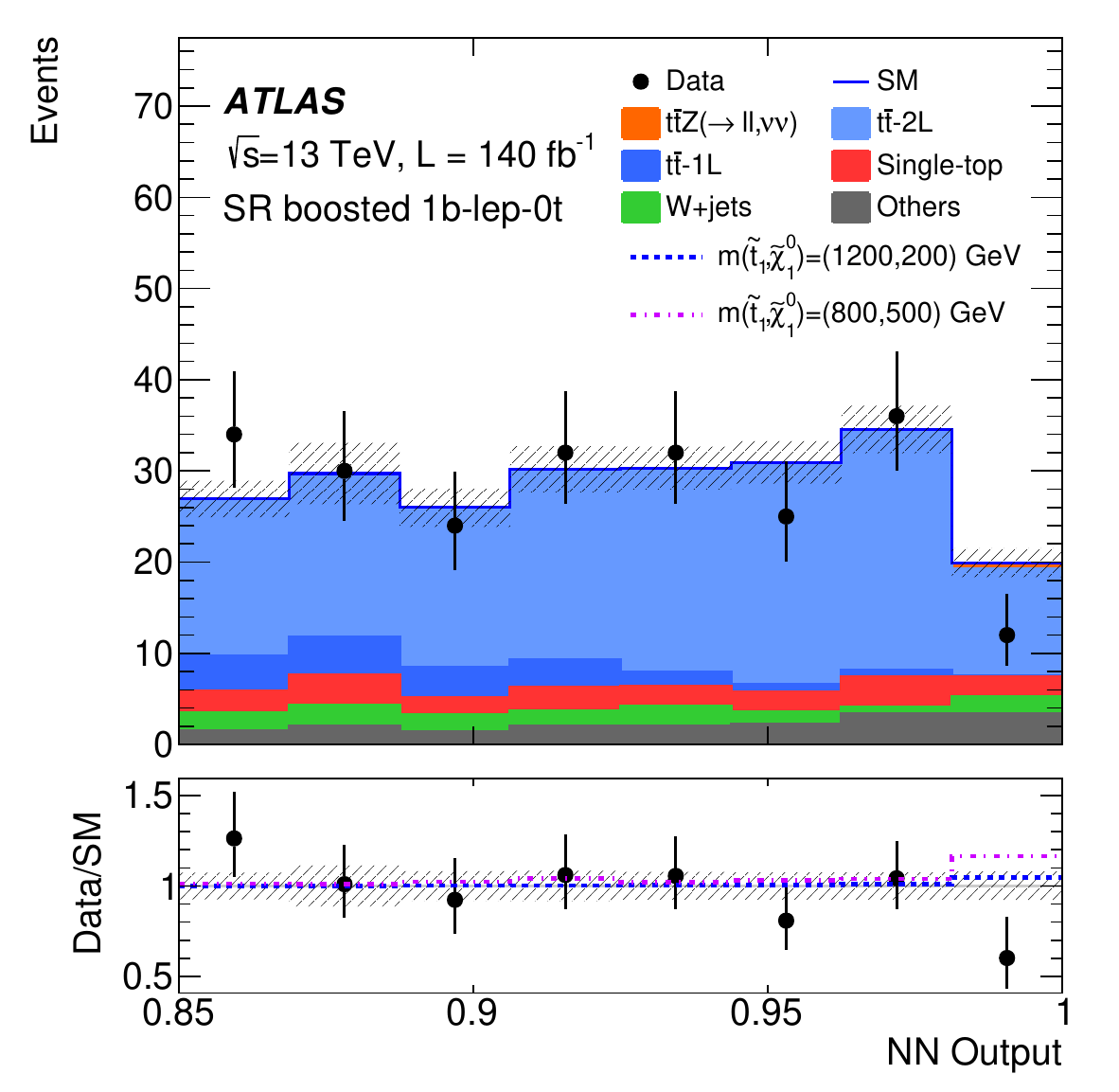}
\caption{Observed and expected  events as a function of  $\mt \times q(\ell)$ in the CR (top) and of the stop-NN output value in the VR (middle row) and the SR (bottom) in the boosted 1b-had-0t (left) and  1b-lep-0t (right) stop regions.  The expected distributions are as determined in the background-only fit with data in CRs and SRs for the $\stopone\stopone$ search.  The hatched area around the total SM prediction
includes statistical and systematic uncertainties.   Uncertainties on predictions in VRs are dominated by  statistical uncertainties on  simulated  samples.   The lower panels show the ratio of the observed data to the expected SM events and the ratio of the sum of the SM and expected signal  events to the SM events.
In CRs,  the last (first) bin contains overflows (underflows).      }
\end{figure}

Good agreement between data and  background predictions is seen in all CRs, VRs and SRs,  although small excesses of events in data compared with predictions are seen in SRs with two $b$-tagged jets in bins at high NN output values.   To quantify the level of agreement between data and predictions and the combined significance of any excesses,  a profile-likelihood-ratio test~\cite{Cowan:2010js} is performed on the result of the full fit,  in which the statistical model includes signal predictions.  In the $\stopone\stopone$ search,   no point in the considered parameter space  leads to a potential non-zero signal contribution  by more than 2.1 standard deviations.  The points with highest significance are those at high $m(\stopone)$ and low $m(\ninoone)$ whose distributions are expected to be strongly peaked at high NN output values and to predominantly populate the boosted 2b-1t category (Figure~\ref{fig:postfit_stop_boosted1}).  For the $\ttbar$+DM search,  no potential non-zero signal contribution is observed with a significance of more than 1.8 standard deviations.  The significance for different models does not show a clear dependence on the mass or parity of the mediator.

The results of the $\stopone\stopone$ and $\ttbar$+DM searches presented in this paper are statistically combined with those from previous searches in  $\ttbar+\met$ final states performed by ATLAS.  For the $\stopone\stopone$ search,  results from this paper are combined with those from the search in the final state with no leptons~\cite{SUSY-2018-12},  which has better sensitivity at high $m(\stopone)$ and low $m(\ninoone)$.  Of the SRs used in Ref.~\cite{SUSY-2018-12},  only SRA and SRB, which target signals  at high $m(\stopone)$,  are included in the combined fit, together with their associated CRs.   For the $\ttbar$+DM search,  the combined fit performed in Ref.~\cite{SUSY-2019-12} with the zero-, one- and two-lepton final states  is updated by replacing the results from the previous one-lepton analysis~\cite{SUSY-2018-07} with the results presented here.  For both combined fits,  it was checked that there is no overlap among the analysed sets of data events.  Fit parameters associated with experimental systematic uncertainties are correlated across all the input analyses,  while NFs and NPs associated with background modelling uncertainties are left uncorrelated.
An alternative fit for the $\stopone\stopone$ search in which all fit parameters across the two analyses besides the parameter of interest were left uncorrelated was found to yield compatible results.

\section{Interpretations}
\label{sec:interpretations}

\begin{figure}[ht]
\centering
\includegraphics[width=0.7\textwidth]{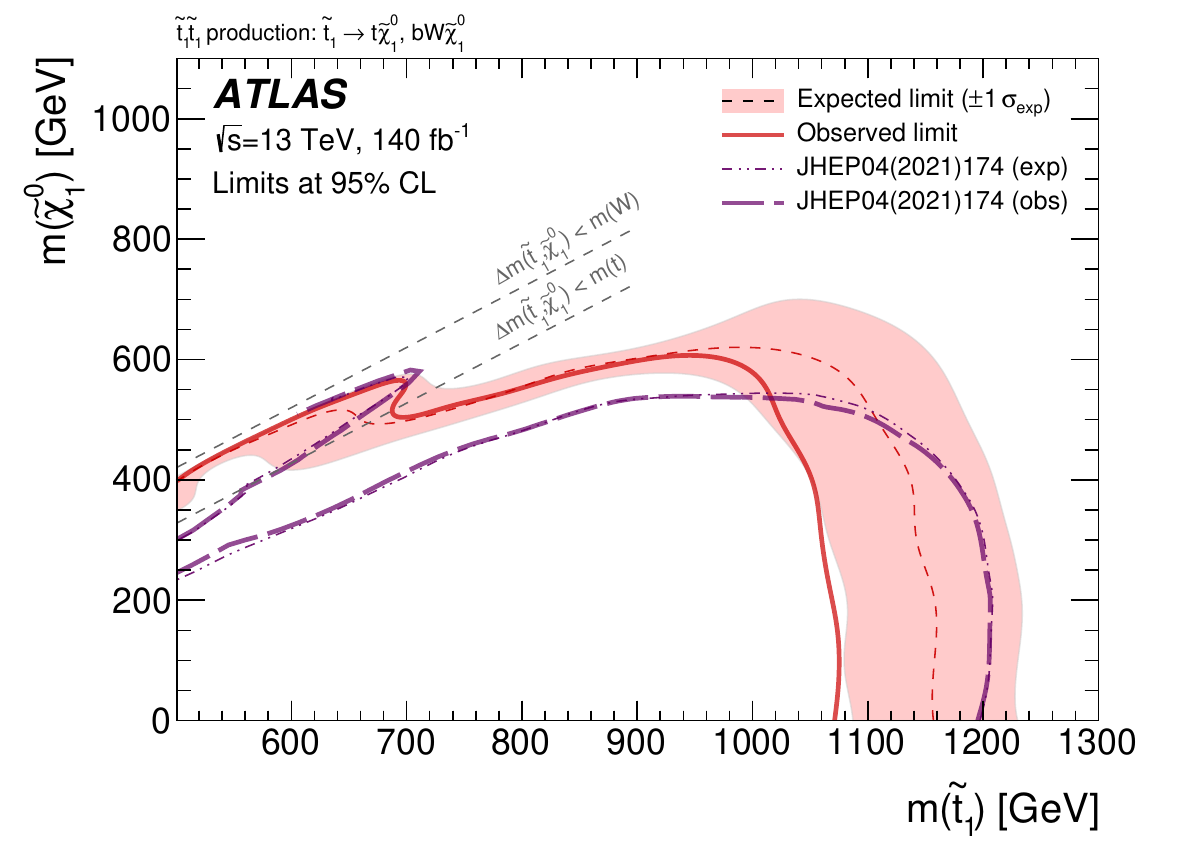}
\caption{Expected and observed 95\% CL excluded regions in the plane of $m(\ninoone)$ and $m(\stopone)$
for $\stopone\stopone$ production
assuming either a $\stop \to t \ninoone$ or $\stop \to bW\ninoone$ decay with a branching ratio of 100\%.  %
Models that lie within the contours are excluded.  The red band indicates the $\pm1\sigma$ variation of the expected limit.  The diagonal dashed lines indicate the kinematic threshold of the stop decay modes.  Also shown is the region excluded by Ref.~\cite{SUSY-2018-07}, which targets a similar final state and uses the same data sample.
}
\label{fig:stop_exclusion}
\end{figure}

\begin{figure}[ht]
\centering
\includegraphics[width=0.7\textwidth]{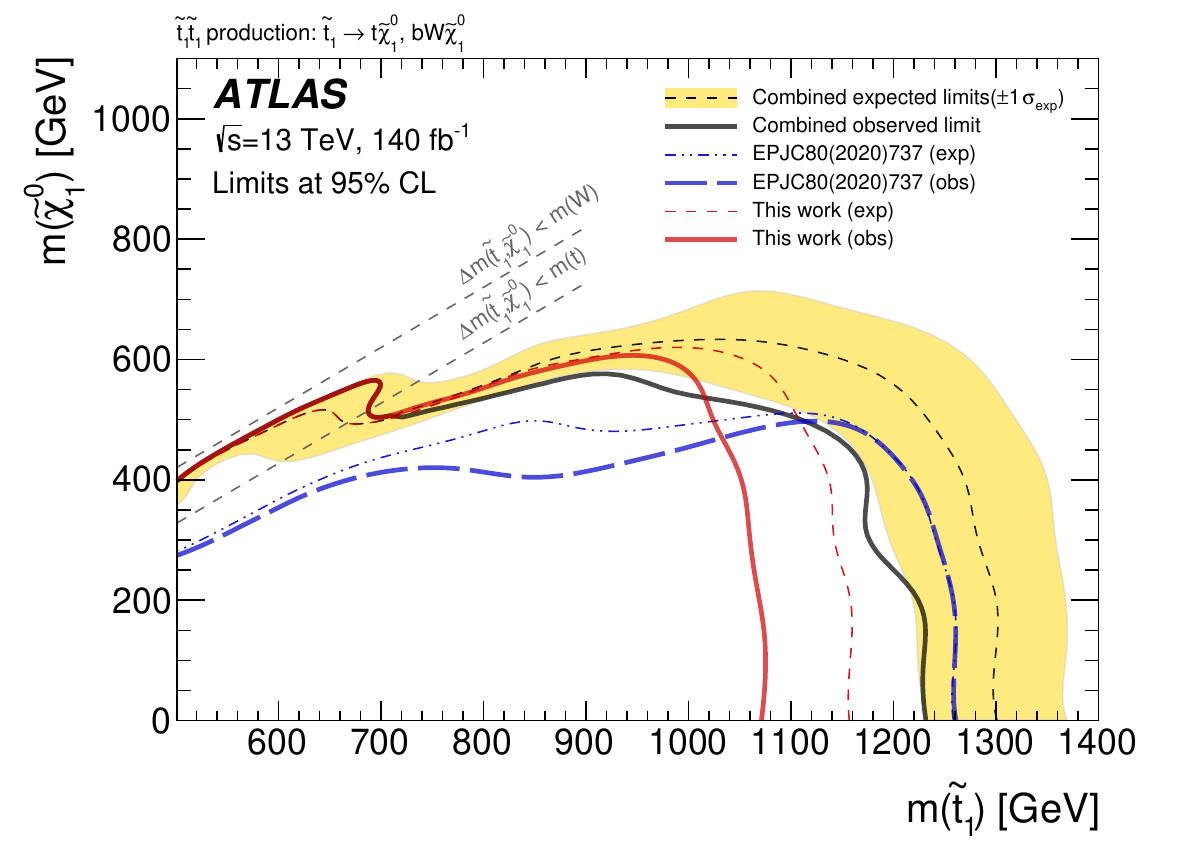}
\caption{Expected and observed 95\% CL excluded regions in the plane of $m(\ninoone)$ and $m(\stopone)$
for $\stopone\stopone$ production
assuming either a $\stop \to t \ninoone$ or $\stop \to bW\ninoone$ decay with a branching ratio of 100\% from the statistical combination of the analysis presented in this paper and Ref.~\cite{SUSY-2018-12}.  %
Models that lie within the contours are excluded.  Uncertainty bands corresponding to the $\pm1\sigma$ variation of the combined expected limit (yellow band)
are also indicated. The diagonal dashed lines indicate the kinematic threshold of the stop decay modes.
}
\label{fig:stop_exclusion_combo}
\end{figure}

As no significant excess is observed,   exclusion limits are calculated using the $\mathrm{CL}_\mathrm{s}$ prescription~\cite{Read:2002hq}  at 95\% confidence level (CL).
Figure~\ref{fig:stop_exclusion} shows the expected and observed exclusion contours as a function of $m(\stopone)$ and    $m(\ninoone)$ for the $\stop \to t \ninoone$ or $\stop \to bW \ninoone$ scenarios.
The $\pm 1\sigma_\mathrm{exp}$ uncertainty band indicates
how much the expected limit is affected by  systematic and statistical uncertainties.
Compared with Ref.~\cite{SUSY-2018-07},  this search significantly improves the sensitivity for high neutralino masses covering the parameter region at $\Delta m(\stopone,\ninoone)\sim m(t)$ which was not probed by the previous search.
At high stop masses where the sensitivity is driven by the statistical power of the recorded data sample,  this analysis reaches a sensitivity close to the one from Ref.~\cite{SUSY-2018-07}.
For signals with three-body decays ($m(W)<\Delta m(\stopone,\ninoone)< m(t)$),  this analysis has a sensitivity  similar to the one from  Ref.~\cite{SUSY-2018-07} even though the stop-NNs have not been trained for these signal events and in Ref.~\cite{SUSY-2018-07} an optimised event selection and an NN-based discriminant were used.
Overall,  the unified approach developed in this search without event categories targetting specific signal models achieves its design target.  It provides more stringent constraints in the challenging signal parameter region at $\Delta m(\stopone,\ninoone)\sim m(t)$ and is almost as sensitive as the approach with dedicated event categories used in the previous search  in the other parameter regions.
Table~\ref{tab:sys_breakdown} shows the main sources of uncertainty on the determination of a potential signal contribution.  Statistical uncertainties are the dominant source,  followed by the theoretical uncertainties on the background predictions.
The exclusion limits determined from the combination of the analysis presented in this paper with Ref.~\cite{SUSY-2018-12} are shown in Figure~\ref{fig:stop_exclusion_combo}.  This combination improves the sensitivity at high $m(\stopone)$ and in the region around $m(\stopone,\ninoone)\sim(1100,500)$~GeV.

\begin{table}[ht]
\caption{Ratios of the signal cross-sections as determined in the fit to the expected signal cross-sections $\mu=\sigma^\mathrm{sig}_\mathrm{fit}/\sigma^\mathrm{sig}_\mathrm{Th}$  for different signal models in the $\stopone\stopone$ and $\ttbar$+DM searches.  Components of the total uncertainty $\sigma(\mu)$ from statistical and major systematic uncertainties are also shown.  The component from statistical uncertainties in data is estimated as the uncertainty on $\mu$ when all nuisance parameters in the fit are fixed.  Components from  systematic uncertainties are estimated as $\sigma_\mathrm{sys}(\mu) = \sqrt{\sigma^2(\mu)-\sigma_\mathrm{fix}^2(\mu)}$,  where $\sigma_\mathrm{fix}(\mu)$ is the uncertainty when the nuisance parameters associated with the systematic uncertainties are fixed.  The components are reported  as percentages relative to the total uncertainty $\sigma(\mu)$ only for positive variations of $\mu$ because  uncertainties on negative variations may be biased by the requirement for $\mu$ to be positive.  Correlations across components are not accounted for.  }
\label{tab:sys_breakdown}
\centering
\renewcommand\arraystretch{1.4}
\begin{tabular}{l|cc|cc}
\toprule
& \multicolumn{2}{c|}{$\stopone\stopone$, $m(\stopone,\ninoone)$ GeV} & \multicolumn{2}{c}{$\ttbar$+DM,  $m(a,\chi)$ GeV} \\
& $(1000,600)$  &$(1200,200)$ & $(50,1)$  & $(150,1)$\\
\midrule
$\mu\pm \sigma(\mu)$ (total uncertainty)  &  $0.25^{+0.42}_{-0.25}$ &  $0.8^{+0.7}_{-0.5}$ &    $0.08^{+0.10}_{-0.08}$  &   $0.12^{+0.13}_{-0.12}$\\
\midrule
Data statistical uncertainty  & $82$ \%   & $74$ \% & $67$ \%  & $69$ \% \\
Background modelling  & $45$ \% & $62$ \% & $51$ \%  & $48$  \%\\
MC statistical uncertainty   & $25$ \%  & $20$ \%  & $34$ \% & $33$ \%   \\
Jet energy scale and resolution  & $20$ \%  & $13$ \%  & $29$ \%  & $28$ \% \\
Flavour tagging efficiency & $18$ \%  & $10$ \%& $21$ \% & $21$ \% \\
\bottomrule
\end{tabular}
\end{table}

\begin{figure}[ht]
\centering
\includegraphics[width=0.48\textwidth]{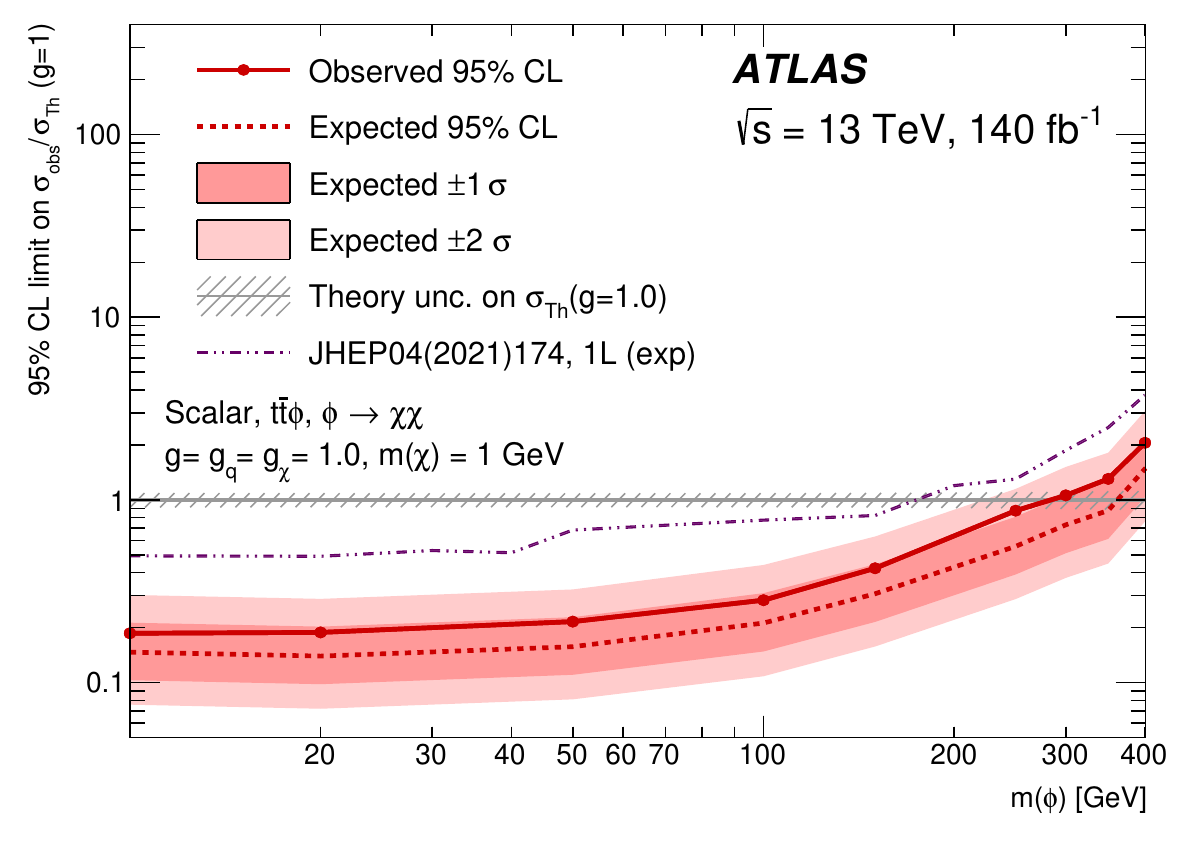}
\includegraphics[width=0.48\textwidth]{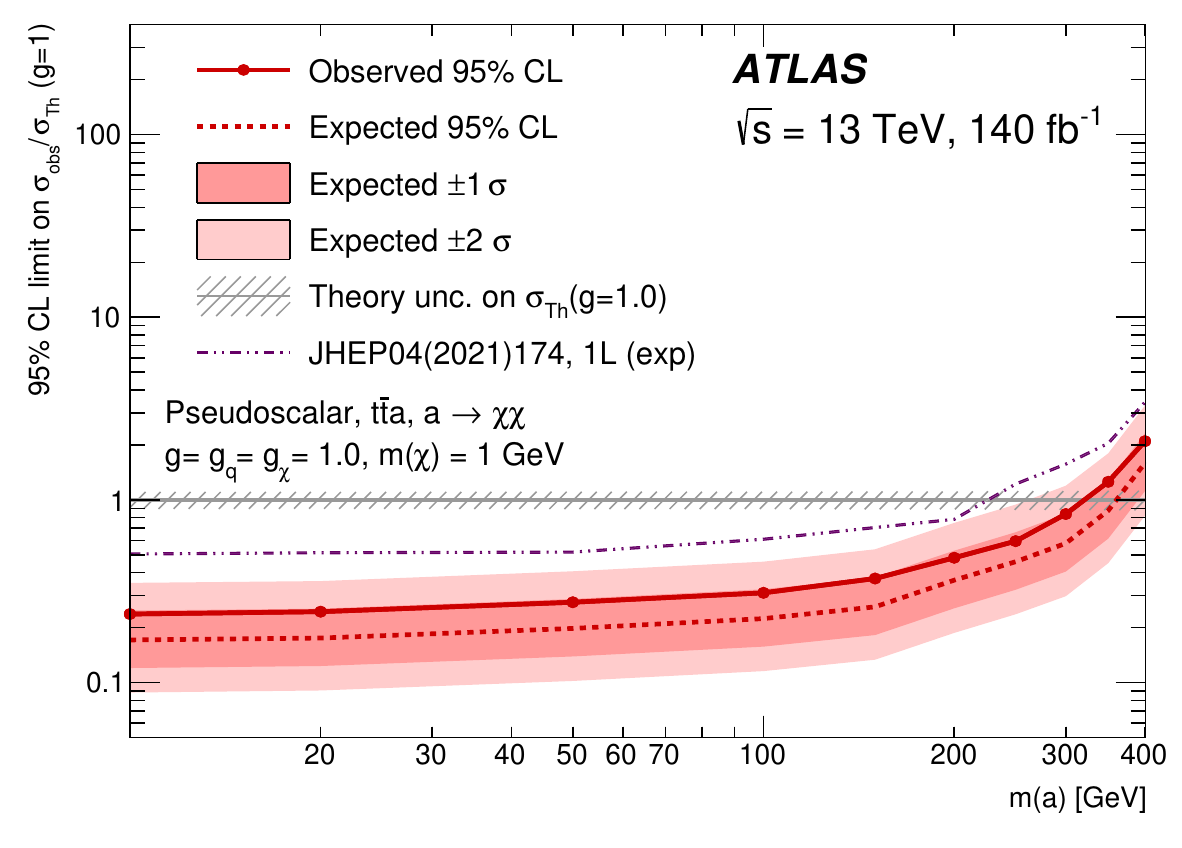}
\includegraphics[width=0.48\textwidth]{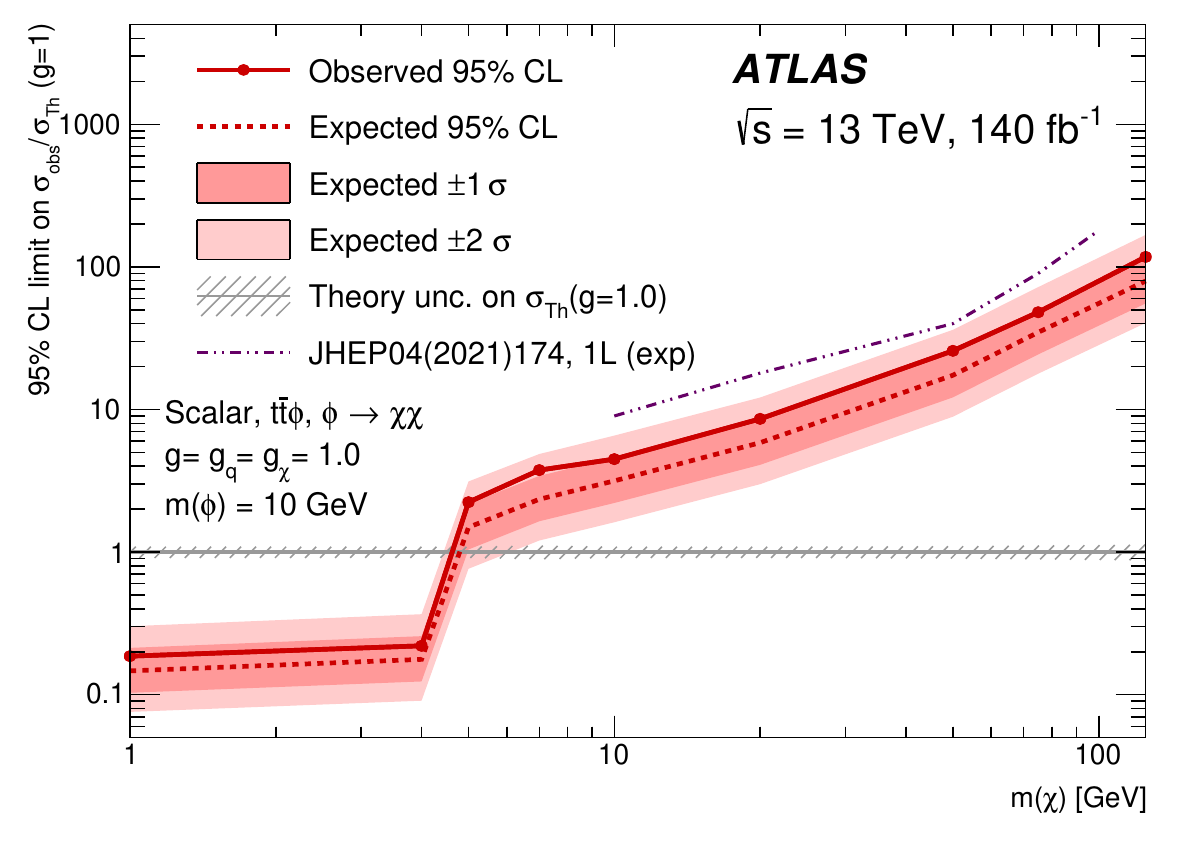}
\includegraphics[width=0.48\textwidth]{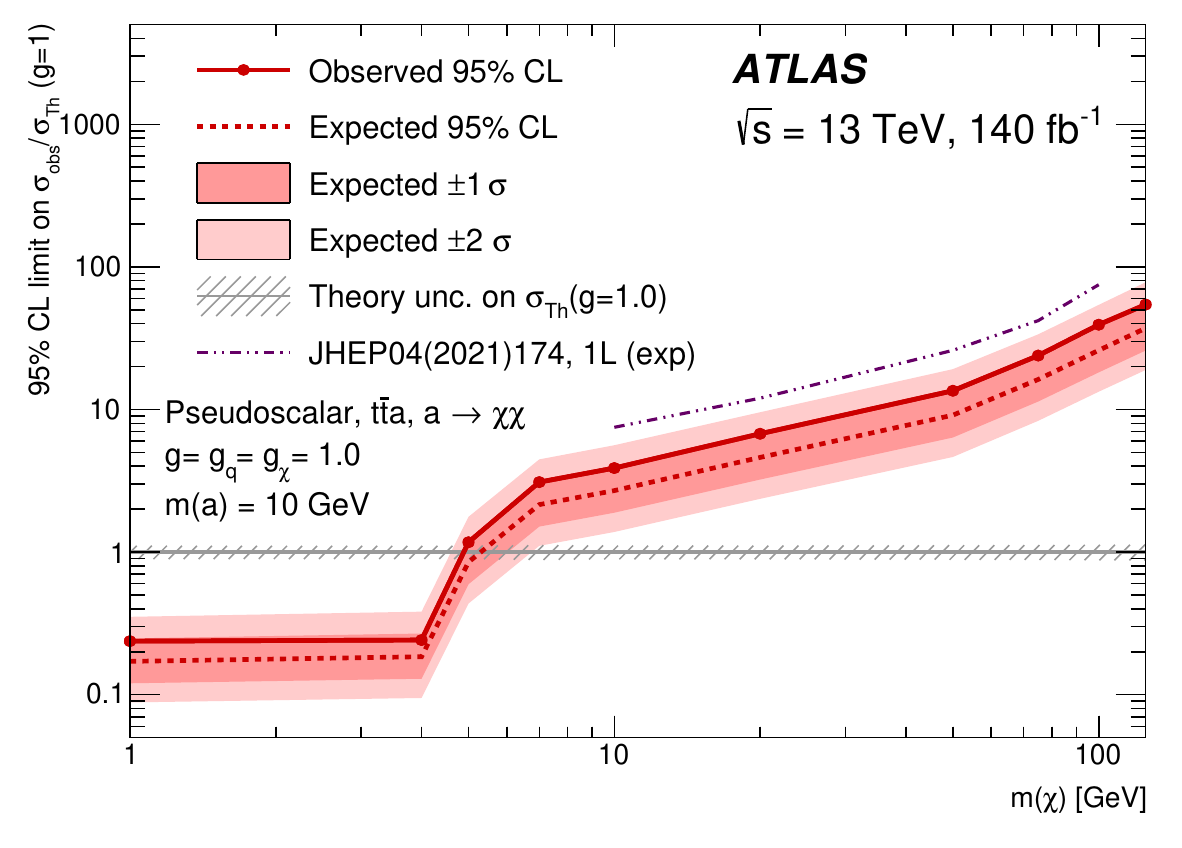}
\caption{Upper limit at 95\% CL on the ratio of the $\ttbar$+DM production cross-section to the theoretical cross-section under the hypothesis of (left) a scalar or (right) a pseudoscalar mediator.  Limits are shown as a function of $m(\phi/a)$ assuming  $m(\chi)=\SI{1}{GeV}$ (top),  or as a function of  $m(\chi)$ assuming  $m(\phi/a)=\SI{10}{GeV}$ (bottom).   All limits are computed assuming $g= 1$.  Limits from Ref.~\cite{SUSY-2018-07} are also shown.
}
\label{fig:DM_exclusion}
\end{figure}

\begin{figure}[ht]
\centering
\includegraphics[width=0.48\textwidth]{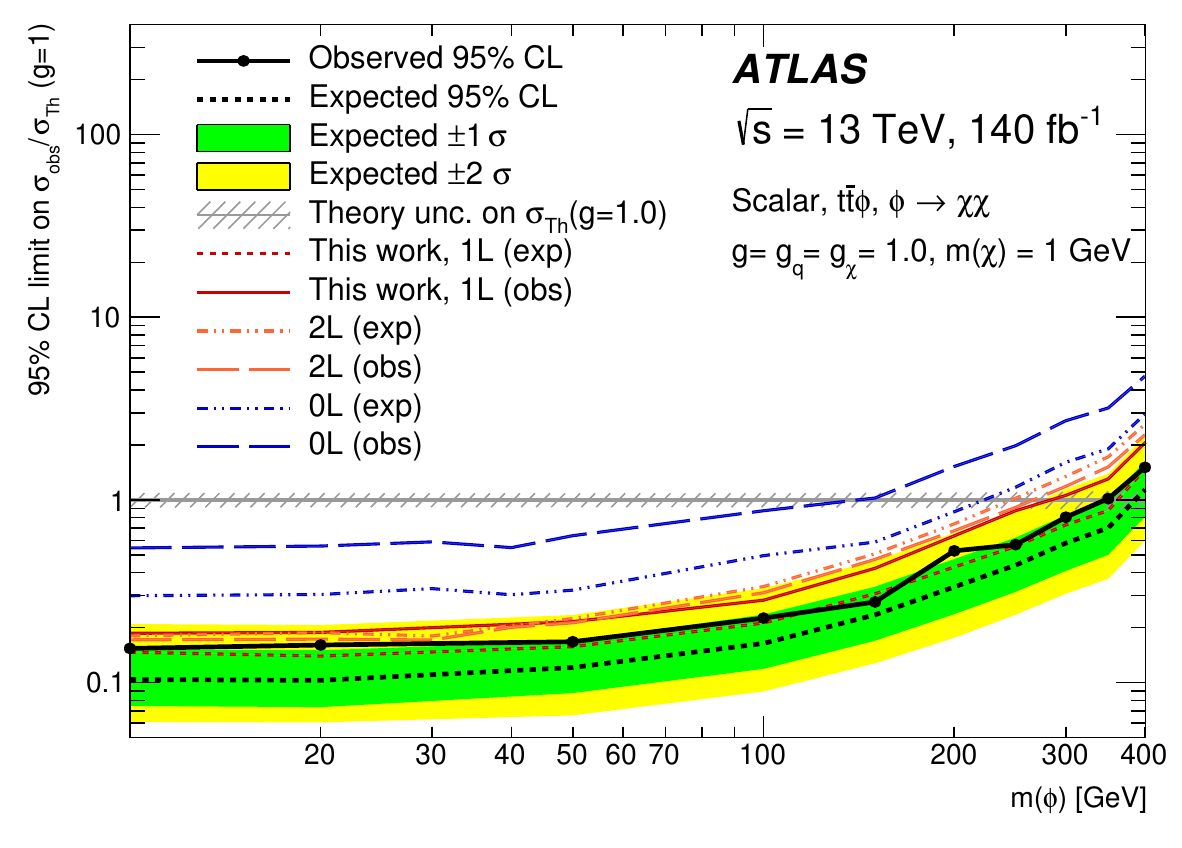}
\includegraphics[width=0.48\textwidth]{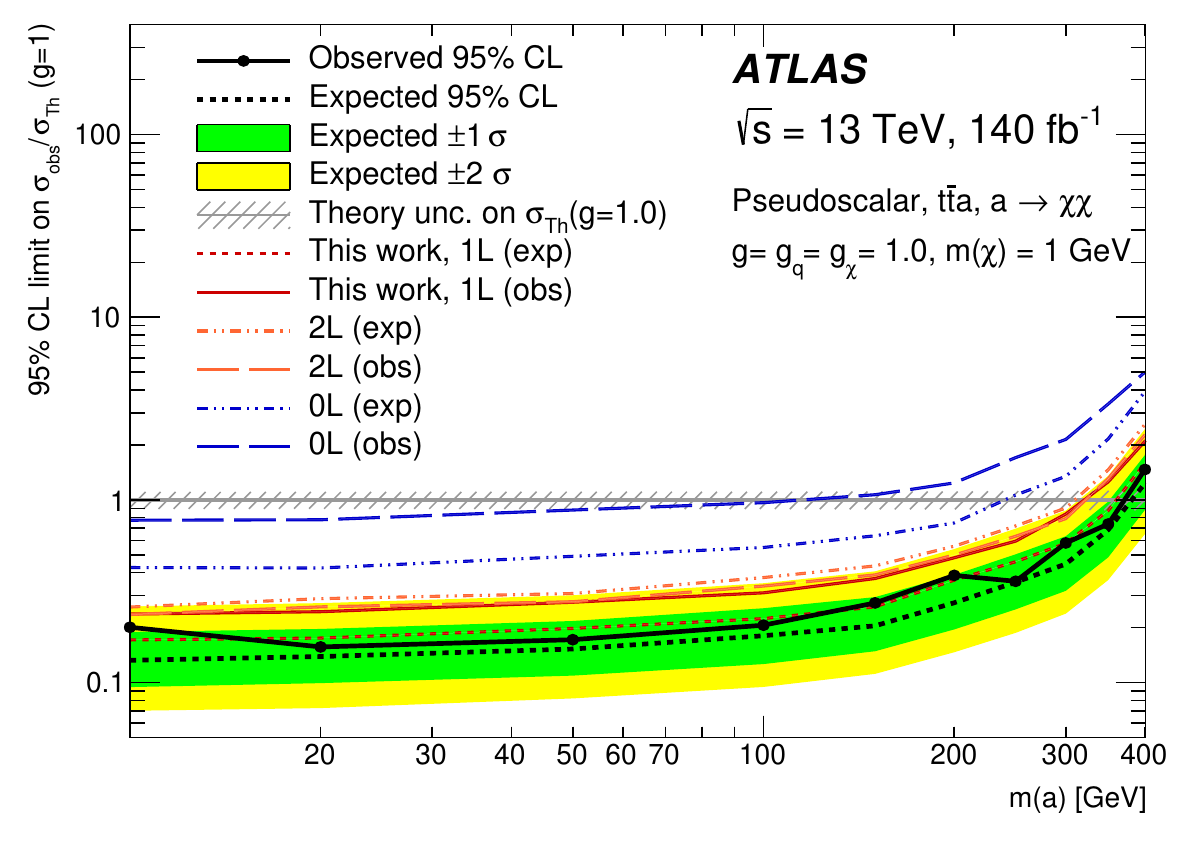}
\caption{Upper limit at 95\% CL on the ratio of the $\ttbar$+DM production cross-section to the theoretical cross-section under the hypothesis of (left) a scalar or (right) a pseudoscalar mediator.  Limits are shown as a function of $m(\phi/a)$ assuming  $m(\chi)=\SI{1}{GeV}$.   All limits are computed assuming $g= 1$.  Combined limits from Ref.~\cite{SUSY-2019-12} are updated to include the analysis in the one-lepton final state presented in this paper.  Limits from the individual analyses included in the combination are also shown.
}
\label{fig:DM_exclusion_combo}
\end{figure}

The exclusion limits for the $\ttbar$+DM search are shown in Figure~\ref{fig:DM_exclusion}.  These limits are computed as upper limits at 95\% CL on the ratio of the observed $\ttbar$+DM production cross-section for the spin-0 mediator model to the theoretical cross-section.   Limits are shown either as a function of $m(\phi/a)$ assuming  $m(\chi)=\SI{1}{GeV}$,  or as a function of  $m(\chi)$ assuming  $m(\phi/a)=\SI{10}{GeV}$.  %
All limits are computed under the hypothesis that
$g=g_\chi=g_q=1$.    All signal uncertainties are included in the fit.  The $\pm 1\sigma$ and $\pm 2\sigma$ uncertainty bands around the expected upper limits account for all statistical and systematic uncertainties.
The upper limits plotted as a function of $m(\chi)$ show a sharp weakening at $m(\chi)=\SI{5}{GeV}$ as this corresponds to the transition from on-shell to off-shell pair production of DM from a mediator with $m(\phi/a)=\SI{10}{GeV}$.
The expected upper limits set by this search supersede  those from the previous search in the one-lepton final state~\cite{SUSY-2018-07} and improve on the combined limits from Ref.~\cite{SUSY-2019-12},  in which the sensitivity is largely dominated by the search in the two-lepton final state.
As shown in Table~\ref{tab:sys_breakdown}, the impact of statistical and systematic uncertainties on the sensitivity of the $\ttbar$+DM search is similar to that of the $\stopone\stopone$ search.
Figure~\ref{fig:DM_exclusion_combo}  shows  the combined limits obtained by updating the previous combination~\cite{SUSY-2019-12} and by replacing the previous analysis of the one-lepton final state with the analysis presented in this paper.

Results from the  $\stopone\stopone$ fit are  used to set constraints on effective $t\bar{t}\nu\bar{\nu}$ contact interactions.
Assuming the magnitude of the Wilson coefficients to be $|V_{ij}|=4\pi$,  lower limits on $\Lambda$ are set at 95\% confidence level depending on the chirality of the top quarks involved in the interaction and the sign of the Wilson coefficient,  as reported in Table~\ref{tab:CI}.
To test the validity of the EFT approach,  limits are also computed considering only subsets of the signal contributions selected by applying upper thresholds on the invariant mass of the two neutrinos involved in the CI ($m_{\nu\bar{\nu}}$).

\begin{table}[ht]
\caption{Constraints on effective $\ttbar\nu\bar{\nu}$ contact interactions involving all three generations of left-handed neutrinos based on the results of the $\stopone\stopone$ search.  Constraints are set independently for different   effective vector operators and for different hypotheses about the sign of the Wilson coefficient which leads to a constructive or destructive interference with $\ttbar Z(\to\nu\bar{\nu})$.  Observed (expected) limits at 95\% CL are reported for $\sqrt{|V_{ij}|}/\Lambda$ and for $\Lambda$, for the full phase space and for specified regions of the true invariant mass of the neutrino pair,
assuming $|V_{ij}|=4\pi$.   Limits corresponding to $\pm 1 \sigma$ variations of the expected limits are also reported. }%
\label{tab:CI}
\centering
\renewcommand\arraystretch{1.4}
\begin{tabular}{l|>{\centering\arraybackslash}p{4cm}>{\centering\arraybackslash}p{4cm}}
\toprule
Wilson coefficient  &	 Observed (Expected) upper limit on $\sqrt{|V_{ij}|}/\Lambda$ [TeV$^{-1}$]	  & Observed (Expected)   lower limit on	$\Lambda$ for $|V_{ij}|=4\pi$ [TeV]  \\
\midrule
$V_{LL}>0$ &		1.59 (1.44$^{1.58}_{1.31}$) &	2.23 (2.47$^{2.71}_{2.25}$) \\
\hfill $m_{\nu\bar{\nu}}<\SI{1}{TeV}$   &	1.84 (1.66$^{1.82}_{1.51}$)  &	1.93 (2.14$^{2.35}_{1.95}$) \\
\hfill $m_{\nu\bar{\nu}}<\SI{2}{TeV}$   &	1.62 (1.46$^{1.61}_{1.36}$)  &	2.18 (2.42$^{2.66}_{2.21}$) \\
\midrule
$V_{LL}<0$  &	1.66 (1.52$^{1.66}_{1.40}$)  &	2.13 (2.33$^{2.53}_{2.14}$) \\
\hfill $m_{\nu\bar{\nu}}<\SI{1}{TeV}$   &	1.96 (1.80$^{1.95}_{1.66}$)   &	1.81  (1.97$^{2.13}_{1.82}$) \\
\hfill  $m_{\nu\bar{\nu}}<\SI{2}{TeV}$   &	1.70 (1.56$^{1.69}_{1.44}$)  &	2.08 (2.28$^{2.47}_{2.10}$) \\
\midrule

$V_{LR}>0$ &		1.67 (1.53$^{1.66}_{1.40}$)  &	2.12 (2.32$^{2.53}_{2.13}$) \\
\hfill $m_{\nu\bar{\nu}}<\SI{1}{TeV}$  &	1.92 (1.78$^{1.94}_{1.64}$)   &	1.84 (1.99$^{2.16}_{1.82}$)\\
\hfill $m_{\nu\bar{\nu}}<\SI{2}{TeV}$   &	1.70 (1.56$^{1.70}_{1.44}$)   &	2.08 (2.27$^{2.47}_{2.08}$) \\
\midrule
$V_{LR}<0$ &		1.63 (1.49$^{1.63}_{1.36}$)  &	2.17 (2.38$^{2.60}_{2.18}$) \\
\hfill $m_{\nu\bar{\nu}}<\SI{1}{TeV}$   &	1.86 (1.72$^{1.89}_{1.58}$)   &	1.91 (2.06$^{2.25}_{1.88}$) \\
\hfill  $m_{\nu\bar{\nu}}<\SI{2}{TeV}$  &	1.66 (1.52$^{1.67}_{1.40}$)  &	2.13 (2.33$^{2.54}_{2.13}$) \\
\bottomrule
\end{tabular}
\end{table}

\FloatBarrier

\section{Discussion and  conclusions}

This paper presents searches for the direct production of a pair of top squarks,  for the production of a spin-0 mediator that decays into DM particles in association with a pair of top quarks, and for effective $t\bar{t}\nu\bar{\nu}$ contact interactions,  using the full Run~2 $pp$ data sample recorded by ATLAS at $\sqrt{s}=\SI{13}{TeV}$.  The searches are conducted with events compatible with the presence of one semileptonic and one hadronic top-quark decay
and with large \met.  The results are combined with previous searches in similar final states but with different lepton multiplicities.

The new searches employ an analysis approach based on neural networks and event categories designed to target final states rather than specific signal model parameter regions.
Event categories are defined  using the  multiplicities and kinematic properties of  jets, $b$-tagged jets and \met. In each category,  neural networks are  trained by considering potential signals across the parameter space under study, so that the resulting signal regions are sensitive to wide ranges of the parameter space.  A novel approach using neural networks is also developed to reconstruct hadronic top-quark decays.

In the $\stopone\stopone$ search,  the sensitivity to  stop pair production for a mass splitting between the stop and the neutralino masses near the top quark mass is significantly improved compared with Ref.~\cite{SUSY-2018-07}, which uses the same data sample.  For models with heavy stops and light neutralinos,  the sensitivity of the new search, which is  not as high as in Ref.~\cite{SUSY-2018-07},  is instead aided by the combination with Ref.~\cite{SUSY-2018-12},  which is performed in final states with no leptons and is the most sensitive study by ATLAS in that parameter region to date.  As a result,  the combined sensitivity for models  with two-body decays
is improved  across the full parameter region and supersedes the sensitivities of  Refs.~\cite{SUSY-2018-07, SUSY-2018-12}.
The sensitivity of the new search  to spin-0 mediators is also improved across the full parameter space.  In addition to enhanced sensitivity,  this analysis approach offers the benefit that no dedicated signal region needs to be optimised for specific regions of the parameter space.  Moreover,  with only one discriminant per category for all signal models under test, the analysis is simpler, since the data and background distributions remain the same while different signal predictions  are tested.  This makes the validation of the background modelling more straightforward,  and the interpretability of results for new signals is likely to be simpler and more  effective.  Another important development included in this paper is the improved modelling of top quark production in both \ttbar and single-top processes.

No statistically significant deviations from the SM expectation are observed and exclusion limits at 95\% confidence level are set by combining the new searches with previous analyses performed in similar final states with different lepton multiplicities requirements.
In the simplified models considered in this paper,  stops are excluded for masses up to 1230~GeV, while neutralinos are excluded for masses  up to  570~GeV.
These limits are slightly weaker than  expected  due to the presence of  small excesses of data over the predicted background.
In the $\ttbar$+DM search,  when assuming $g=g_{q}=g_\chi=1$ for the strengths of the couplings of the mediator to DM and SM particles,  models with scalar (pseudoscalar)  mediators with masses up to 350 (370)~GeV  are excluded.  At lower mediator masses,  models with production cross-sections as small as 0.15 (0.16) times the nominal predictions are excluded.  %
Results from the analysis presented in this paper are also interpreted in the context of a search for effective vector contact interactions between top quarks and all three generations of left-handed neutrinos.  Assuming the Wilson coefficient to satisfy $|V_{ij}|=4\pi$, lower limits on  the energy scale of new physics $\Lambda$ at 95\% confidence level are between \SI{2.12}{TeV} and \SI{2.23}{TeV},  depending on the chirality of the top quarks involved in the CI and the sign of the Wilson coefficient.

\section*{Acknowledgements}
%

% The next lines are included from the .//acknowledgements/Acknowledgements.tex input file
%
%

%
%

We thank CERN for the very successful operation of the LHC, as well as the
support staff from our institutions without whom ATLAS could not be
operated efficiently.

We acknowledge the support of
ANPCyT, Argentina;
YerPhI, Armenia;
ARC, Australia;
BMWFW and FWF, Austria;
ANAS, Azerbaijan;
CNPq and FAPESP, Brazil;
NSERC, NRC and CFI, Canada;
CERN;
ANID, Chile;
CAS, MOST and NSFC, China;
Minciencias, Colombia;
MEYS CR, Czech Republic;
DNRF and DNSRC, Denmark;
IN2P3-CNRS and CEA-DRF/IRFU, France;
SRNSFG, Georgia;
BMBF, HGF and MPG, Germany;
GSRI, Greece;
RGC and Hong Kong SAR, China;
ISF and Benoziyo Center, Israel;
INFN, Italy;
MEXT and JSPS, Japan;
CNRST, Morocco;
NWO, Netherlands;
RCN, Norway;
MEiN, Poland;
FCT, Portugal;
MNE/IFA, Romania;
MESTD, Serbia;
MSSR, Slovakia;
ARRS and MIZ\v{S}, Slovenia;
DSI/NRF, South Africa;
MICINN, Spain;
SRC and Wallenberg Foundation, Sweden;
SERI, SNSF and Cantons of Bern and Geneva, Switzerland;
MOST, Taipei;
TENMAK, T\"urkiye;
STFC, United Kingdom;
DOE and NSF, United States of America.
In addition, individual groups and members have received support from
BCKDF, CANARIE, CRC and DRAC, Canada;
PRIMUS 21/SCI/017 and UNCE SCI/013, Czech Republic;
COST, ERC, ERDF, Horizon 2020, ICSC-NextGenerationEU and Marie Sk{\l}odowska-Curie Actions, European Union;
Investissements d'Avenir Labex, Investissements d'Avenir Idex and ANR, France;
DFG and AvH Foundation, Germany;
Herakleitos, Thales and Aristeia programmes co-financed by EU-ESF and the Greek NSRF, Greece;
BSF-NSF and MINERVA, Israel;
Norwegian Financial Mechanism 2014-2021, Norway;
NCN and NAWA, Poland;
La Caixa Banking Foundation, CERCA Programme Generalitat de Catalunya and PROMETEO and GenT Programmes Generalitat Valenciana, Spain;
G\"{o}ran Gustafssons Stiftelse, Sweden;
The Royal Society and Leverhulme Trust, United Kingdom.

The crucial computing support from all WLCG partners is acknowledged gratefully, in particular from CERN, the ATLAS Tier-1 facilities at TRIUMF/SFU (Canada), NDGF (Denmark, Norway, Sweden), CC-IN2P3 (France), KIT/GridKA (Germany), INFN-CNAF (Italy), NL-T1 (Netherlands), PIC (Spain), RAL (UK) and BNL (USA), the Tier-2 facilities worldwide and large non-WLCG resource providers. Major contributors of computing resources are listed in Ref.~\cite{ATL-SOFT-PUB-2023-001}.

%
%

% End of text imported from the .//acknowledgements/Acknowledgements.tex input file

%
%
%
\printbibliography
\clearpage
\input{atlas_authlist}
%

%
%

%

%
%
%
%

\end{document}

%% file: atlas_authlist.tex
% ATLAS Collaboration author list
% Reference date of SUSY-2023-22 is 2023-11-10
% Author list last updated on date 22-JAN-24
% Data extracted on 22-Jan-2024 for paper reference SUSY-2023-22
% at 2:17pm
 
\begin{flushleft}
\hypersetup{urlcolor=black}
{\Large The ATLAS Collaboration}

\bigskip

\AtlasOrcid[0000-0002-6665-4934]{G.~Aad}$^\textrm{\scriptsize 103}$,
\AtlasOrcid[0000-0001-7616-1554]{E.~Aakvaag}$^\textrm{\scriptsize 16}$,
\AtlasOrcid[0000-0002-5888-2734]{B.~Abbott}$^\textrm{\scriptsize 121}$,
\AtlasOrcid[0000-0002-1002-1652]{K.~Abeling}$^\textrm{\scriptsize 55}$,
\AtlasOrcid[0000-0001-5763-2760]{N.J.~Abicht}$^\textrm{\scriptsize 49}$,
\AtlasOrcid[0000-0002-8496-9294]{S.H.~Abidi}$^\textrm{\scriptsize 29}$,
\AtlasOrcid[0009-0003-6578-220X]{M.~Aboelela}$^\textrm{\scriptsize 44}$,
\AtlasOrcid[0000-0002-9987-2292]{A.~Aboulhorma}$^\textrm{\scriptsize 35e}$,
\AtlasOrcid[0000-0001-5329-6640]{H.~Abramowicz}$^\textrm{\scriptsize 152}$,
\AtlasOrcid[0000-0002-1599-2896]{H.~Abreu}$^\textrm{\scriptsize 151}$
% ...
\bigskip
\\

% ...
$^{16}$Department for Physics and Technology, University of Bergen, Bergen; Norway.\\
% ...
$^{29}$Physics Department, Brookhaven National Laboratory, Upton NY; United States of America.\\
% ...
$^{35}$$^{(a)}$Facult\'e des Sciences Ain Chock, R\'eseau Universitaire de Physique des Hautes Energies - Universit\'e Hassan II, Casablanca;$^{(b)}$Facult\'{e} des Sciences, Universit\'{e} Ibn-Tofail, K\'{e}nitra;$^{(c)}$Facult\'e des Sciences Semlalia, Universit\'e Cadi Ayyad, LPHEA-Marrakech;$^{(d)}$LPMR, Facult\'e des Sciences, Universit\'e Mohamed Premier, Oujda;$^{(e)}$Facult\'e des sciences, Universit\'e Mohammed V, Rabat;$^{(f)}$Institute of Applied Physics, Mohammed VI Polytechnic University, Ben Guerir; Morocco.\\
% ...
$^{44}$Physics Department, Southern Methodist University, Dallas TX; United States of America.\\
% ...
$^{49}$Fakult\"{a}t Physik , Technische Universit{\"a}t Dortmund, Dortmund; Germany.\\
% ...
$^{55}$II. Physikalisches Institut, Georg-August-Universit\"{a}t G\"ottingen, G\"ottingen; Germany.\\
% ...
$^{103}$CPPM, Aix-Marseille Universit\'e, CNRS/IN2P3, Marseille; France.\\
% ...
$^{121}$Homer L. Dodge Department of Physics and Astronomy, University of Oklahoma, Norman OK; United States of America.\\
% ...
$^{151}$Department of Physics, Technion, Israel Institute of Technology, Haifa; Israel.\\
$^{152}$Raymond and Beverly Sackler School of Physics and Astronomy, Tel Aviv University, Tel Aviv; Israel.\\
% ...
$^{a}$ Also Affiliated with an institute covered by a cooperation agreement with CERN.\\
$^{b}$ Also at An-Najah National University, Nablus; Palestine.\\
$^{c}$ Also at Borough of Manhattan Community College, City University of New York, New York NY; United States of America.\\
$^{d}$ Also at Center for High Energy Physics, Peking University; China.\\
$^{e}$ Also at Center for Interdisciplinary Research and Innovation (CIRI-AUTH), Thessaloniki; Greece.\\
$^{f}$ Also at Centro Studi e Ricerche Enrico Fermi; Italy.\\
% ...

\end{flushleft}

% Created with Glance <Atlas.Glance@cern.ch>